\documentclass[usenatbib]{mn2e}
\usepackage{graphicx}
\usepackage{url}
\usepackage{amsmath}
\usepackage{array} 
\usepackage{threeparttable}
\usepackage{hyperref}
\usepackage{caption3}
\usepackage[all]{hypcap}        
\usepackage[capitalise]{cleveref}
\crefformat{equation}{Eq.~#2#1#3}
\usepackage{soul} 

\usepackage{xcolor} 
\definecolor{darkblue}{rgb}{0,0,0.5}
\hypersetup{
  colorlinks   = true, 
  urlcolor     = blue, 
  linkcolor    = blue, 
  citecolor    = darkblue   
}
\defcitealias{lyman14}{LBJ14}

\newcolumntype{H}{@{}>{\lrbox0}l<{\endlrbox}}
\newcommand{\Ha}{H$\alpha$}

\newcommand{\kms}{km~s$^{-1}$}                         

\newcommand{\simlt}{{\small\raisebox{-0.6ex}{$\,\stackrel{\raisebox{-.2ex}{$\textstyle <$}}{\sim}\,$}}}
\newcommand{\simgt}{{\small\raisebox{-0.6ex}{$\,\stackrel{\raisebox{-.2ex}{$\textstyle >$}}{\sim}\,$}}}   
\newcommand{\iraf}{{\sc IRAF}}
\newcommand{\rsun}{\mbox{\,R$_\odot$}}         
\newcommand{\msun}{\mbox{\,M$_\odot$}}        
\newcommand{\Lbol}{$\mathrm{L}_\mathrm{bol}$}                       
\newcommand{\Mbol}{$\mathrm{M}_\mathrm{bol}$}                       
\newcommand{\mbol}{$\mathrm{m}_\mathrm{bol}$}
\newcommand{\dmbol}{$\Delta m_{15,\mathrm{bol}}$}

\newcommand{\nic}{$\mathrm{^{56}Ni}$}
\newcommand{\cob}{$\mathrm{^{56}Co}$}
\newcommand{\ek}{$E_\mathrm{K}$}
\newcommand{\mej}{$M_\mathrm{ej}$}
\newcommand{\mni}{$M_\mathrm{Ni}$}
\newcommand{\vph}{$v_\mathrm{ph}$} 
\newcommand{\kopt}{$\kappa_{\textrm{opt}}$}
\newcommand{\II}{~{\sc ii}}
\newcommand{\I}{~{\sc i}}
\newcommand{\tpk}{$t_\mathrm{peak}$}
\newcommand{\mpk}{$M_\mathrm{peak}$}
\newcommand{\ks}{K-S test}

\newcommand{\taum}{$\tau_\mathrm{m}$}

\begin{document}

\title[Bolometric light curves of SE~SNe]{Bolometric light curves and explosion parameters of 38 stripped-envelope core-collapse supernovae}

\author[Lyman et al.]
{\parbox{\textwidth}{J. D. Lyman$^{1,2}$\thanks{E-mail: J.D.Lyman@warwick.ac.uk},
		     D. Bersier$^2$,
		     P. A. James$^2$,
		     P. A. Mazzali$^2$,
		     J. J. Eldridge$^3$,
		     M. Fraser$^4$,
		     E. Pian$^{5,6}$}
\vspace{0.4cm}\\
$^1$Department of Physics, University of Warwick, Coventry CV4 7AL, UK.\\
$^2$Astrophysics Research Institute, Liverpool John Moores University, Liverpool, L3 5RF, UK\\
$^3$The Department of Physics, The University of Auckland, Private Bag 92019, Auckland, New Zealand\\
$^4$Institute of Astronomy, University of Cambridge, Madingley Road, Cambridge CB3 0HA, UK\\
$^5$INAF-Istituto di Astrofisica Spaziale e Fisica Cosmica, Via P. Gobetti 101, I-40129 Bologna, Italy\\
$^6$Scuola Normale Superiore di Pisa, Piazza dei Cavalieri 7, I-56126 Pisa, Italy\\
}

\date{Accepted . Received ; in original form }

\pagerange{\pageref{firstpage}--\pageref{lastpage}} \pubyear{2014}

\maketitle

\label{firstpage}

\begin{abstract}
Literature data are collated for 38 stripped-envelope core-collapse supernovae (SE~SNe; i.e.\ SNe IIb, Ib, Ic and Ic-BL) that have good light curve coverage in more than one optical band. Using bolometric corrections derived in previous work, the bolometric light curve of each SN is recovered and template bolometric light curves provided. Peak light distributions and decay rates are investigated; SNe subtypes are not cleanly distinguished in this parameter space, although some grouping of types does occur and there is a suggestion of a Phillips-like relation for most SNe~Ic-BL. The bolometric light curves are modelled with a simple analytical prescription and compared to results from more detailed modelling. Distributions of the explosion parameters shows the extreme nature of SNe~Ic-BL in terms of their \nic{} mass and the kinetic energy, however ejected masses are similar to other subtypes.
SNe Ib and Ic have very similar distributions of explosion parameters, indicating a similarity in progenitors. SNe~IIb are the most homogeneous subtype and have the lowest average values for \nic{} mass, ejected mass, and kinetic energy. Ejecta masses for each subtype and SE~SNe as a whole are inconsistent with those expected from very massive stars. The majority of the ejecta mass distribution is well described by more moderately massive progenitors in binaries, indicating these are the dominant progenitor channel for SE~SNe.
\end{abstract}


\section{Introduction}

Core-collapse supernovae (CCSNe) are the endpoints of massive stars. The observable signatures of these very luminous events exhibit strong heterogeneity, with several main subtypes existing amongst further, poorly known or peculiar events. The main classifications are made based on spectral information. Type II SNe (SNe~II) exhibit strong, long-lasting hydrogen features in their spectra with type Ib SNe (SNe~Ib) being hydrogen deficient and type Ic SNe (SNe~Ic) being hydrogen and helium deficient.\footnote{Although this does not exclude the presence of small amounts of hydrogen and/or helium in the pre-SN progenitors \citep{hachinger12}.} Intermediate to this broad Ibc/II division are type IIb SNe (SNe~IIb), with SN~1993J as the prototypical example of this class. SNe~IIb initially show the strong hydrogen features that warrant a SN~II classification, but subsequently evolve as a SN~Ib after a period of one to a few weeks \citep[see][for a review of the spectral classification of SNe]{filippenko97}. A further designation of `broad-line' (BL) is attached should the observed spectra reveal very large velocities for the ejecta (although there is no exact value for the dividing velocity, BL is generally assigned when a substantial fraction of the ejecta has velocities above that of ordinary SNe Ib/c, i.e. $\simgt 14000$\kms{}). SNe~Ic-BL have gathered much interest in recent years given their association to long-duration gamma-ray-bursts \citep[GRBs; e.g.][see \citealt{hjorth12} for a review]{galama98, stanek03}. The lack of detected GRBs coincident with some SNe~Ic-BL, such as SNe 2002ap \citep[e.g.][]{galyam02}, 2003jd \citep{valenti08}, 2009bb \citep{pignata11} and 2010ah \citep{corsi11,mazzali13}, would suggest relativistic jet formation is not required to power BL SNe; although the case of associated off-axis GRBs is possible \citep{mazzali05}, for one event the geometry of the explosion makes this a less favourable scenario \citep{pignata11}. SNe IIb, Ib, Ic and Ic-BL are considered stripped-envelope SNe (SE~SNe) since they display little to no hydrogen, indicating a massive hydrogen envelope is not present at the point of explosion, in contrast to SNe~IIP.

The various observable signatures of CCSNe are linked to the state of the progenitor star upon explosion. The retention or not of a massive hydrogen envelope is dependent on the amount of mass loss the progenitor experiences during its lifetime. Higher mass stars or stars in binary systems are expected to be able to shed their outer envelopes (and therefore explode as SE~SNe), either via intrinsically high mass loss rates for very high mass stars, or enhanced mass loss due to a binary companion for more modest mass stars. Metallicity and rotation of the progenitor are also likely to increase the mass loss, and hence the observed SN type. Observationally, SNe~II are seen to be explosions from stars at the lower end of the massive star range \citep[$\sim8-16$~M$_{\odot}$; see][for a review]{smartt09}\footnote{The observed upper bound of this mass range may be uncertain due to mistreatment of reddening \citep{walmswell12,kochanek12}. This may reconcile observations with hydrodynamical modelling of SNe~IIP, which generally infer higher masses \citep[e.g.][]{utrobin08,utrobin09,bersten11}.}, which have not suffered sufficient mass loss to remove their hydrogen envelopes. The introduction of binaries as the progenitors of at least a fraction of SE~SNe, as backed up observationally \citep[e.g.][]{maund04,ryder06,smith11b,folatelli14b,fox14} and theoretically \citep[e.g.][]{pod92,pod93,pols02,eldridge08,yoon10,claeys11,eldridge13,benvenuto13,bersten14} would act to wash out any clear distinction between the masses of the progenitors, by allowing much more modestly massive stars to be stripped of their envelopes and making the binary parameters the dominant influence on the observed SN type. The current understanding of mass loss from massive stars is encompassed in the review of \citet{smith14}, which includes a discussion of recent SN studies that favour the majority of SE~SNe arising from binaries, with single, very massive Wolf-Rayet stars contributing a small fraction at most. This is consistent with studies showing most massive-stars are expected to have some form of interaction (mass transfer, merging etc.) with a companion during their evolution \citep[e.g.][]{sana12}; this binary companion is also likely to be a massive star \citep{kobulnicky07,sana11}, exacerbating the dearth of truly single very massive stars.

Direct detection studies in archival imaging can reveal the properties of the SN progenitor \citep[e.g.][]{vandyk03,smartt04,li07,galyam09,vandyk12a,maund11,fraser11,fraser12,vandyk13}, but only for a very limited number of events. The requisite proximity of the SN and existence of pre-explosion deep, high-resolution imaging of the SN location restrict qualifying SN numbers severely and currently there is no confirmed progenitor detection of a SN Ib, Ic or Ic-BL \citep[e.g.][]{maund05,crockett07,eldridge13}. However the studies of \citet{yoon12} and \citet{groh13} show that even massive pre-SN stars are likely to be faint in optical bands, where direct detection studies occur, meaning present limits cannot be used to conclusively rule out very massive progenitors for SNe Ib/c. Although a high mass star (M$_\textrm{ZAMS}\simeq 30-40$~\msun{}) was proposed as the progenitor of the SN~Ib iPTF13bvn \citep{cao13,groh13} based on detections in pre-explosion imaging, the recent works of \citet{fremling14,bersten14} and \citet{eldridge14} argue against a single very massive progenitor, and favour a binary origin with a lower mass progenitor from both the pre-explosion imaging and studies of the SN light curve. Confirmation that the progenitor has faded in post-explosion imaging, possibly revealing the putative binary companion, is awaited at the time of writing to confirm the nature of the progenitor system. 

Due to the severely restricted numbers of direct detection studies, other methods of SN investigation have provided new and complementary insights due to the much larger samples that can be obtained. Analysis of the hosts of SNe \citep[e.g.][]{prieto08,arcavi10,svensson10,kelly14}, the environments and metallicities of SN locations within the hosts \citep[e.g.][]{anderson08,boissier09,anderson10,modjaz11,leloudas11,kelly12,anderson12,crowther13}, the stellar populations surrounding SNe \citep[e.g.][]{kuncarayakti13,williams14}, and SN rates \citep[e.g.][]{li11,smith11b} all provide alternate means of probing the underlying progenitors populations and investigating differences in age, metallicity etc. between the subtypes. Results providing constraints on the ages/masses of SNe, generally to point to a mass sequence of increasing (\emph{average}) initial mass for the SN subtypes of II~$\rightarrow$~IIb~$\rightarrow$~Ib~$\rightarrow$~Ic~$\rightarrow$~Ic-BL, albeit with significant overlap in the mass ranges. Rates of SE~SNe compared to SNe~II appear too high to only arise from the most massive stars, when considering the stellar initial mass function (IMF).

The luminous transient signature of a SN is also a rich source of information about the exploding star, as well as obviously informing on the manner of the explosion itself. Spectra are required in the first instance to type the SN, and, alongside spectral modelling codes \citep[e.g.][]{mazzali93,jerkstrand11} and long term monitoring, they can be used to infer the structure and stratification of the ejecta and reveal bulk parameters of the SN with good precision \citep[e.g.][]{mazzali07,mazzali13,shivvers13,jerkstrand14}. As well as spectral information, the bolometric light curve of a SN is a useful tool for determining the nature of the explosion itself, and is used to correctly scale spectral analyses. Hydrodynamical modelling of SE~SN bolometric light curves can be used to reveal the nature of the exploding stars giving constraints on the physical properties of the star upon explosion and the explosion parameters \citep[e.g.][]{blinnikov98,tanaka09,bersten14,fremling14,folatelli14}.

Unlike host galaxy/environmental analyses, study of the luminous transient signature from SNe is a time critical analysis. In order to properly infer the properties that led to the observed explosion, quick photometric and spectroscopic follow-up must occur (and continue for several months to tightly constrain models). Given the large SN discovery rates currently made possible by dedicated transient surveys (e.g. PTF and iPTF, Pan-STARRS, SkyMapper, La Silla-QUEST), such intense monitoring can only be afforded to a select few of the most interesting and observationally favourable events. This has meant a trade-off exists between studying individual objects with large data sets, or investigating a large number of SNe with poorer follow-up \citep[e.g.][]{hamuy03,drout11}. \citet{cano13} addressed this issue by rescaling optical light curves of SNe Ib/c, Ic-BL and gamma-ray burst-SNe (GRB-SNe) based on their relative peak brightness and light curve width to SN~1998bw, a very unusual SN. These rescaled light curves provided a means to model other SNe by scaling the explosion parameters found from modelling of SN~1998bw; the assumption being made that all SNe in the sample evolve spectrally and photometrically in a similar manner to SN~1998bw. It was found that SNe Ib and Ic are very similar in their explosion properties, with SNe Ic-BL and GRB-SNe exhibiting larger explosion energies and higher ejected masses, pointing to a different progenitor channel for these SNe. They find results in broad agreement with previous studies, indicating scaling relations of SN~1998bw may be applied to other SE~SNe, although the use of average photospheric velocities (see \cref{sect:lit_model}) for over half of the sample of SNe incurs large uncertainties in the results.

Here we utilise a method that allows a SE~SN to be modelled utilising only two-filter optical follow-up and a peak-light spectrum. The bolometric corrections (BCs) to core-collapse SNe of \citet[][hereafter \citetalias{lyman14}]{lyman14} are used to provide a consistent, robust and accurate method of creating fully bolometric light curves for a large number of SE~SNe from their optical colours. An analytical model \citep{arnett82,valenti08} is used to provide estimates of the bulk parameters of the explosions, fitted to each SN individually, which discern the nature of the progenitor system upon explosion.

\cref{sect:method} introduces the sample and describes the method of creating the bolometric light curves and the analytical method employed. Results are then presented and discussed in \cref{sect:results,sect:discuss} respectively.

\section{Method}
\label{sect:method}

\subsection{SN sample and data}
\label{sect:sn_sample}

Bolometric corrections to CCSNe subtypes (SE and SNe~IIP) are presented in \citetalias{lyman14}. The spectral energy distributions of a sample of literature SNe with ultra-violet, optical and near infrared coverage (with corrections for flux outside the observed wavelength regime) were integrated and correlated with optical colours of the SN, to provide an accurate (rms $< 0.1$~mag) method of determining the bolometric light curve of a SN with a polynomial fitted to optical colour.
Since a method of creating bolometric light curves from just optical colours has been formulated, the creation of bolometric light curves is now not limited to those SNe with extended photometric coverage, but rather is possible for a SN with coverage in just two optical bands. Additional to this, distance and reddening determinations are needed in order to convert to luminosity and correct for the effects of dust.

The sample consists of literature SE~SNe that have good light curve coverage in at least two bands from which to construct a colour, which is in turn used to derive the BC. Here the requirement of `good' coverage refers to capturing at least the peak of the light curve (and preferably epochs on the rise) and at least several epochs within the next $\sim$60~days, extending at least 15~days past peak. Additionally, an optical spectrum near peak is required. We restrict our SNe to low redshift in order to avoid the need for $K$-corrections \citepalias[see][]{lyman14}. The sample consists of 9 IIb, 13 Ib, 8 Ic and 8 Ic-BL. The SN names, types, reddening values, distance moduli (to host) and colour(s) used for the BCs are presented in \cref{tab:sn_sample_mod}.

\begin{table*}
\hspace*{3cm}
\begin{threeparttable}  
\centering
     \caption{Data for SNe used to create bolometric light curves.}
  \begin{tabular}{llcccc}
      \hline
SN name  & Type  & $E(B-V)_\mathrm{tot}$ & Distance modulus & Colour used &  Refs.\\
         &       &  (mag)         & (mag)  &             &       \\
\hline
1993J    & IIb   & 0.194          & $27.81\pm0.12$    & $B-I$       & 1--3   \\
1994I    & Ic    & 0.3            & $29.60\pm0.10$     & $B-I$       & 4   \\
1996cb   & IIb   & 0.03           & $31.00\pm0.31$\tnote{a}& $B-R$   & 5   \\
1998bw   & Ic-BL & 0.065          & $32.89\pm0.15$    & $B-I$       & 6   \\
1999dn   & Ib    & 0.10           & $32.95\pm0.11$    & $B-I$       & 7   \\
1999ex   & Ib    & 0.3            & $33.42\pm0.25$    & $B-I$       & 8   \\ 
2002ap   & Ic-BL & 0.09           & $29.50\pm0.14$    & $B-I$       & 9--16   \\
2003bg   & IIb   & 0.02           & $31.68\pm0.14$    & $B-I$       & 17   \\
2003jd   & Ic-BL & 0.144          & $34.46\pm0.20$    & $B-I$       & 18   \\
2004aw   & Ic    & 0.37           & $34.17\pm0.19$    & $B-I$       & 19  \\
2004dk   & Ib    & 0.337          & $31.81\pm0.18$    & $V-R$       & 20 \\
2004dn   & Ic    & 0.568          & $33.54\pm0.16$    & $V-R$       & 20 \\
2004fe   & Ic    & 0.315          & $34.29\pm0.15$    & $V-R$       & 20 \\
2004ff   & IIb\tnote{b}   & 0.302 & $34.82\pm0.16$    & $V-R$       & 20 \\
2004gq   & Ib    & 0.253          & $32.07\pm0.45$    & $V-R$       & 20 \\
2005az   & Ic\tnote{b}    & 0.441 & $32.96\pm0.21$    & $V-R$       & 20 \\
2005bf   & Ib-pec& 0.045          & $34.50\pm0.27$     & $B-V$       & 21   \\
2005hg   & Ib    & 0.685          & $34.67\pm0.15$    & $B-i$       & 22   \\
2005kz   & Ic-BL & 0.514          & $35.30\pm0.15$    & $V-R$       & 20 \\
2005mf   & Ic    & 0.398          & $35.27\pm0.15$    & $B-i$       & 22   \\
2006T    & IIb   & 0.075\tnote{c} & $32.58\pm0.19$    & $B-i$       & 22 \\
2006aj   & Ic-BL & 0.142          & $35.81\pm0.10$    & $B-I$       & 23   \\
2006el   & IIb   & 0.303          & $34.23\pm0.21$    & $V-R$, $V-i$\tnote{d}       & 20 \\
2006ep   & Ib    & 0.035\tnote{c} & $33.84\pm0.20$    & $B-i$       & 22,24   \\
2007C    & Ib    & 0.682          & $31.99\pm0.25$    & $V-R$       & 20 \\
2007Y    & Ib    & 0.112          & $31.36\pm0.14$    & $B-i$       & 25   \\
2007gr   & Ic    & 0.092          & $29.84\pm0.16$    & $B-I$       & 26   \\
2007ru   & Ic-BL & 0.27           & $34.15\pm0.10$    & $B-I$       & 27   \\
2007uy   & Ib    & 0.63           & $32.40\pm0.15$    & $B-V$, $B-I$\tnote{e}       & 28   \\
2008D    & Ib    & 0.6            & $32.46\pm0.15$    & $B-I$       & 29   \\
2008ax   & IIb   & 0.4            & $29.92\pm0.29$    & $B-I$       & 30,31,32   \\
2009bb   & Ic-BL & 0.58           & $33.01\pm0.15$    & $B-I$       & 33   \\
2009jf   & Ib    & 0.117          & $32.65\pm0.10$    & $B-I$       & 34   \\
2010bh   & Ic-BL & 0.507          & $36.90\pm0.15$    & $g-i$       & 35   \\
2011bm   & Ic    & 0.064          & $34.90\pm0.15$    & $B-I$       & 36   \\
2011dh   & IIb   & 0.07           & $29.46\pm0.10$    & $B-I$       & 37   \\
2011hs   & IIb   & 0.17           & $31.85\pm0.15$    & $B-I$       & 38   \\
iPTF13bvn& Ib    & 0.07           & $31.76\pm0.30$    & $B-I$       & 39   \\
\hline
\end{tabular}
\label{tab:sn_sample_mod}
\begin{tablenotes}
 \item [a]{Taken from NED.}
 \item [b]{Updated SN classifications, presented in \citet{modjaz14}, are used.}
 \item [c]{Galactic extinction only.}
 \item [d]{The $V-R$ correction was used for the rise to peak, with $V-i$ used for the peak and decline.}
 \item [e]{The $B-V$ correction was used for early \emph{Swift} data, with $B-I$ used for subsequent ground-based data.}
\end{tablenotes}
\vspace{0.3cm}
References:
(1) \citet{richmond94}; (2) \citet[][and IAU circulars within]{matthews02}; (3) \citet{matheson00};
(4) \citet{richmond96};
(5) \citet{qui99};
(6) \citet{clocchiatti11};
(7) \citet{benetti11};
(8) \citet{stritzinger02};
(9) \citet{mattila02}; (10) \citet{hasubick02}; (11) \citet{riffeser02};(12) \citet{motohara02}; (13) \citet{galyam02}; (14) \citet{takada02}; (15) \citet{yoshii03}; (16) \citet{foley03};
(17) \citet{hamuy09};
(18) \citet{valenti08};
(19) \citet{taubenberger06};
(20) \citet{drout11};
(21) \citet{tominaga05};
(22) \citet{modjaz07};
(23) \citet{mirabal06};
(24) \citet{bianco14};
(25) \citet{stritzinger09};
(26) \citet{hunter09};
(27) \citet{sahu09};
(28) \citet{roy13};
(29) \citet{modjaz09};
(30) \citet{taubenberger11}; (31) \citet{pastorello08}; (32) \citet{tsvetkov09};
(33) \citet{pignata11};
(34) \citet{valenti11};
(35) \citet{olivares12};
(36) \citet{valenti12};
(37) \citet{ergon14};
(38) \citet{bufano14}
(39) \citet{fremling14}.
\end{threeparttable}
\end{table*}

Photometric data were extracted from the literature for SNe in the sample (see references in \cref{tab:sn_sample_mod}). The same methods of light curve dereddening and interpolation were employed as in \citetalias{lyman14}, in order to obtain values of simultaneous observations in the chosen filters, which give the colour. Extrapolations were not used for this analysis.

The reddening-corrected values of the chosen colour (see \cref{tab:sn_sample_mod}) were then fed into the BC polynomial fits of \citetalias{lyman14} in order to recover the bolometric light curve. Some SNe were observed in a combination of Johnson-Cousins and Sloan optical filters -- new BCs for these combinations were calculated following the method and data of \citetalias{lyman14}, the parameters for these new fits can be found in \cref{tab:new_BC}, with the corresponding fits for SNe~II also presented for interest. 

The resulting BC was then applied to the appropriate SN light curve (e.g. for colour $B-I$, the BC is applied to the $B$-band light curve). Using the distance modulus we can convert \mbol{} to \Mbol{} and finally to \Lbol{}.
For clarity in plotting, nearly contemporaneous data have been combined by averaging any epochs within 0.2~days of each other.

The results provide the largest sample of bolometric light curves for SE~SNe thus far, on which a simple analytical model can be applied, in order to extract estimates for the explosion parameters.

\begin{table*}
 \centering
 \caption{BCs for new filter combinations following method of \citetalias{lyman14}.}

  \begin{tabular}{cc@{\hskip 1.1cm}crrrr@{\hskip 1.1cm}crrrr}
  &   & \multicolumn{5}{c}{SE~SNe} & \multicolumn{5}{c}{SNe~II} \\ 
$x$ & $y$ & $x-y$ range &{$c_0$} & {$c_1$} & {$c_2$} &  {rms}& $x-y$ range & {$c_0$} & {$c_1$} & {$c_2$} & {rms} \\ 
\hline
$B$ & $i$ & $-$0.581 to 1.769 & $-0.186$ & $-0.412$ & $-0.172$ & 0.061 &  $-$0.392 to 2.273 & $-0.155$ & $-0.450$ & $-0.167$ & 0.023 \\
$V$ & $i$ & $-$0.933 to 0.504 & $0.095$  & $-0.320$ & $-0.102$ & 0.093 &  $-$0.391 to 0.658 & $0.181$  & $-0.212$  & $-1.137$ & 0.044 \\ 
\hline
\label{tab:new_BC}
\end{tabular}
\end{table*}

\subsection{The analytical model}
\label{sect:lit_model}

The analytical model is based on that of \citet{arnett82},
and is appropriate for SE~SNe, where the light curve is powered predominantly by the decay of \nic{} (i.e. excluding interacting SNe). 
The model is fit over the photospheric phase, during the optically thick phase of the ejecta (up to $\sim$ 1--2 months after explosion, depending on the evolution-speed of the SN).
The bolometric output is described by the model, and as such the model should be fitted to a bolometric light curve, with an additional constraint required in the form of a characteristic velocity of the ejecta (\cref{sect:scale_vely}). From this simple analytical fitting, estimates of the mass of nickel synthesised (\mni{}) and the mass and kinetic energy of the ejecta (\mej{} and \ek{}, respectively) can be made.

Naturally an analytical approximation requires some simplifying assumptions. These are listed in \citet{arnett82,valenti08}, with brief discussion given here (see also discussion in \citealp{cano13}).

\begin{itemize}
\item The radius at is small at the onset of explosion. Although this is appropriate for most SE~SNe progenitors, which have radii $\sim$~\rsun{}, up to $\sim 10$~\rsun{} (but see \citealt{yoon10}), it may not be appropriate for some cases where an extended, low-mass envelope is present (e.g.\ SN~1993J, 2011hs), which can affect the light curve shortly after explosion. To minimise the impact of this very early signature on the overall light curve model, very early data are not fit in our method.
 \item Homologous expansion with spherical symmetry (\mbox{$V \propto R$}). SE~SNe show evidence for some degree of asphericity of the ejecta, as gleaned from double-peaked nebular emission features \citep{maeda08}. 
 \item A constant opacity (\kopt{}). In reality this is dictated by the density and composition of the ejecta, and should therefore evolve with time. Here, \kopt{} is set to be 0.06~cm$^2$~g$^{-1}$ \citep[e.g.][]{maeda03,valenti11}.
 \item Centrally concentrated \nic{}. The amount of mixing will affect the rise time of the SNe since radiation from high-velocity (i.e.\ further out in radius, given homologous expansion) \nic{} will have a shorter diffusion time \citep{maeda03,dessart12}. 3D modelling has shown that a small fraction of high-velocity \nic{} is not uncommon in SNe, although the bulk is generally located close to the centre \citep[e.g.][]{hammer10}.  
 \item The decay of \nic{} and \cob{} power the light curve. These radioactive isotopes represent the main source of energy of a SE~SN, dominating the luminosity evolution for many months.
\end{itemize}

In the model, the photospheric-phase luminosity of a SN is described by equation A1 of \citet{valenti08}, an update of the original \citet{arnett82} model. The equation is fitted with two light curve parameters, \mni{} and \taum{}, which is given by:
\begin{equation}
 \tau_\mathrm{m}^2 = \frac{2\kappa_\mathrm{opt} M_\mathrm{ej}}{\beta c v_\mathrm{sc}}\textrm{\! .}
 \label{eq:model_taum}
\end{equation} 
$\beta \simeq 13.8$ is a constant, and $c$ is the speed of light. $v_\mathrm{sc}$ is the \emph{scale velocity} of the SN, which is observationally set as the photospheric velocity at maximum light (\vph{}). As also noted by \citet{wheeler15}, the initial relation for the scale velocity given by \citet[][eq. 54]{arnett82} is incorrect due to a typographical error -- this error has been carried over into some subsequent studies using the model. With the simplifying assumptions of a constant density sphere undergoing homologous expansion, the relation should be
\begin{equation}
 v_\mathrm{sc}^2 \equiv v_\mathrm{ph}^2 = \frac{5}{3}\frac{2E_\mathrm{K}}{M_\mathrm{ej}}\textrm{\! .}
 \label{eq:vsc}
\end{equation}
The direct relation between \taum{}, \mej{} and \ek{} is therefore given by
\begin{equation}
 \tau_\mathrm{m} = \left(\frac{\kappa_{\mathrm{opt}}}{\beta c}\right)^{0.5}\left(\Lambda\frac{M_{\mathrm{ej}}^3}{E_\mathrm{k}}\right)^{0.25}\textrm{\! ,}
 \label{eq:model_taum_m_e}
\end{equation}
where $\Lambda = 6/5$. In \citet{valenti08}, $\Lambda$ is incorrectly given as $10/3$, through propagation of the incorrect numerical factor (cf. their equation 2 and \cref{eq:vsc} -- note that an exponent of two is also missing on \vph{} in their equation). Although \mej{}, determined from the photospheric phase, is related directly from the observed values of \vph{} and \taum{}, and thus not affected by this discrepancy, calculating \ek{} is. Where the typographical error is propagated in the literature, estimates of \ek{} from the model should be revised down by a factor $25/9$ ($=\frac{5}{3}$/$\frac{3}{5}$).
However, this then results in a change (increase) in the ratio \mej{}/\ek{} determined during the photospheric-phase. 
This further means that results from the two-zone \emph{nebular phase} model ($>$60~days post-explosion; see \citealt{valenti08} section 9 and appendix A and references therein for a description of this model) for SE~SNe is affected. The nebular phase model accounts for a high-density inner component of the ejecta that is not emergent during the photospheric phase, \emph{in addition} to the contribution of the ejecta component determined by the photospheric model. 
The timescale of decay for the model is dictated by \mej{}/\ek{} \citep[][eqs A10 and A11]{valenti08}. The equations in the nebular model are separate from those of the photospheric model and thus their form is not explicitly affected by the typographical error. However, since a component of the nebular model is reliant on \mej{}/\ek{} determined during the photospheric phase, the effects of the typographical error affect the fitting of the nebular phase model.
We do not consider the nebular component here as we could not obtain satisfactory fits using the corrected formula. The primary issue was that the nebular phase model remained too bright (even when neglecting a further contribution from the inner, high-density component). This arises since an increase in \mej{}/\ek{} means the timescale for the evolution of the incomplete trapping of gamma rays becomes extended. For this reason we take only values for \mni{}, \mej{} and \ek{} derived from the photospheric-phase.

The photospheric phase of each light curve was fitted for \mni{} and \taum{} (\cref{eq:model_taum}). \taum{} was decomposed to \mej{} and \ek{} using \vph{} of the SN and \cref{eq:vsc}. Photospheric data were only fit beginning $\simlt$10~days before peak, determined by early light curve coverage. When fitting the model, data prior to 10~days before peak are not included in the fit since the assumptions in the model may not be appropriate. For example, an extended envelope can imprint on the rising light curve through cooling emission, post shock-breakout. By restricting the time-range fitted we become largely insensitive to the progenitor radius, and model only the \nic{} and \cob{} powered emission, but we note that the contribution of any thin, extended envelope would not be included in our resulting parameters. The fitting of the analytical functions to the light curve data was done via the {\sc curve\_fit} function in {\sc scipy}.

\subsubsection{Determining scale velocity of SNe}
\label{sect:scale_vely}

SNe exhibit strong P-Cygni line profiles in their spectra due to the fast moving ejecta. This causes absorption that is blue-shifted by the velocity of the absorbing material relative to the rest wavelength of the spectral line. Due to the stratification of the ejecta and homologous expansion, elements towards the outer layers of the ejecta (e.g. helium and calcium) can exhibit large velocities compared to heavier, more centrally concentrated elements. 
Two elements chosen to better trace the photospheric velocity are silicon and iron, with Si\II{} $\lambda$6355 and the Fe\II{} set of lines clustered around 4500-5200~\AA{} used. 

Practically, the measurements of \vph{} consist of a simple Gaussian fitting procedure to the absorption features of a wavelength- and flux-calibrated spectrum of the SN taken on or near peak, obtained from WISeREP \citep{yaron12}\footnote{\url{http://www.weizmann.ac.il/astrophysics/wiserep/}} and \citet{modjaz14}, performed in the \iraf{} package {\sc splot}. This was performed for individual Fe\II{} lines (Fe\II{} $\lambda$4924, Fe\II{} $\lambda$5018, Fe\II{} $\lambda$5169), before averaging these values to obtain a value of \vph{}. In the case where Fe\II{} lines could not be accurately measured (e.g. strong line blending or no spectral coverage at those wavelengths), the Si\II{}~$\lambda$6355 feature was measured. 
When data were not available for a SN (i.e. we could not obtain a spectrum to analyse), we relied on values for \vph{} that were found by other authors in the literature. These literature \vph{} values were typically found using a similar Gaussian fitting technique or through spectral fitting. Measurements from this simple Gaussian-fitting method were found to agree well with those from detailed spectral fitting codes in the cases where a comparison was possible. Both methods have uncertainties of $\sim$ a thousand \kms{}, which we take as a fiducial minimum uncertainty on our measurements (an additional uncertainty based on the epoch of the spectrum relative to peak is also added, see \cref{sect:uncert}). This uncertainty arises from the data quality as well as the broad-featured characteristic of SNe spectra at maximum light -- velocities of at least several thousands of \kms{}, as is seen for CCSNe, make line blending an issue and it is often the case that one cannot attribute a single absorption feature to one specific transition. Similarly, nearby emission from other transitions will also impact on the shape of the absorption feature, affecting the Gaussian fit and, ultimately, the photospheric velocity derived. 
Finally, in the absence of appropriate data or literature value for a SN, as was the case for SN~2005kz, \vph{} was taken to be the mean \vph{} for the SN's type from the rest of the sample. The average \vph{} values are presented in \cref{tab:sn_av_mod}. \citet{branch02} provide a power law fit to their estimates of the variation with time of \vph{} values for SNe~Ib (as determined from Fe\II{} lines). The value at peak of this power law,$\sim$ 9000~\kms{}, is in good agreement with the average SN~Ib \vph{} of 9900~\kms{} found here, given the uncertainties. The values of \mej{} and \ek{} derived for SN~2005kz using this average \vph{} are clearly susceptible to a larger systematic uncertainty, but the \nic{} mass is unaffected.

\section{Results}
\label{sect:results}

\subsection{Bolometric light curves}
\label{sect:bololc}

\begin{figure*}
\centering
 \includegraphics[width=0.85\textwidth]{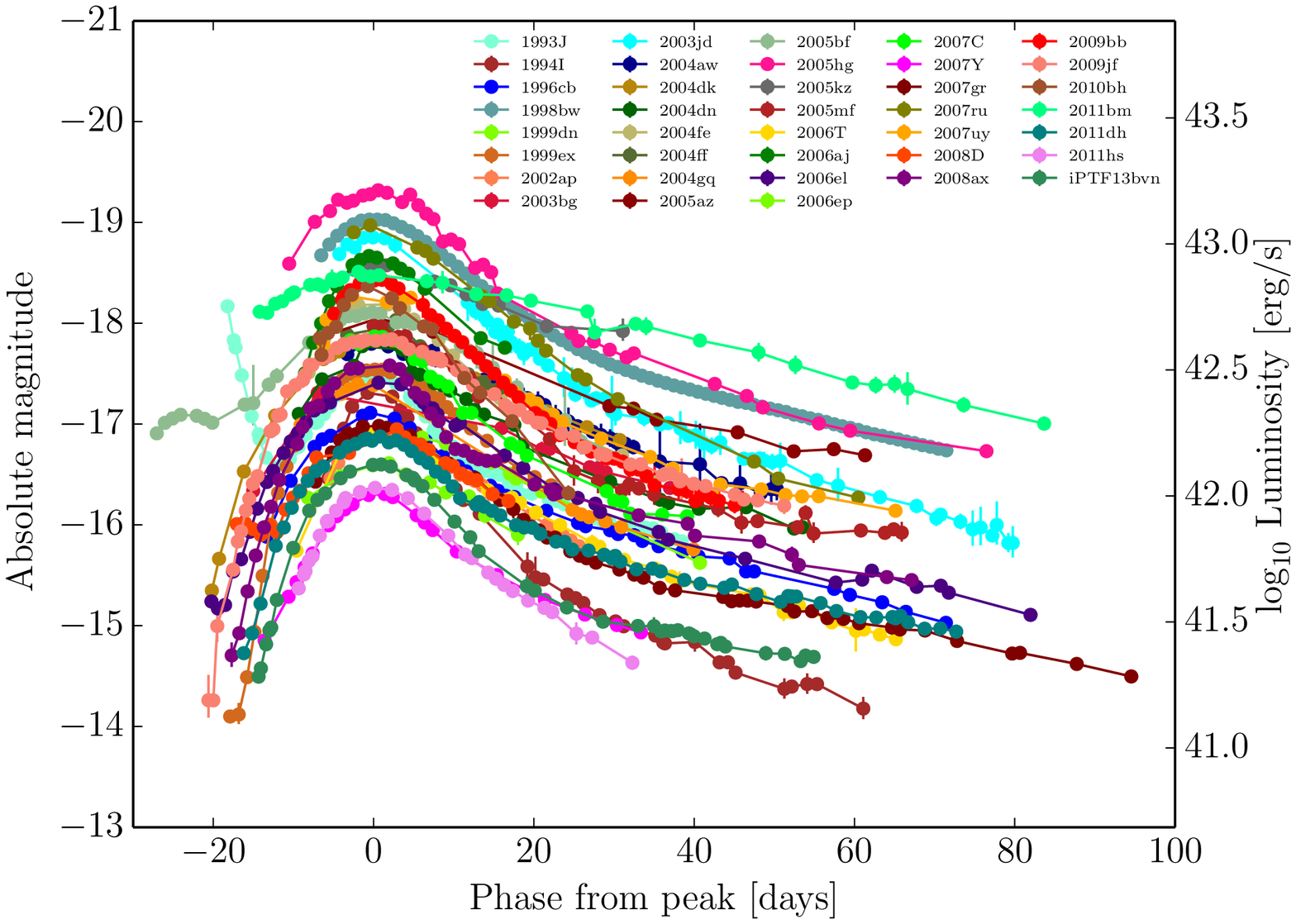}\\
 \includegraphics[width=0.85\textwidth]{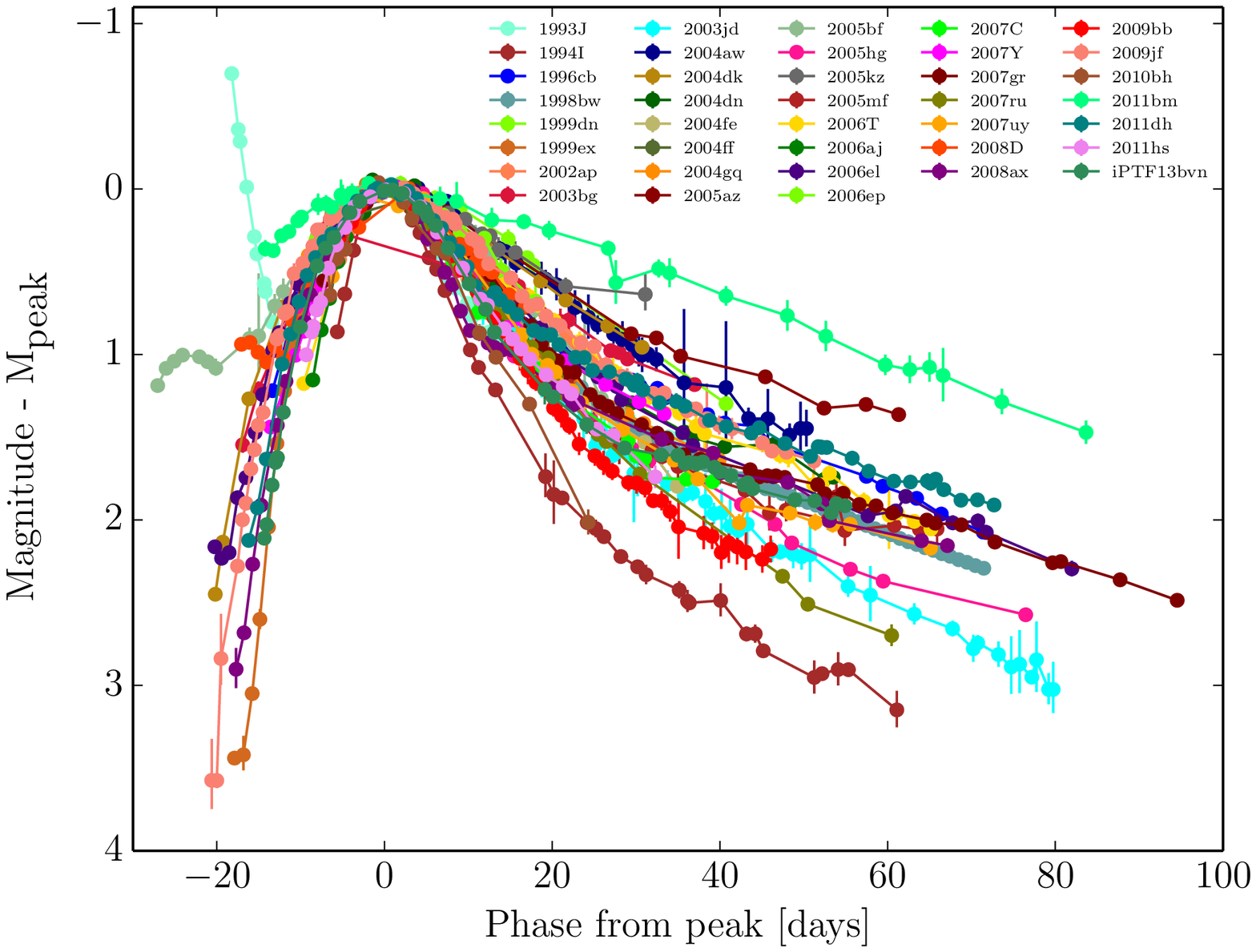}
\caption{Bolometric light curves of SE~SNe (top). The peak-normalised light curves are also displayed (bottom). Error bars are indicative of the uncertainty of the BC only, which is found by taking the uncertainty in the colour and translating that as an error on the BC fits. Distances, for example, will be a source of uncertainty in the top plot.}
\label{fig:sn_bololc}
\end{figure*}

The bolometric light curves are presented in \cref{fig:sn_bololc}. The time and peak of each light curve (\tpk{} and \mpk{}, respectively) were found with a low-order polynomial, fitted to data around peak. The diversity of SE~SNe becomes apparent from these plots. 
A roughly continuous spread over $\sim3$~magnitudes is observed in \mpk{}; interestingly, two SNe~Ib encompass the extremes of the spread, ranging from SN~2007Y\footnote{The classification of SN~2007Y as a Ib has been questioned by \citet{maurer10,folatelli14}, where detections of \Ha{} would favour a IIb classification} at \mpk{}$\sim-16.3$ to SN~2005hg at \mpk{}$\sim-19.2$.\footnote{The classification of SN~2005hg was originally made as a SN~Ic \citep{cbet267}, before the detection of He lines by \citet{cbet271}.} This spread in \mpk{} is similar to that found by \citet{drout11} in their $V$- and $R$-band light curves when considering the overlapping sample. It should be noted that the photometry of SNe taken from \citet{drout11} may be systematically brighter than the intrinsic brightness of the SN since host-subtraction is not performed \citep{bianco14}, however we include only photometry from their `gold' sample here, which appear more in agreement with host-subtracted photometry \citep{bianco14}, and remove very late time data for SNe 2004dn and 2005mf, where flattening of the light curves was observed. As such, we consider this potential contamination to not significantly affect results or discussion. 

The decline rates of the sample vary greatly; the speed of the evolution is parameterised by \dmbol{}, which is the number of magnitudes from peak the bolometric light curve has declined by 15~days after peak (values were found from polynomial fits to the light curves). SN~1994I, despite being often cited as a `prototypical' SN~Ic, has unusually fast evolution, as has been previously noted, with \dmbol{}$= 1.37$ calculated here. SN~2011bm displays the slowest evolution, with \dmbol{}$ = 0.20$. The evolution speeds appear to form a continuum, as is evident from the bottom panel of \cref{fig:sn_bololc}, although SNe 1994I and 2011bm are noticeably displaced from the extremities of the distribution. Perhaps unexpectedly, a XRF-SN, SN~2010bh, is exceeded only by SN~1994I in terms of speed of evolution. This extremely fast evolution was noticed by \citet{cano11} and \citet{olivares12}, but is highlighted here when compared to many other SE~SNe. Such fast evolution is at odds with the perception of GRB/XRF-SN progenitors being very massive when considering the analytical form of SE~SN light curves, since the timescale of the evolution is directly related to \mej{} (\cref{eq:model_taum}). The \dmbol{} values here are similar to the spread of preliminary values found for $V$- $r$- and $i$-bands by \citet{bianco14} \citep[see also][]{walker14}, whereas the values they find for bluer (redder) filters are systematically larger (smaller) than the average value for the bolometric light curves, indicative of the relative decline rates of these individual bands when compared to the bolometric light curve. No statistical distinction of the various subtypes can be made in \dmbol{}, as was shown for the $V$- and $R$-band by \citet{drout11}. 

\begin{figure}
\centering
 \includegraphics[width=\columnwidth]{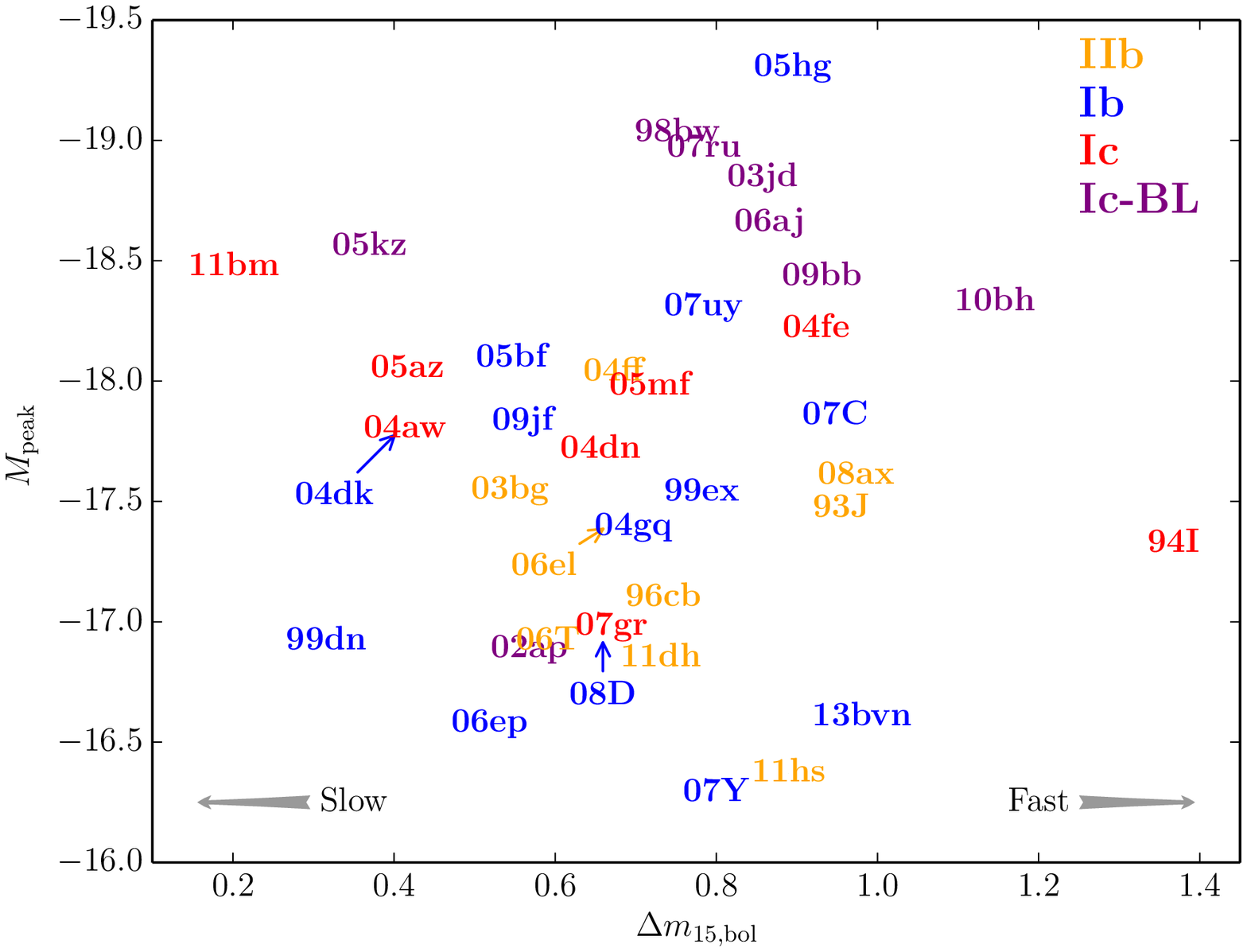}\\
\caption[Evolution speed versus peak for SE~SNe]{The peak magnitude of the evolution of the bolometric light curves against speed (parameterised as \dmbol{}, see text); the direction of light curve evolution speed is denoted by the labelled grey arrows. The thin, coloured arrows indicate the true positions of SNe 2004dk, 2006el and 2008D, which have been offset for clarity. SNe are colour-coded according to their type. }
\label{fig:m15peak}
\end{figure}

To investigate any possible correlation between light curve peak and decline rate in SE~SN bolometric light curves, \dmbol{} values are plotted against \mpk{} in \cref{fig:m15peak}. There appears to be a dearth of bright, slowly evolving and fast-and-faint SN, but the Spearman's rank coefficient (0.633) is not significant enough to reject the case of no correlation. There appears to be no reason why bright, slowly evolving SNe would be missed in surveys compared to quicker evolving events at similar luminosities, and it may indeed indicate that events such as SNe~2005kz and 2011bm are intrinsically rare.\footnote{Another slowly-evolving literature example is SN~1997ef -- a very energetic SN~Ic-BL \citep{mazzali00}, however this SN was not particularly bright, with $M_R \sim -17.2$ \citep{iwamoto00}. It is omitted in this study since a bolometric light curve could not be constructed using the BC method presented.} Conversely, fainter, quickly evolving SNe (such as events similar to SNe 2005ek, \citealt{drout13}, and 2010X, \citealt{kasliwal10})  are most likely to have been missed from detection (particularly prior to peak, which is one of the criteria imposed on this sample). As this is a literature based sample, the selection effects cannot be analysed beyond these qualitative statements. 
The time and value of the light curve peak and \dmbol{} value are listed for each SN in \cref{tab:sn_lcparams}.

\begin{table}
\centering
  \caption{Bolometric light curve properties for SE~SNe.}
\begin{tabular}{lccc}
SN        & \tpk{}   &  \mpk{} &\dmbol{}\\
          &  (MJD)   &  (mag)  & (mag)  \\
\hline
1993J     & 49094.4 &  -17.5 &    0.96\\
1994I     & 49450.1 &  -17.3 &    1.37\\
1996cb    & 50061.2 &  -17.1 &    0.73\\
1998bw    & 50944.0 &  -19.0 &    0.75\\
1999dn    & 51418.2 &  -16.9 &    0.32\\
1999ex    & 51499.0 &  -17.5 &    0.78\\
2002ap    & 52312.1 &  -16.9 &    0.57\\
2003bg    & 52717.0 &  -17.5 &    0.54\\
2003jd    & 52942.5 &  -18.8 &    0.86\\
2004aw    & 53088.6 &  -17.8 &    0.41\\
2004dk    & 53239.5 &  -17.8 &    0.41\\
2004dn    & 53229.7 &  -17.7 &    0.66\\
2004fe    & 53318.3 &  -18.2 &    0.92\\
2004ff    & 53313.6 &  -18.0 &    0.67\\
2004gq    & 53361.4 &  -17.4 &    0.70\\
2005az    & 53473.9 &  -18.1 &    0.42\\
2005bf    & 53497.8 &  -18.1 &    0.55\\
2005hg    & 53681.7 &  -19.3 &    0.89\\
2005kz    & 53710.5 &  -18.6 &    0.37\\
2005mf    & 53733.4 &  -18.0 &    0.72\\
2006T     & 53780.0 &  -16.9 &    0.59\\
2006aj    & 53793.8 &  -18.7 &    0.87\\
2006el    & 53983.6 &  -17.4 &    0.67\\
2006ep    & 53988.5 &  -16.6 &    0.52\\
2007C     & 54117.3 &  -17.9 &    0.95\\
2007Y     & 54163.6 &  -16.3 &    0.80\\
2007gr    & 54337.4 &  -17.0 &    0.67\\
2007ru    & 54439.1 &  -19.0 &    0.78\\
2007uy    & 54477.4 &  -18.3 &    0.78\\
2008D     & 54492.5 &  -16.9 &    0.66\\
2008ax    & 54548.8 &  -17.6 &    0.97\\
2009bb    & 54920.9 &  -18.4 &    0.93\\
2009jf    & 55120.6 &  -17.8 &    0.56\\
2010bh    & 55279.3 &  -18.3 &    1.15\\
2011bm    & 55677.2 &  -18.5 &    0.20\\
2011dh    & 55732.1 &  -16.9 &    0.73\\
2011hs    & 55888.8 &  -16.4 &    0.89\\
iPTF13bvn & 56475.1 &  -16.6 &    0.98\\
\end{tabular}
\label{tab:sn_lcparams}
\end{table}

It is clear the various SN types do not inhabit exclusive regions of this parameter space, although some clustering of SNe~IIb and SNe~Ic-BL (with two exceptions) occurs. All SNe~IIb in the sample occur within a small region of roughly average evolution speed and have modest-to-low peak magnitudes when compared to the entire sample. The decline rates of SNe~IIb were noted to be distinct from the more slowly declining SNe~IIP and IIL in the $R$-band by \citet{arcavi12}, and indeed found to be similar to those of SNe~Ib and Ic. This is confirmed here for the bolometric light curve decline rates also. SNe~Ic-BL are all luminous when compared to the rest of the sample with the exception of SN~2002ap, and all have fast evolution with the exceptions of SNe~2002ap and 2005kz.

\begin{figure}
\centering
\includegraphics[width=0.8\columnwidth]{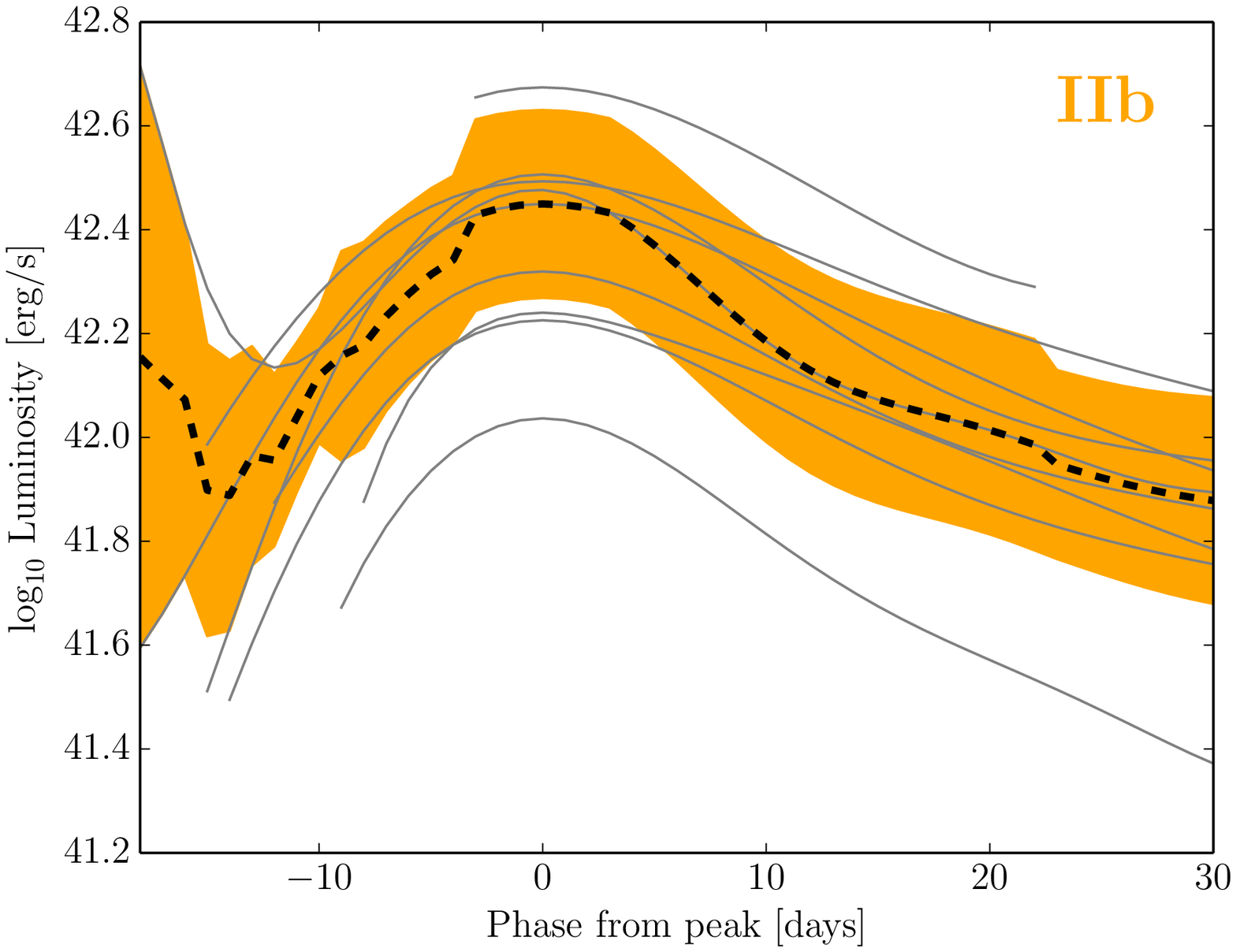}\\
\includegraphics[width=0.8\columnwidth]{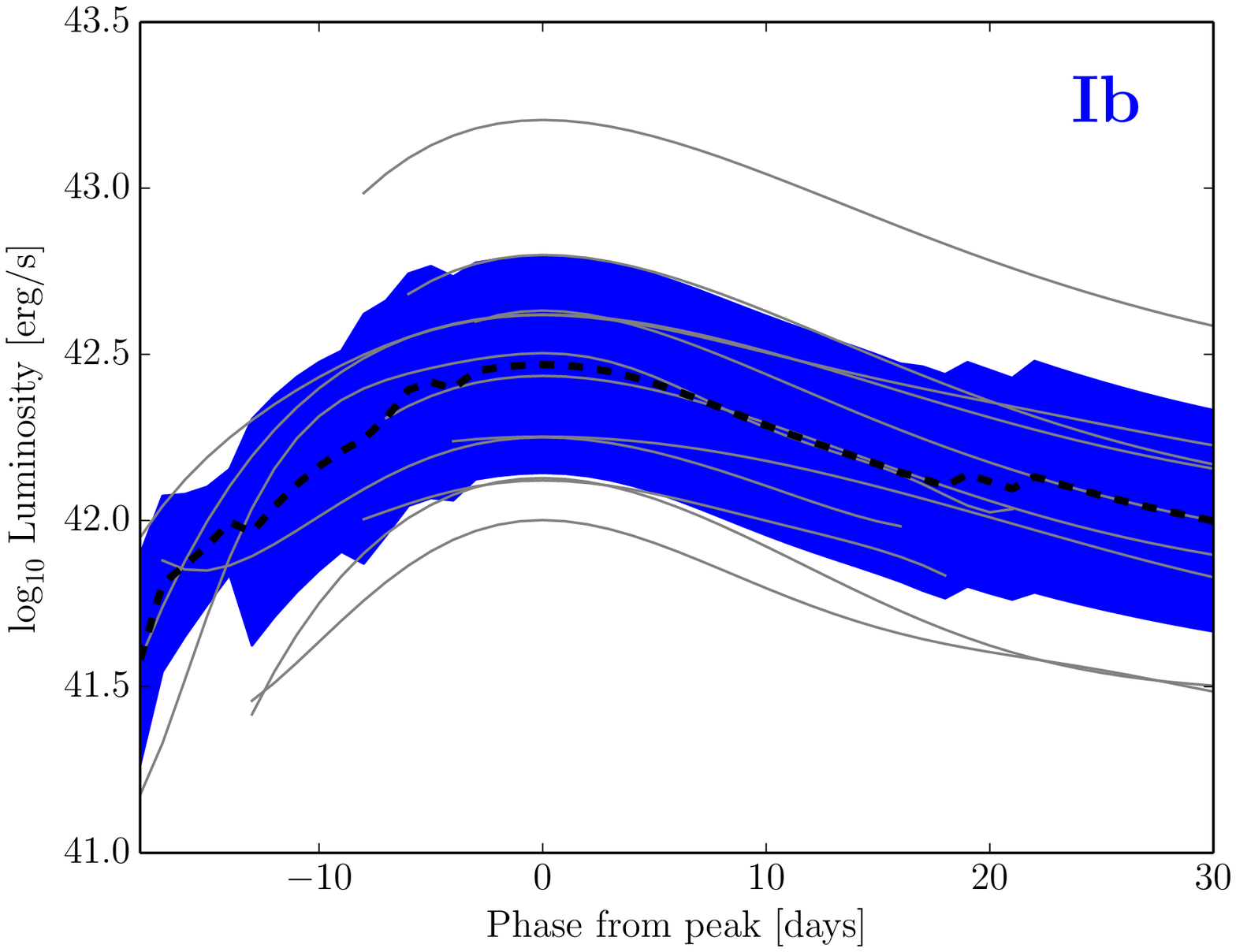}\\
\includegraphics[width=0.8\columnwidth]{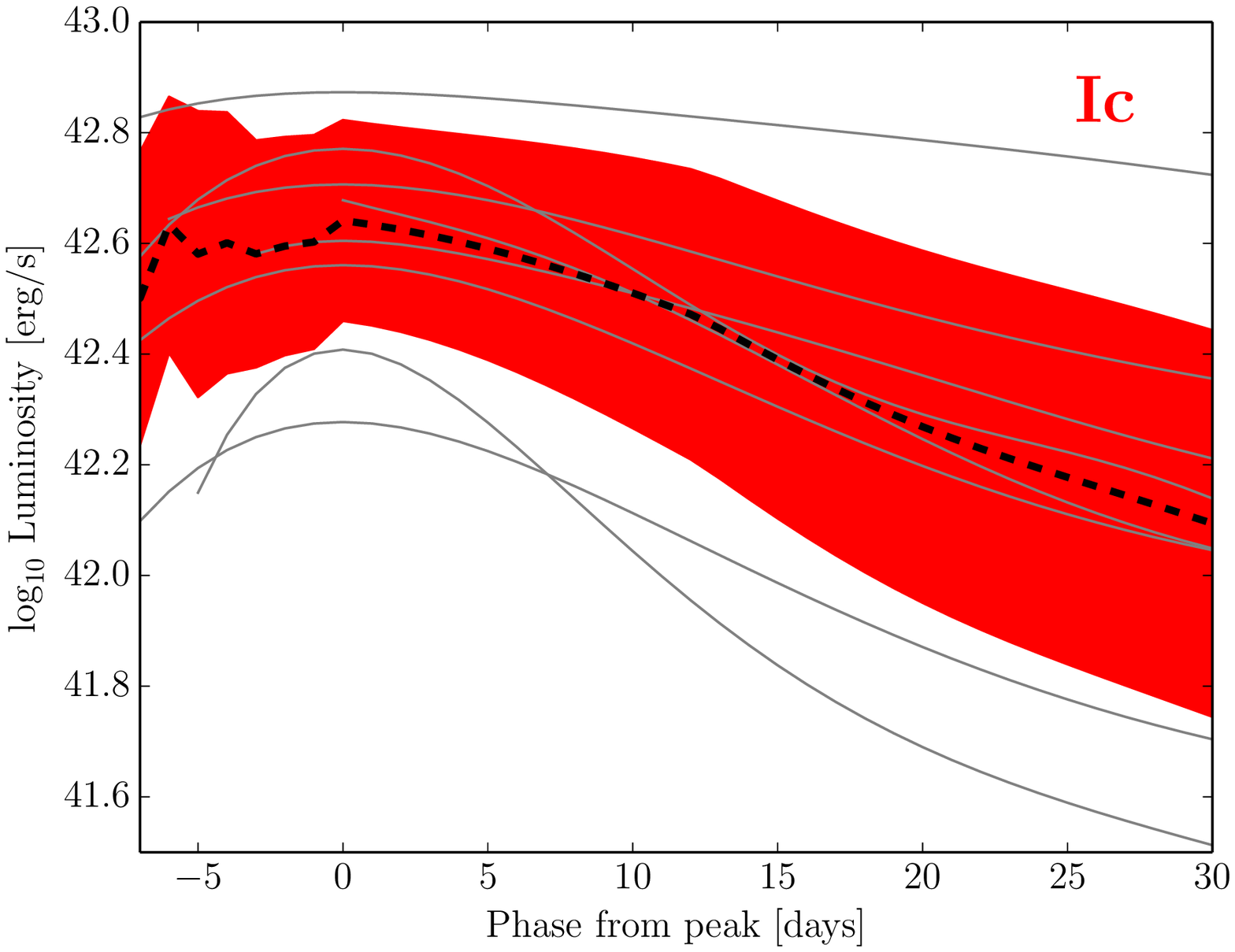}\\
\includegraphics[width=0.8\columnwidth]{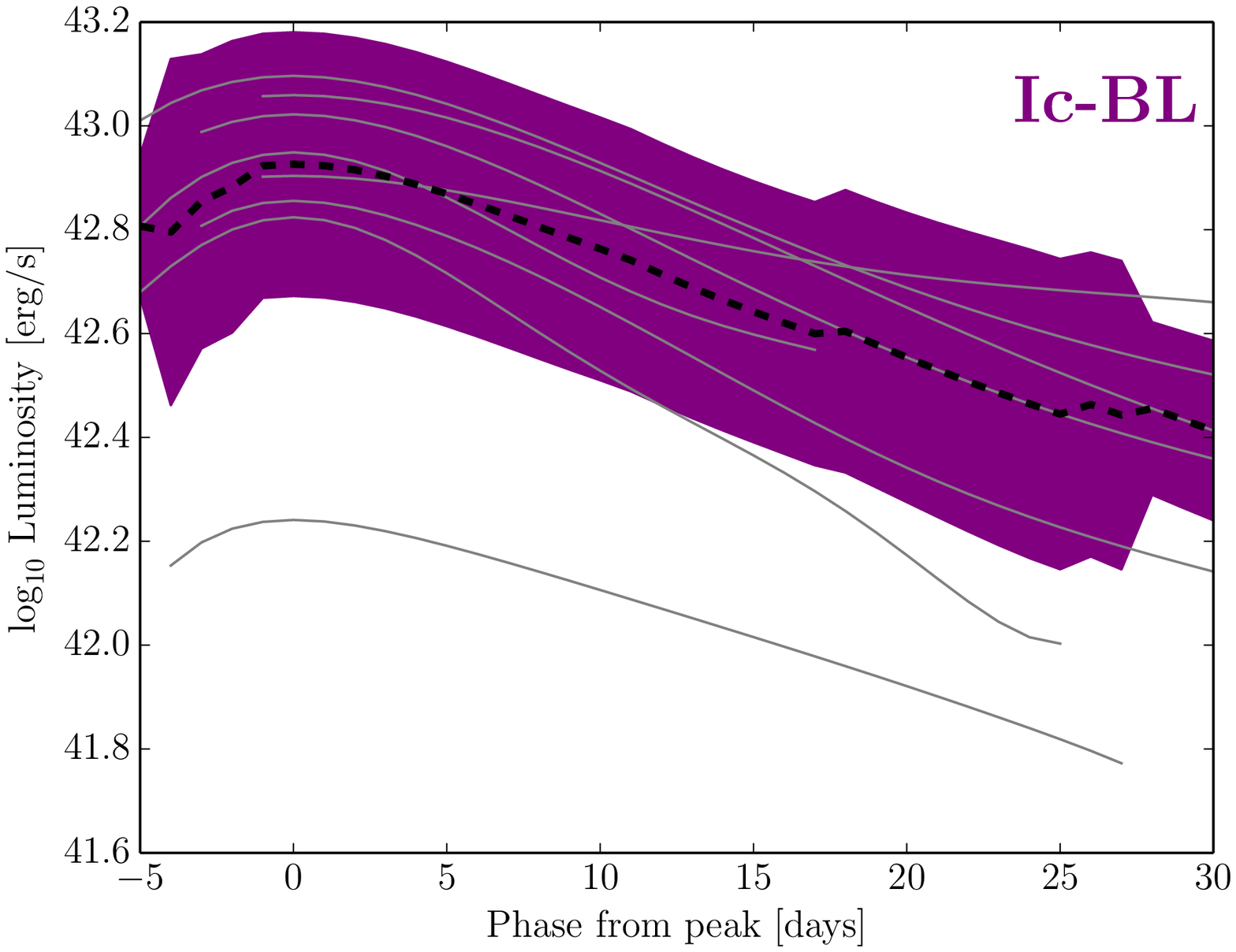}\\
\caption[Template bolometric light curves for SE~SNe]{Template bolometric light curves for SE~SNe, indicating the median value (black dashed line) and the standard deviation of the sample of light curves for that phase (coloured filled regions). Individual light curves are shown as thin solid lines. }
\label{fig:lc_templates}
\end{figure}

Template bolometric light curves for SE~SNe are presented in \cref{fig:lc_templates}, with dashed lines showing the median value and the coloured regions the standard deviation. The data for these templates are presented in \cref{tab:sn_temp} with the phases being relative to the peak of \Lbol{}. These were found by sampling interpolations of the bolometric light curves and calculating the median and standard deviation of those SNe with a light curve covering that particular phase. (The median traces exhibit some mildly erratic behaviour due to the relatively small samples and the limited and non-uniform temporal coverage of the samples.) The templates reveal notable outliers in each case, emphasising the heterogeneity of events even in these well-established SN types.

\subsection{Photospheric velocities}
\label{sect:vphot}

The \vph{} values are presented in \cref{tab:sn_vph_mod}. These values were used to break the degeneracy in \ek{} and \mej{} (\cref{sect:lit_model}). Velocity determinations were found to agree well with literature values that were determined from both the Gaussian-fitting technique and also spectral modelling. Where linear interpolations of \vph{} were relied upon between epochs to obtain an estimate at peak, these were found to differ from using a \citet{branch02} power law by less than our assumed errors, reaching $\sim$~700~\kms{} in the worst case for SN~2006T, but more typically $\sim$100-200~\kms{}. A correlation between the $V$-band magnitude and \vph{} 50~days after explosion was found for SNe~IIP by \citet{hamuy03}, whereas no correlation of \vph{} with peak \Lbol{} was found in the analysis of SE~SNe here (this is discussed in \cref{sect:correlations} in the form of \mni{} against \vph{}). Additionally, no correlation between \vph{} and light curve decline rate (\dmbol{}) was found here. Note that SN~2007uy was found to have a high \vph{}, on the border of BL regime, however the spectrum analysed was taken 3 days prior to peak and therefore has a more uncertain lower limit (as discussed later), and is thus considered a SN~Ib in this study.

\begin{table*}
\begin{threeparttable}
     \caption{Photospheric velocity measurements for SE~SNe}
       \begin{tabular}{llcrcp{9.5cm}}
      \hline
SN name  & Type  & Line(s) used & \vph{}    & Phase\tnote{a} & Notes \\
         &       &              &   (\kms{})  & (days)                       &       \\
\hline
1993J    & IIb   &   Fe\II{}      & 8000$\pm{1000}$      &     0                   & Agrees with Fe\II{} velocities found by \citet{ohta94,barbon95,pastorello08}.      \\
1994I    & Ic    &   Fe\II{}      & 11500$\substack{+1000 \\ -1400}$     &     $-1$                & Agrees with spectral modelling value of \vph{} in \citet{sauer06} and Fe\II{} velocity in \citet{clocchiatti96}.      \\
1996cb   & IIb   &   Fe\II{}      & 8500$\substack{+1300 \\ -1000}$      &     1                   & \citet{qui99}, however, find a velocity of He\I{} of 8870\kms{} at epoch $-$15~days which would indicate extremely low \vph{} at peak. Inconsistency was found when measuring velocities on the same spectrum, and the Fe\II{} measurement found here is preferred.\\
1998bw   & Ic-BL &   Si\II{}      & 19500$\substack{+1700 \\ -1000}$     &     1                   & Fe\II{} lines are largely blended. \vph{} agrees with value found by \citet{patat01} and is similar to Si\II{} velocity found by \citet{pignata11}.      \\                                          
1999dn   & Ib    &   Fe\II{}      & 10500$\pm{1000}$     &     0                   & Taken from \citet{benetti11} \\                                            
1999ex   & Ib    &   Fe\II{}      & 8500$\substack{+1300 \\ -1000}$      &     1                   & Agrees with velocities given in \citet{hamuy02}  \\                                             
2002ap   & Ic-BL &   Fe\II{}      & 13000$\substack{+2000 \\ -1000}$     &     2                   & \citet{galyam02} find the velocity of Si\II{} to be 15000\kms{} at peak. Features blended somewhat.  \\                                             
2003bg   & IIb   &   (Fe\II{})    & 8000$\pm{1000}$      &     0                   & The value of \vph{} from the spectral modelling of \citet{mazzali09} is used, this is consistent with a Fe\II{} velocity found from a spectrum near peak.   \\                                           
2003jd   & Ic-BL &   Si\II{}      & 13500$\pm{1000}$     &     0                   & Taken from \citet{valenti08}.  \\                                  
2004aw   & Ic    &   Fe\II{}      & 11000$\substack{+1000 \\ -1900}$     &     $-$2                & \citet{taubenberger06} show a contemporaneous Si\II{} velocity of 12500\kms{}. \\                                             
2004dk   & Ib    &   Si\II{}      & 9200$\substack{+1400 \\ -1000}$      &     1                   & Taken from \citet{harutyunyan08}.   \\                                          
2004dn   & Ic    &   Si\II{}      & 12500$\substack{+1500 \\ -1000}$     &     1                   & Taken from \citet{harutyunyan08}.   \\                                         
2004fe   & Ic    &   Fe\II{}      & 11000$\pm{1000}$     &     0                   & --    \\                                            
2004ff   & IIb   &   Fe\II{}      & 11000$\substack{+1000 \\ -2700}$     &  $-$4                   & --    \\                                           
2004gq   & Ib    &   Fe\II{}      & 13000$\substack{+1000 \\ -1500}$     &     $-$1                & \citet{modjaz07} show a He\I{} velocity of 14000\kms{} at peak.  \\                                            
2005az   & Ic    &   Si\II{}      & 9500$\substack{+1400 \\ -1000}$      &     1                   & --                                     \\         
2005bf   & Ib-pec&   Fe\II{}      & 7500$\substack{+1800 \\ -1000}$      &     3                   & Matches value for Fe\II{} lines found by \citet{folatelli06}.   \\                                       
2005hg   & Ib    &   Fe\II{}      & 9000$\pm{1000}$      &     0                   & \citet{modjaz07} show a He\I{} velocity of 10000\kms{} at peak.  \\                                           
2005kz   & Ic-BL &   n/a\tnote{b} & 19100$\pm{2500}$     &     --                  & \citet{iauc8639} report a spectral similarity to SNe~1998bw and 2002ap  \\
2005mf   & Ic    &   Fe\II{}      & 10000$\substack{+1000 \\ -1800}$     &   $-$2                  & -- \\
2006T    & IIb   &   Fe\II{}      & 7500$\pm{1000}$      &     0                   & Found from a linear interpolation of the Fe\II{} velocities at $-$11 and +7 days from the spectra of \citet{modjaz14}. \\
2006aj   & Ic-BL &   (Si\II{})    & 18000$\pm{1000}$     &     0                   & The value of \vph{} presented in \citet{pian06} is used as the spectrum is noisy and heavily blended. This value agrees with that found by \citet{pignata11} from measuring Si\II{}.    \\                                          
2006el   & IIb   &   Fe\II{}      & 11000$\substack{+1000 \\ -2700}$     &     $-$4                & A velocity of H$\beta$, somewhat past peak, is given as 11500\kms{} in \citet{cbet626}.                             \\                                                        
2006ep   & Ib    &   Fe\II{}      & 9500$\pm{1000}$      &     0                   & Found from a linear interpolation of the Fe\II{} velocities at $-$8 and +8 days from the spectra of \citet{modjaz14}. \\                                                      
2007C    & Ib    &   Fe\II{}      & 11000$\substack{+1000 \\ -1400}$     &   $-$1                  & --                             \\                                                      
2007Y    & Ib    &   Fe\II{}      & 9000$\substack{+1000 \\ -1700}$      &     $-$2                & Matches the values for Fe\II{} velocities found by \citet{stritzinger09} and \citet{valenti11}.  \\                                                                     
2007gr   & Ic    &   Fe\II{}      & 10000$\pm{1000}$     &     0                   & Agrees with values from spectral modelling presented in \citet{hunter09}.  \\                                                                             
2007ru   & Ic-BL &   Si\II{}      & 19000$\pm{1000}$     &     0                   & Taken from \citet{sahu09}.   \\                                                   
2007uy   & Ib    &   Fe\II{}      & 14000$\substack{+1000 \\ -2600}$     &     $-$3                & \citet{roy13} find the velocity of He\I{} to be 15200\kms{} at the same epoch.   \\                                                                            
2008D    & Ib    &   Fe\II{}      & 9500$\substack{+2100 \\ -1000}$      &     3                   & \citet{tanaka09} determine a value of \vph{} from spectral modelling in good agreement. \\                                                                 
2008ax   & IIb   &   Fe\II{}      & 7500$\substack{+2100 \\ -1000}$      &     4                   & Matches the values for Fe\II{} velocities found by \citet{pastorello08} and \citet{taubenberger11}. \\                                                     
2009bb   & Ic-BL &   Fe\II{}      & 17000$\substack{+2900 \\ -1000}$     &     3                   & \citet{pignata11} find Si\II{} velocities at this epoch to be 18000\kms{} and find Fe\II{} lines to be at 17000\kms{} using a spectral modelling code.      \\                                                 
2009jf   & Ib    &   Fe\II{}      & 9500$\substack{+2100 \\ -1000}$      &     3                   & Matches values found by \citet{valenti11}.  \\                                                                                                                                                                                                    
2010bh   & Ic-BL &   Si\II{}      & 30000$\pm{1000}$     &     0                   & Taken from \citet{chornock10} with linear interpolation between $\sim -3$ and +13 days to get velocity at \tpk{}.  \\
2011bm   & Ic    &   Fe\II{}      & 9000$\pm{1000}$      &     0                   & Taken from \citet{valenti11}.   \\                                                    
2011dh   & IIb   &   Fe\II{}      & 7000$\pm{1000}$      &     0                   & Taken from \citet{bersten12}.    \\                                                   
2011hs   & IIb   &   Fe\II{}      & 8000$\pm{1000}$      &     0                   & Taken from \citet{bufano14} with linear interpolation $\sim -2$ and +7 days to get velocity at \tpk{}.\\
iPTF13bvn& Ib    &   Fe\II{}      & 8000$\pm{1000}$      &     0                   & Taken from \citet{fremling14} with linear interpolation $\sim -2$ and +1 days to get velocity at \tpk{}.\\
\hline
\end{tabular}
\label{tab:sn_vph_mod}
\begin{tablenotes}
 \item [a]{Approximate phase measured relative to the bolometric light curve peak of spectrum used to measure \vph{}.}
 \item [b]{No value of \vph{} could be measured or was available in the literature. The average \vph{} for the SN type was used (see \cref{tab:sn_av_mod}).}
\end{tablenotes}
\end{threeparttable}
\end{table*}

\subsection{Explosion parameters}
\label{sect:exp_params}

\begin{figure*}
\centering
 \includegraphics[width=0.85\columnwidth]{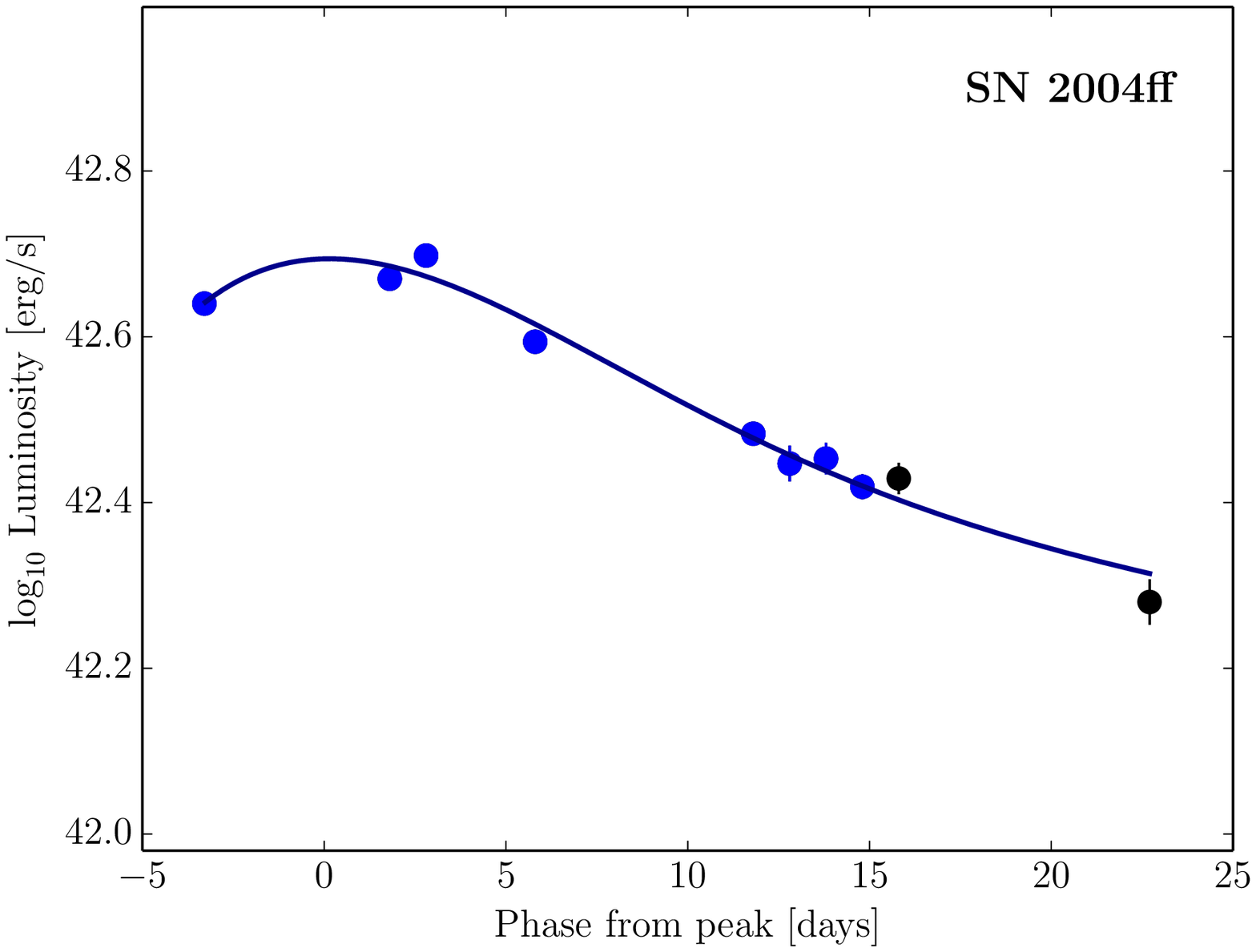}\includegraphics[width=0.85\columnwidth]{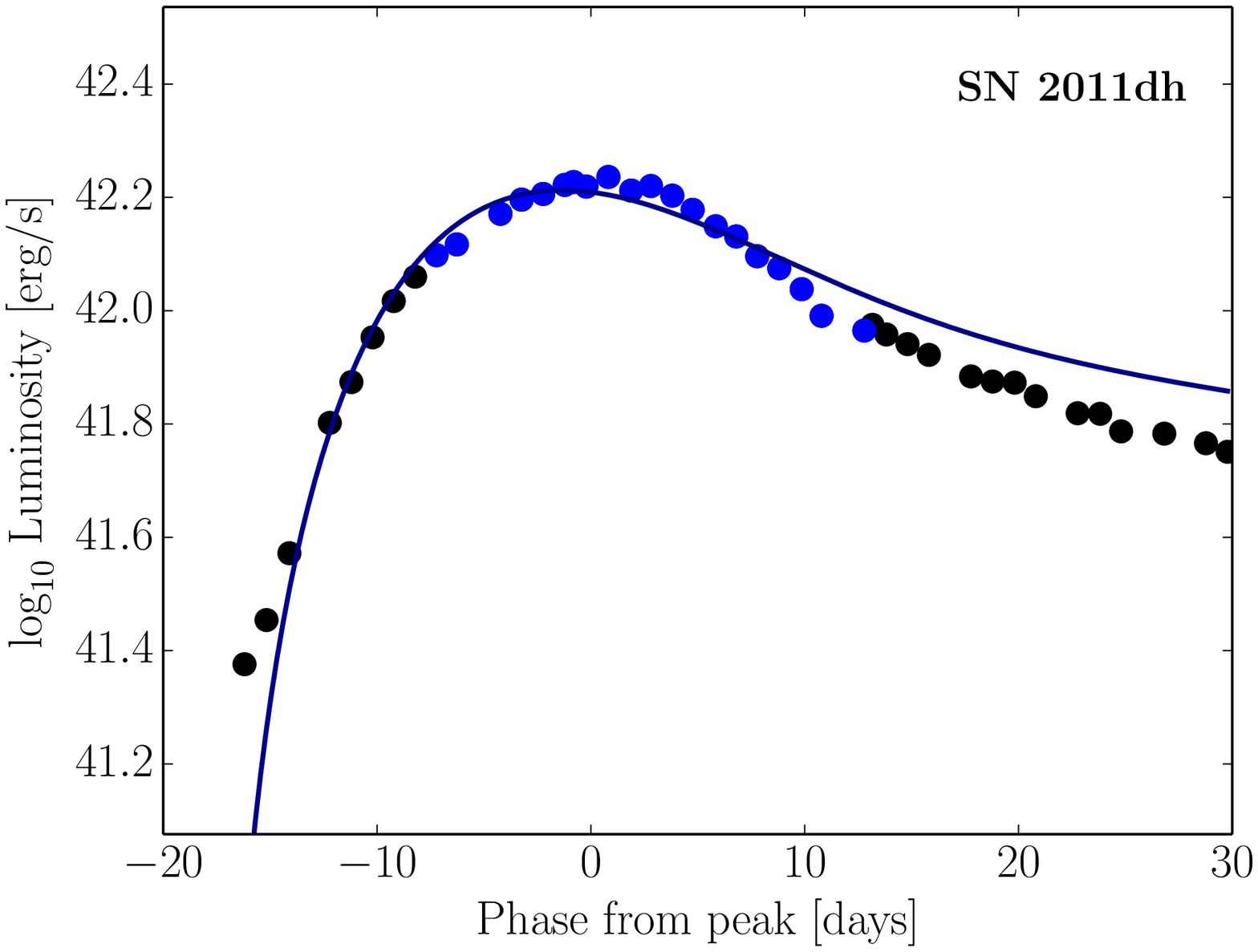}\\
 \includegraphics[width=0.85\columnwidth]{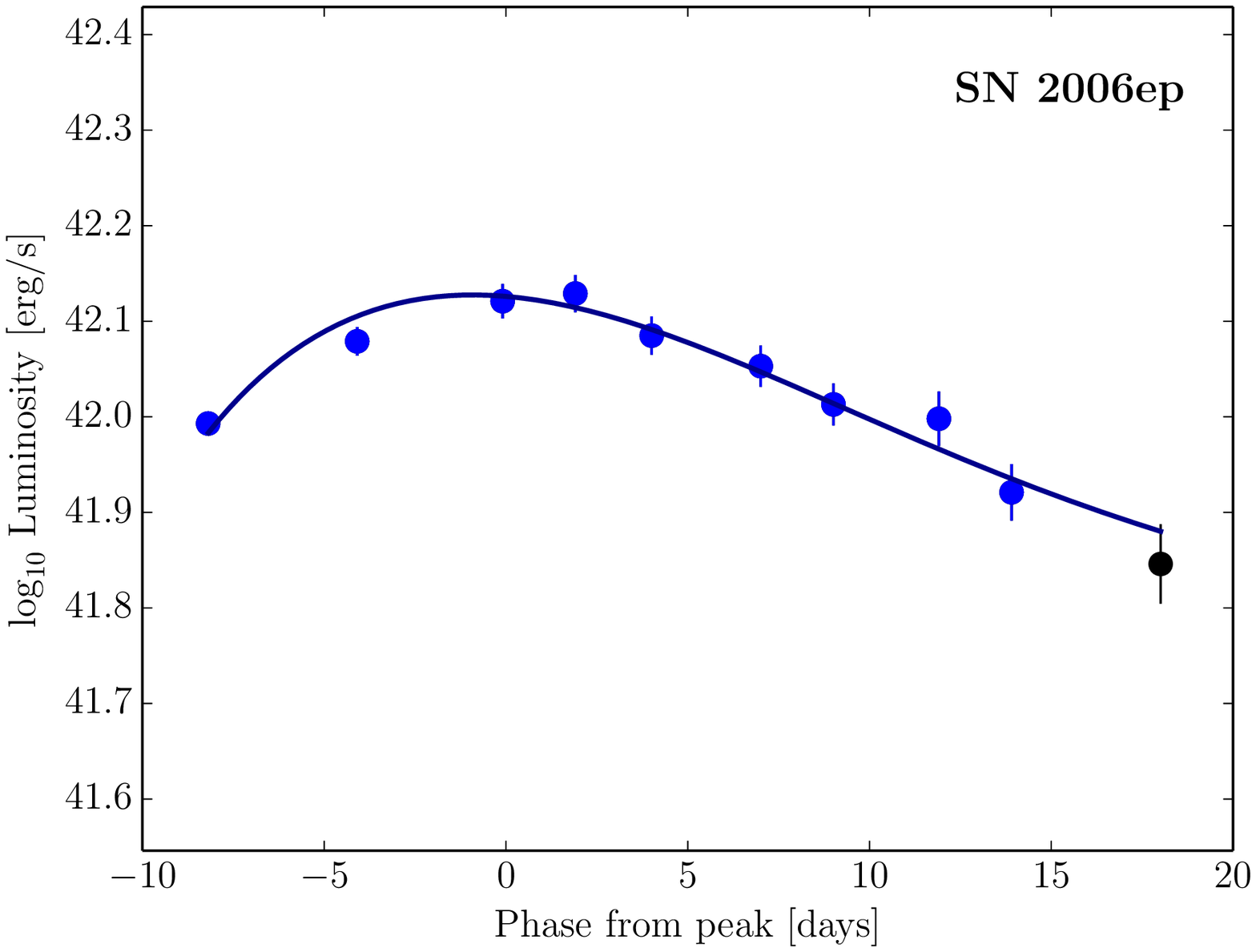}\includegraphics[width=0.85\columnwidth]{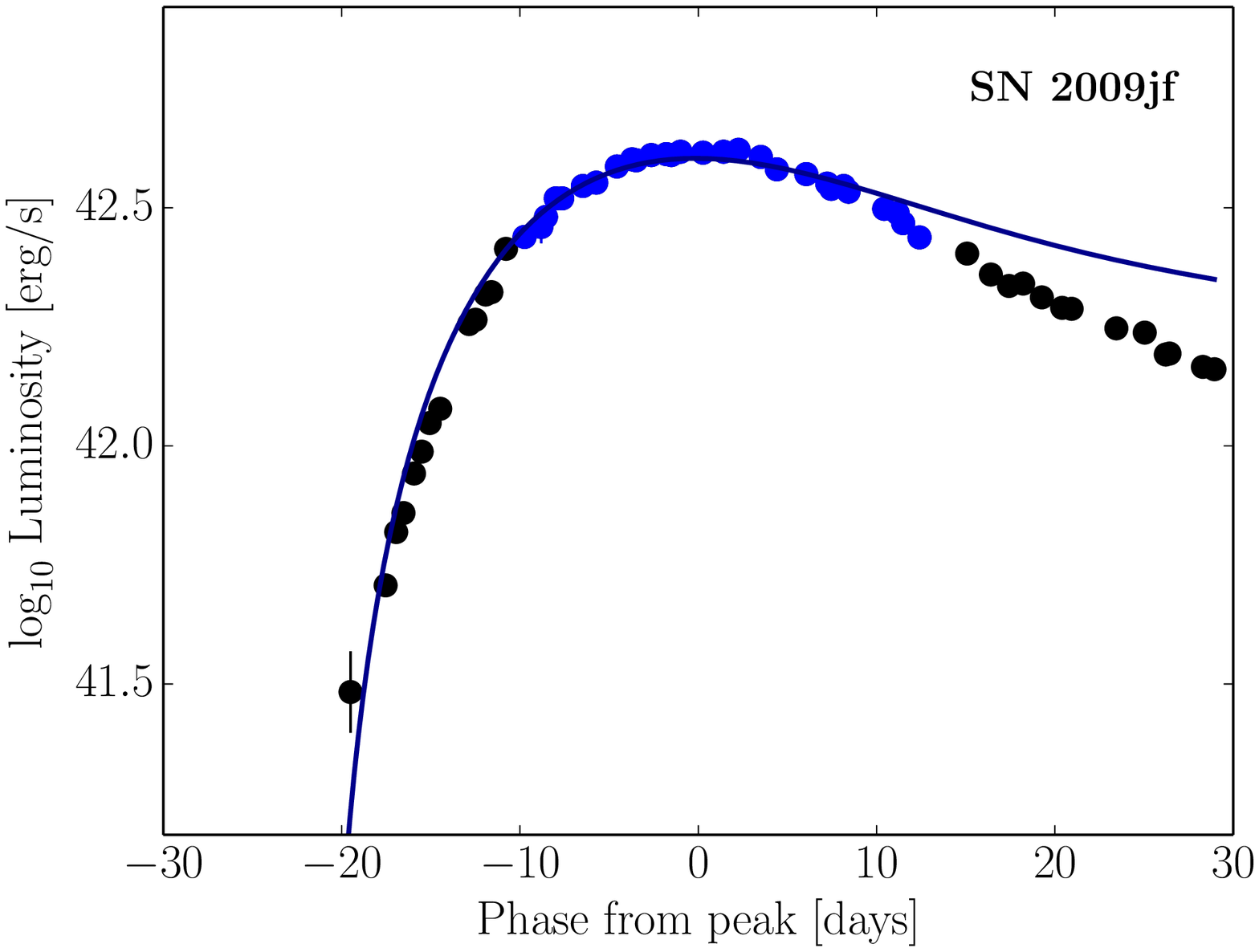}\\
 \includegraphics[width=0.85\columnwidth]{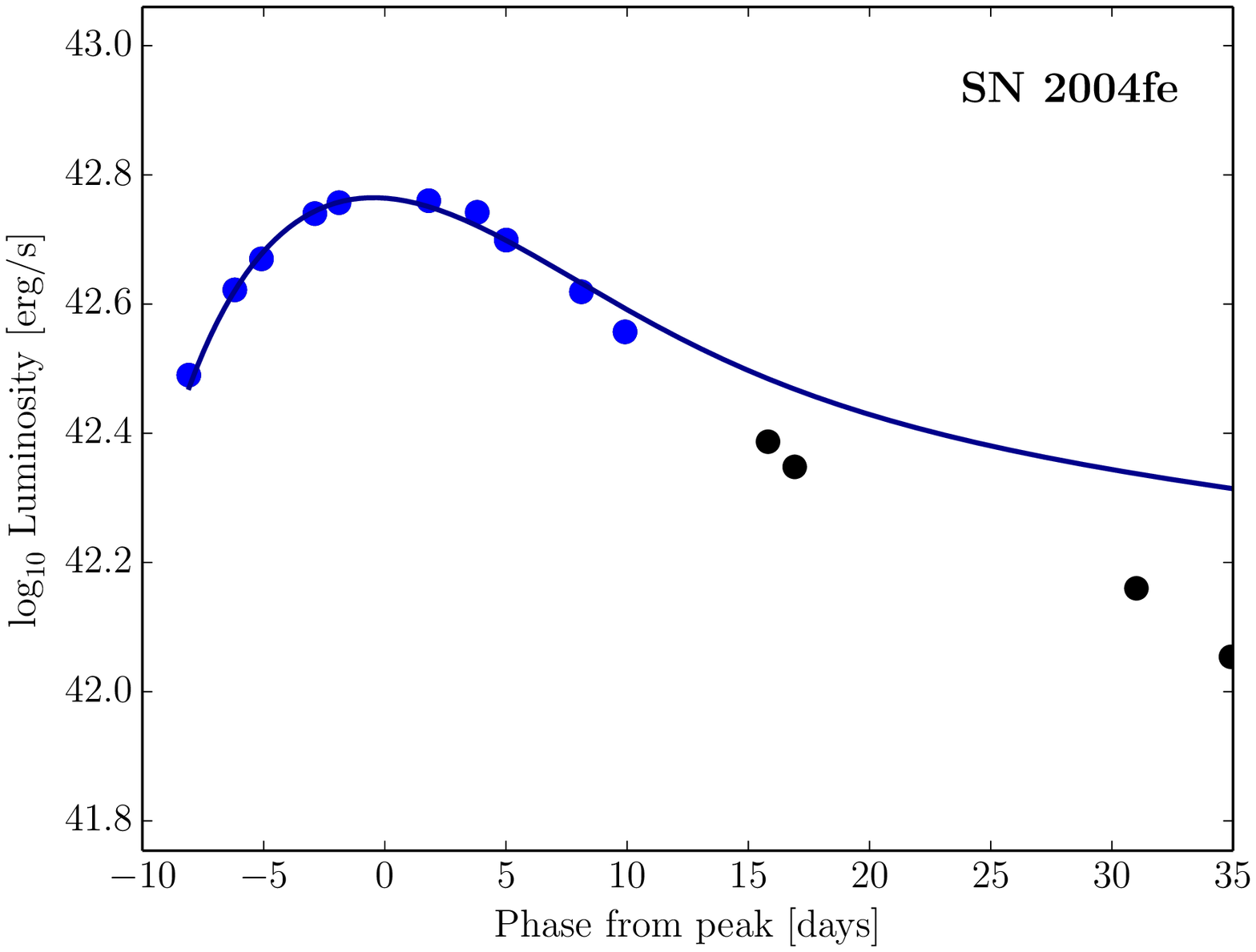}\includegraphics[width=0.85\columnwidth]{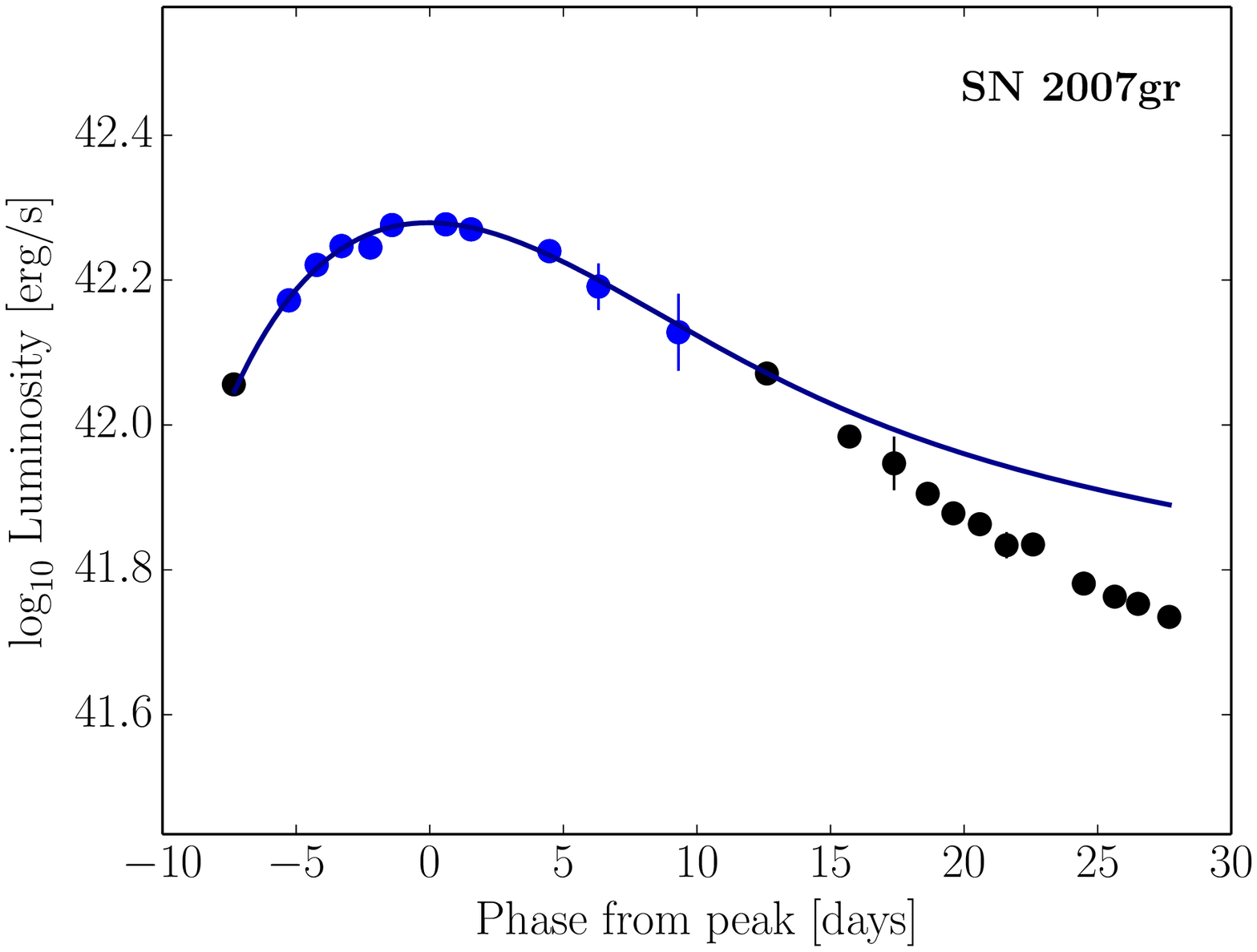}\\
 \includegraphics[width=0.85\columnwidth]{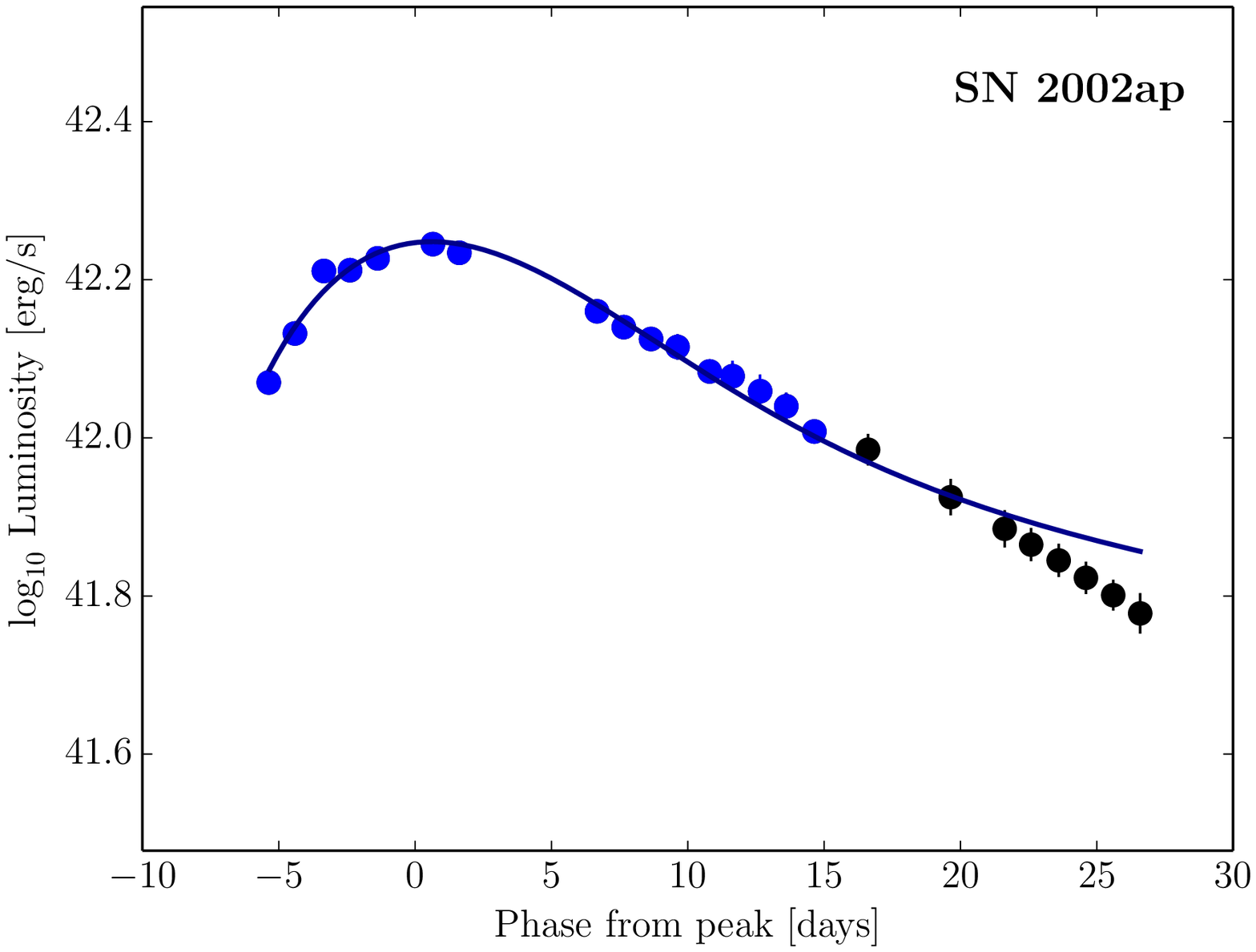}\includegraphics[width=0.85\columnwidth]{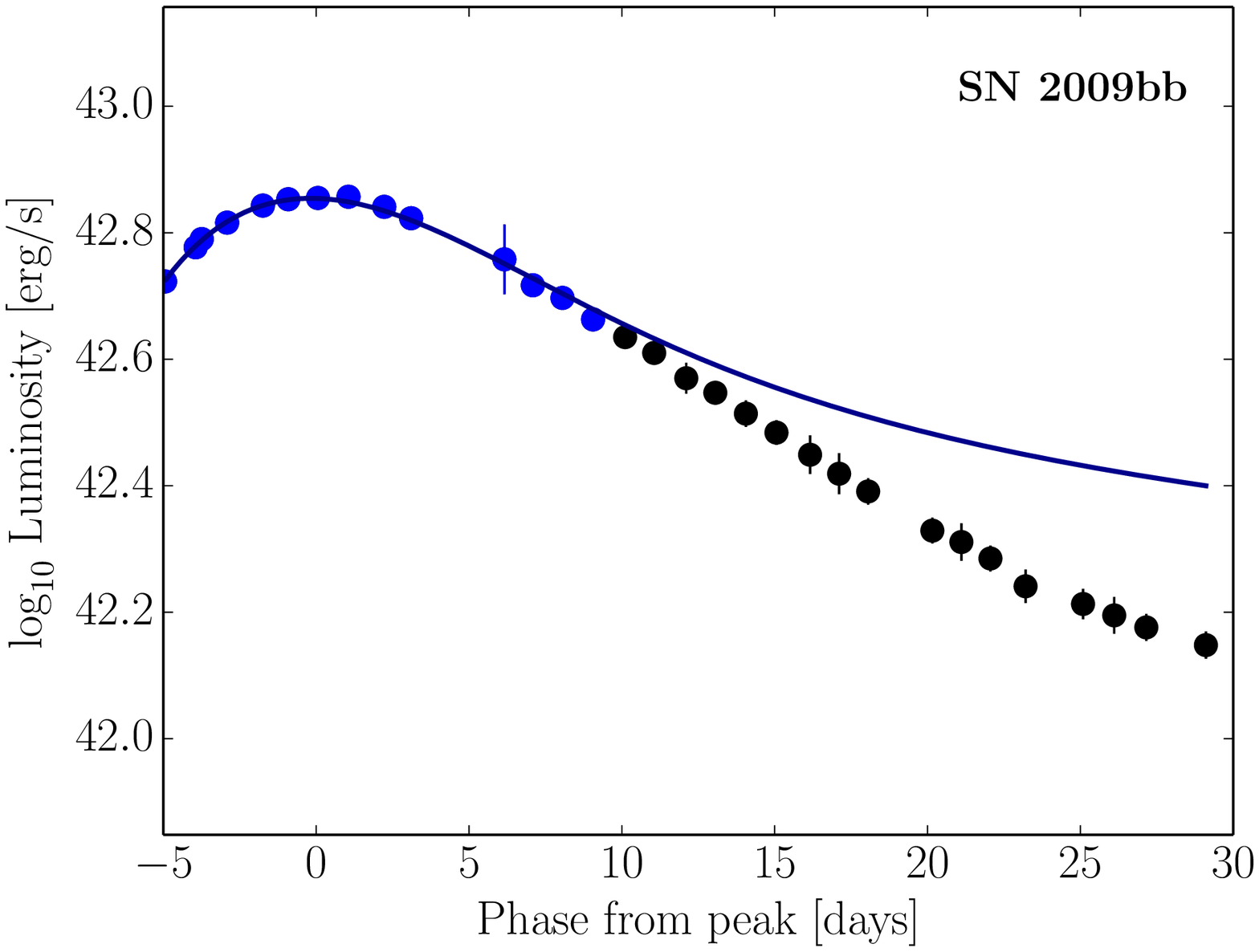}\\
\caption[Examples of fits to SE~SN bolometric light curves]{Some examples of the fits found when modelling the bolometric light curves with the analytical prescription (examples of SNe IIb, Ib, Ic and Ic-BL from top to bottom). Data points in blue indicate those that were used in the fitting routine.}
\label{fig:sn_fits_mod}
\end{figure*}

The results of the analytical modelling are given in \cref{tab:sn_results_mod}, and some example fits are shown in \cref{fig:sn_fits_mod}. 

As mentioned in \cref{sect:lit_model}, the propagation of a typographical error in the literature means some estimates of \ek{}, those that used the incorrect form, will be larger, with \mej{} values also affected, depending on the specific model employed. Explosion parameter estimates here were found to broadly agree with those in the literature where an analytical model was applied to the SN \citep[e.g.][]{taubenberger06,valenti08,soderberg08,benetti11,pignata11,drout11,taubenberger11,valenti11,cano13,roy13,taddia14} modulo the differences arising from choices of numerical factors. Differing values of $\Lambda$ in \cref{eq:model_taum_m_e} are used, for example \citet{cano13} have $\Lambda = 1$ as they use \vph{}$^2$~$=$~2\ek{}/\mej{}, whereas we use $\Lambda = 6/5$ given our form of \vph{} (\cref{eq:vsc}). This numerical factor difference is likely partly responsible for the somewhat smaller (albeit consistent) \ek{} values for the subtype averages we find compared to \citet{cano13}, as discussed later.
Differing numerical factor choices will affect the absolute parameter values but not the relative differences (e.g. between subtype averages). Previous literature modelling was done either through a direct fitting of the model, as is done here, or using scaling relations for the peak and width of the light curve and appropriately scaling the values of a better-studied SN by assuming similarity in the other properties of the explosion. 
Although agreement between the results is reassuring, we stress that estimates from such modelling are subject to sizeable uncertainties and differences in estimates are largely driven by the respective choices of \vph{}, \kopt{} etc., making a detailed consistency analysis of limited value.

For the events where such studies have been performed, we present in \cref{tab:sn_results_mod} values for the SN explosion parameters determined by more detailed spectral or hydrodynamical modelling in order to make comparisons. Generally, adopted values of distance and reddening match those in the literature works, when they are specified, however see \cref{sect:compare} for a discussion directly comparing results from the two methods.
On the whole, given the simplifications inherent in the analytical model, the estimates are in reasonable agreement for the majority of events and thus the analytical prescription provides an inexpensive method to obtain population statistics for SNe, however there are some notable departures between the two methods. For example, the \mej{} and \ek{} for SNe 1998bw and 2008D are lower than estimates from hydrodynamical modelling. \cref{sect:compare} contains direct comparison between results from the model employed here and those of more detailed modelling, with further discussion of the discrepancies. The average values for each SN type are shown in \cref{tab:sn_av_mod}.
SN~2005bf was a very unusual event that displayed a double-humped light curve. There have been various models proposed for the SN with different energy sources powering the second, brighter hump. Among these, \nic{} decay has been proposed, and, given the high peak luminosity, $\sim5\times10^{42}$~erg~s$^{-1}$, this requires \mni{} $\sim$ 0.32\msun{} to power it \citep{tominaga05}. However, \citet{maeda07}, from nebular spectral modelling, find that \simlt{}~0.08~\msun{} of \nic{} was synthesised, inconsistent with a \nic{}-powered explanation for the second peak. Here, the analytical model was used over the first `precursor' hump, which reveals a \nic{} mass (\mni{}~$\sim 0.07$~\msun{}) that is in good agreement to the value derived from nebular spectral modelling, suggesting that the second hump indeed has some other power source \citep[e.g. magnetar,][]{maeda07}. We note that the particularly unusual nature of this SN may compromise the \mej{} and \ek{} determinations from such simple modelling.

\begin{figure}
\centering
 \includegraphics[width=\columnwidth]{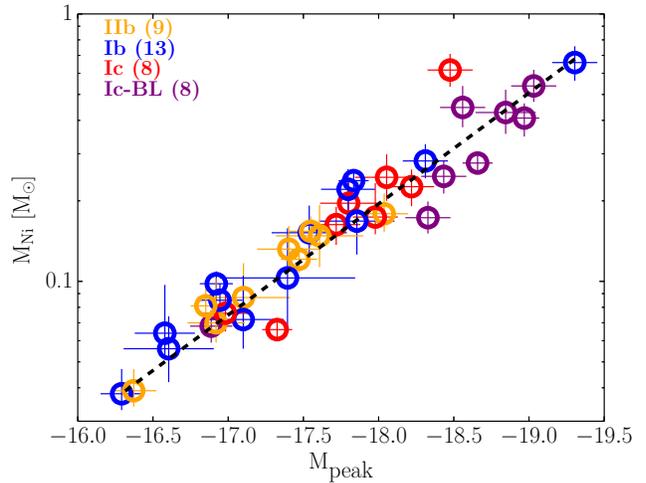}\\
\caption{Relationship between bolometric peak, \mpk{}, and the \mni{} value derived from analytical modelling for SE~SNe. Note the `peak' of SN~2005bf is taken as the precursor at \mpk{}$ \sim -17.1$, see text. The best-fitting log-linear relation is shown (\cref{eq:mnimk}). SNe are colour coded by type.}
\label{fig:lbol_mni}
\end{figure}

A fit to \mpk{} and \mni{} values found from this modelling is presented in \cref{fig:lbol_mni}, with the best-fitting relation given by: 
\begin{equation}
 \log_{10}M_\mathrm{Ni} = -0.415 \times M_\mathrm{peak} - 8.184.
 \label{eq:mnimk}
\end{equation}

The rms of \mni{} values around the fit is 0.064~\msun{}. This is analogous to the relations presented in \citet{perets10} and \citet{drout11}, here in terms of the bolometric luminosity.

\subsubsection{Uncertainties in derived parameters}
\label{sect:uncert}
In this section we describe the contributing factors to the uncertainties given in \cref{tab:sn_results_mod}\footnote{Statistical uncertainties quoted by the fitting procedure were found to be much less than the errors detailed and as such are not included.}

The value primarily affecting \mni{} is the peak of the light curve and it is thus dependent on the distance determination. Literature uncertainties for $\mu$ to each SN host were used (see references in \cref{tab:sn_sample_mod}), or, where no literature value existed, the spread in the distance modulus determinations for the host, as given in the NASA/IPAC Extragalactic Database (NED), was used. Typical uncertainties were 0.1--0.2~mag, i.e. an uncertainty of $\sim$10--20~per~cent in luminosity and thus corresponding uncertainty in \mni{}. Uncertainties on $E(B-V)$ affect the BC used, however the change in the BC is small for SE~SNe colours at peak (see further discussion in \citetalias{lyman14}). Since many reddening values did not have an accompanying uncertainty, we simply use the estimate for each event that is given in the literature. The determinations of \mni{} for SNe~2006T and 2006ep have more uncertain upper limits, found by assuming that they suffer the median $E(B-V)_\mathrm{host}$ of the sample. Reddening and distance uncertainties mainly represent a scaling uncertainty on the light curve \citep[e.g.][]{ergon14} and thus affect \mej{} and \ek{} little.

\mej{} and \ek{} are susceptible to uncertainties arising from a number of sources. One such source is the uncertainty on \vph{}. The errors on \vph{} were found by taking into account both intrinsic uncertainties in the fitting method ($\sim$1000~\kms{}) as well as accounting for the fact that not all spectra were directly observed at peak. For example, a \vph{} determination before peak could overestimate \vph{} at peak and similarly underestimate it for a determination after peak. Therefore \vph{} values derived from spectra before (after) peak had an additional component to the lower (upper) error budget. The power law of \citet{branch02} ($v_\mathrm{ph} \propto t^{-2/(n-1)}$, where $n = 3.6$)  was used as a gauge of the size of this potential offset from the \vph{} at peak, with a fiducial peak time of 20 days. Fe\II{} lines were assumed to trace \vph{}, although in some cases Si\II{} had to be used due to strong blending of Fe\II{} lines, which may be systematically offset due to contamination from other species. The error on the \vph{} value of SN~2005kz, which was assigned the average \vph{} for its type is taken to be 2500~\kms{}. 

Additionally, the choice of \kopt{} directly affects \mej{} and \ek{} for a given \taum{} (\cref{eq:model_taum}), in that it acts to scale these values. The choice of constant opacity is a limitation of this simple modelling scheme, whereas, as mentioned previously, this will evolve with time based on the composition and temperature of the ejecta. We take a 20~per~cent uncertainty in our choice of \kopt{} $= 0.06$~cm$^2$~g$^{-1}$; 
previous studies have assumed values of 0.05 \citep{drout11}, 0.06 \citep{maeda03, valenti08}, 0.07 \citep{cano13,taddia14} and 0.08~cm$^2$~g$^{-1}$ \citep{pignata11}, largely driven by the values of \kopt{} near peak from results of spectroscopic modelling \citep[e.g.][Mazzali et al., in prep]{mazzali00, chugai00, mazzali13}. 
These line-based opacities (i.e. neglecting continuum opacity, see \citealt{mazzali01}) include time-dependant evolution due to, e.g., the temperature of the ejecta. Furthermore, different composition of the ejecta, in particular when considering different SE~SN subtypes, will affect the opacity, although variations in ejecta abundances (e.g. CO/He) amongst SNe is not well known at present. As such, a single choice of opacity represent a simplifying assumption of the model, which deserves further investigation to assess its impact on results for different SE~SNe subtypes across the parameter space of SN explosions.
With other values fixed, this uncertainty contributes an uncertainty of $+25/-17$~per~cent in \mej{} and \ek{} determinations (since both have the same dependence on \kopt{}\footnote{Using \cref{eq:model_taum,eq:vsc}: $M_\mathrm{ej}^3/E_\mathrm{K} \propto \kappa_{\textrm{opt}}^{-2} \rightarrow M_\mathrm{ej} \propto \kappa_{\textrm{opt}}^{-1} v_\mathrm{ph}$, given $E_\mathrm{K}/M_\mathrm{ej} \propto v_\mathrm{ph}^2$ -- and therefore $E_\mathrm{K} \propto \kappa_{\textrm{opt}}^{-1} v_\mathrm{ph}^3$.}). An in-depth study of the evolution of the opacity for SE~SNe is beyond the scope of this paper, but the results of such a study would be useful to constrain the applicability of such analytical models where a constant value of \kopt{} is used.

\begin{figure}
\centering
 \includegraphics[width=\columnwidth]{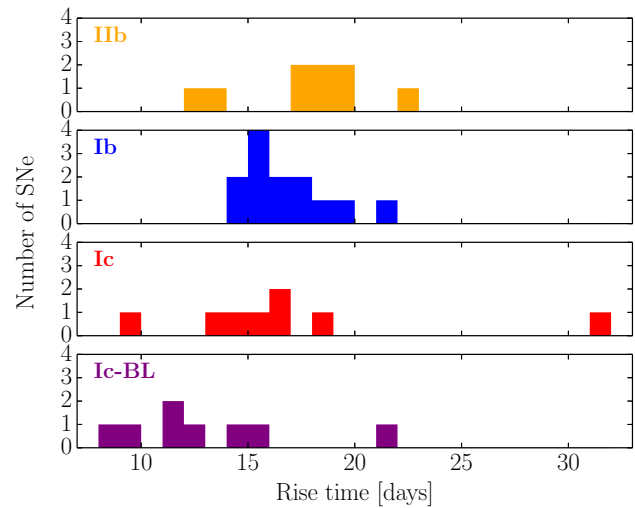}\\
\caption{Rise times of SE~SNe subtypes in the sample as estimated from the fitting of the analytical model. SNe IIb, Ib and Ic share similar average rise times (17.6, 16.7, 16.8 days, respectively), with SNe~Ic-BL somewhat faster (13 days).}
\label{fig:risetimes}
\end{figure}

Finally, the analytical model requires an initial starting time, $t_0$, which affects the value of \taum{} that is fitted and also \mni{}. Where appropriate (e.g. in the case of GRB-SNe), this additional uncertainty did not factor since $t_0$ is known. Where $t_0$ was very poorly constrained, the model was manually fitted for a variety of \taum{} values where the model still reasonably reproduced the observed light curve. The range of $t_0$ used for each SN is shown in \cref{tab:sn_results_mod}. We note in passing, in agreement with the concurrent work of \citet{taddia14} (but here regarding the bolometric rise), we find, for our subtype averages, SNe~Ic-BL exhibit the shortest rise times to peak, with SNe~Ib and IIb having similar rise times. We also find the average SNe~Ic to be similar to the rise times to SNe~Ib and IIb, but this is complicated by SN~2011bm, which \citet{taddia14} exclude from their sample. The distribution of rise times are plotted in \cref{fig:risetimes}, however the interested reader is directed to \citet{taddia14} for a more thorough discussion of rise times. The error on \mni{} arising from varying $t_0$ was \simlt{}~0.01~\msun{} where it was varied over $\sim$1--2~days. For the least constrained events the \mni{} uncertainty was 10--15~per~cent. \mej{} and \ek{} errors were 10--30~per~cent for reasonably well constrained events (1--2~days) but +(30 to 40)/$-$(15 to 25)~per~cent for the more unconstrained events (e.g. SN 2011hs).

Total uncertainties on the parameters were found by refitting the model for all varying parameters and adding in quadrature the uncertainty from each parameter, these are given in \cref{tab:sn_results_mod}. 

We note that the asphericity of the explosions, which breaks the assumption of spherical symmetry in the model (\cref{sect:lit_model}), contributes a systematic uncertainty in our results. It appears some degree of asphericity is near-ubiquitous in SE~SNe around peak light \citep[see the review of][]{wang08}. One may expect the very energetic SNe~Ic-BL (and GRB-SNe) to display the strongest asymmetries, although their global asymmetries appear to be \simlt{}~15~per~cent \citep[e.g.][for SNe 2002ap and 2006aj respectively]{wang03,maund07b} and indeed a normal SN~Ic, SN~1997X showed one of the strongest degrees of polarisation, indicating a high degree of asymmetry \citep{wang01}. The uncertainty due to asphericity is higher in the more stripped SNe~Ic and Ic~BL, where stronger asymmetries in the deep ejecta \citep{wang08} can influence the photosphere during evolution around peak light. It is likely to be less pronounced in SNe~IIb around peak owing to the presence of the hydrogen envelope \citep{maund07a}. Our results, based on a spherically symmetric model, could be described as the isotropic-equivalent values for the SNe \citep{wang03}.

\subsubsection{Direct comparison to detailed modelling}
\label{sect:compare}

Here a recent subset of SNe with explosion parameters derived from hydrodynamical modelling of the light curve, to which we can compare our results, is presented. In order to make a comparison, external factors common to both methods such as distance, reddening and the time of explosion were set to those of the comparison works. 
Due to the inherent differences in the models, such as the lack of treatment for evolution of \kopt{} or \vph{} in the analytical model, just the best-fitted parameters are given for direct comparison (i.e. neglecting our uncertainties in these values).

A note must also be made regarding the new bolometric light curve creation method of \citetalias{lyman14}, used here. Although this has been used for some recent events, other methods of forming bolometric light curves have been used by other studies. For example, discrepancies between our derived \mni{} to that of, e.g. \citet{utrobin94}, where the bolometric light curve was created from $BVRI$ photometry alone, is dominated by the light curve creation method -- only $\sim50-60$~per~cent of the bolometric flux from a SE~SN is emitted in these bands \citepalias{lyman14}. As such we restrict comparisons to those where a good approximation of the bolometric light curve is used -- where appropriate, the method used to create this is highlighted in the discussion. We also note that many of the SNe in the current sample also formed part of the original SN sample of \citetalias{lyman14}, as such we can be confident our method is not introducing some large systematic uncertainty in the resulting bolometric light curves.

\paragraph*{SN~2008D}

We use the same distance and $t_0$ (time of the x-ray flash) as those of \citet{tanaka09}. Our $E(B-V) = 0.6$~mag is also consistent with the value used by these studies of 0.65~mag at a level where derived parameters will not be affected beyond the precision quoted. \citet{tanaka09} obtain parameters of \mni{} $\sim$~0.07~\msun{}, \mej{} = $5.3\pm1.0$~\msun{} and \ek{} = $6.0\pm2.5\times10^{51}$~ergs through hydrodynamical modelling. Our results give \mni{} = 0.09~\msun{}, \mej{} = $2.9$~\msun{} and \ek{} = $1.6\times10^{51}$~ergs. \mej{} and \ek{} estimate here are much lower than those from hydrodynamical modelling. Similarly lower estimates were also made using analytical relations by \citet{soderberg08}. The cause of this disagreement is discussed in \citet{tanaka09} as being symptomatic of using a single opacity and \vph{} value in the model. In contrast \citeauthor{tanaka09} shows SN 2008D displayed strong evolution in \vph{}, with broad-lines early on, before becoming more normal SN Ibc-like around peak. Since the \vph{} in the model is that at peak, the presence of this early, highly energetic evolution has no impact on the derived parameters. A lower value of \kopt{} in the model would work to bring values in better agreement.
The bolometric light curve of \citet{tanaka09} was constructed directly from UV-optical-NIR photometry. There is overall good agreement in the shape of the light curves but a $\sim 0.1$~mag increase in brightness of the light curve created using the BCs here -- this is likely to be a contributing factor to the higher \mni{} value we find here. Additionally we note our \mej{} is significantly less than the $\sim$~7~\msun{} derived by \citet{mazzali08} through spectral modelling.

\paragraph*{SN~2011dh}

For comparison we use the parameters of \citet{ergon14}, which revise those used in \citet{bersten12} where initial modelling of SN~2011dh was performed, in order to compare to their hydrodynamical modelling results. \citet[][updated in \citealt{ergon14}]{bersten12} find  \mni{} = 0.075~\msun{}, \mej{} = 1.8--2.5~\msun{} and \ek{} = 0.6--1.0 $\times10^{51}$~ergs, where the modelling presented here gives \mni{} = 0.08~\msun{}, \mej{} = $2.2$~\msun{} and \ek{} = $0.7\times10^{51}$~ergs. Estimates of each parameter are in good agreement.
The bolometric light curve modelled by previous work is constructed directly from photometry of SN~2011dh from UV--MIR. Our reconstructed bolometric light curve is in very good agreement, which is not unexpected since SN~2011dh formed part of the sample used in the construction of the BCs.

\paragraph*{SN~2011hs}

The distance and reddening to SN~2011hs were set as in \citet{bufano14}, however the authors note a large uncertainty on the time of explosion: $t_0 = 2455872\pm4$ JD. \citet{bufano14} find \mni{} = 0.04~\msun{}, \mej{} = 1.8--2.5~\msun{} and \ek{} = 0.8--0.9 $\times10^{51}$~ergs using the midpoint of the $t_0$ range. When fixing $t_0$ to 2455872 JD, we obtain values of \mni{} $\sim$ 0.05~\msun{}, \mej{} = 2.3~\msun{} and \ek{} = 0.9 $\times10^{51}$~ergs. Although this gives values in very good agreement, it must be stressed the overall fit is poor. The model appears better fitted with a later $t_0$, which explains the difference between our estimates in \cref{tab:sn_results_mod} and those when fixing to $t_0$ here, furthermore, an earlier $t_0$ is favoured by radio observations \citep{bufano14}. Given the extended nature of the progenitor star, the model here may not be as appropriate for such explosions. The bolometric light curve of SN~2011hs was created by \citet{bufano14} through direct integration of photometry covering a wide wavelength range (UV-optical-NIR), and we find good agreement to our light curve.  

\paragraph*{iPTF13bvn}

The reddening and distance in \cref{tab:sn_sample_mod} for iPTF13bvn are those of \citet{fremling14}, and here $t_0$ is fixed to their adopted value of 2456459.25 JD. They obtain parameters of \mni{} = 0.04--0.07~\msun{}, \mej{} = 1.3--2.4~\msun{} and \ek{} = 0.5--1.4$\times10^{51}$~ergs. The results of the modelling presented here gives \mni{} = 0.06~\msun{}, \mej{} = 1.9~\msun{} and \ek{} = 0.7$\times10^{51}$~ergs.
When adopting the values presented in \citet{bersten14}, namely $E(B-V) = 0.21\pm0.03$~mag and $\mu = 32.04\pm0.2$~mag, the results are \mni{} = 0.11~\msun{}, \mej{} = 1.8~\msun{} and \ek{} = 0.7$\times10^{51}$~ergs compared to the values of \citet{bersten14}: \mni{} $\sim 0.1$~\msun{}, \mej{} $\sim 2.3$~\msun{} and \ek{} $\sim 0.7\times10^{51}$~ergs.
The estimates from this simple modelling agree well with each of those found from hydrodynamical modelling by two groups. For \citet{bersten14}, the bolometric light curve modelled was created using the BCs of \citetalias{lyman14}, as is done here, whereas \citet{fremling14} used optical photometry of iPTF13bvn alongside a UV and IR correction derived from SN~2011dh. There is good agreement between the bolometric light curves created through each method.

\subsection{SN type distributions}
\label{sect:sn_type}

\begin{figure}
\centering
 \includegraphics[width=\columnwidth]{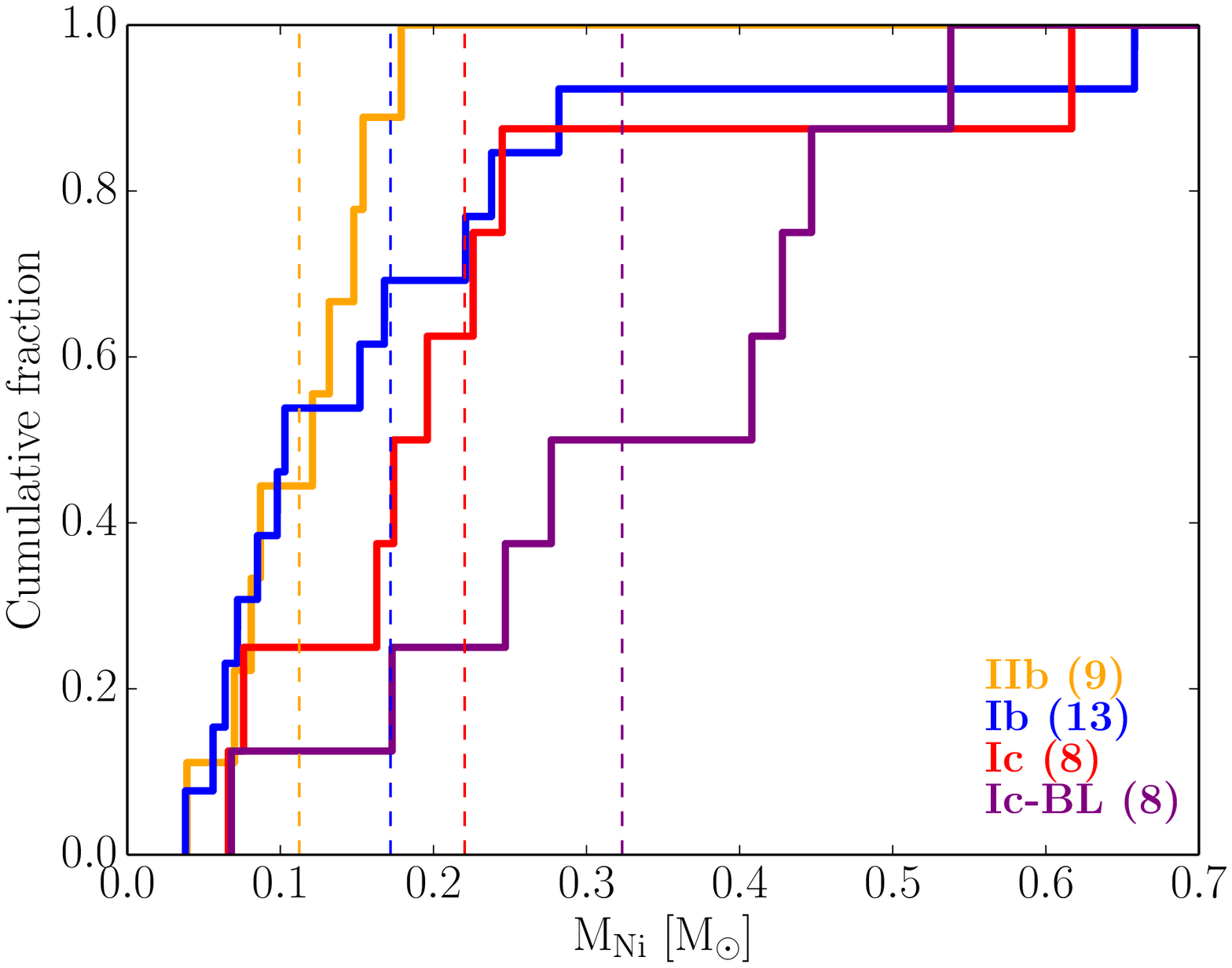}\\
  \includegraphics[width=\columnwidth]{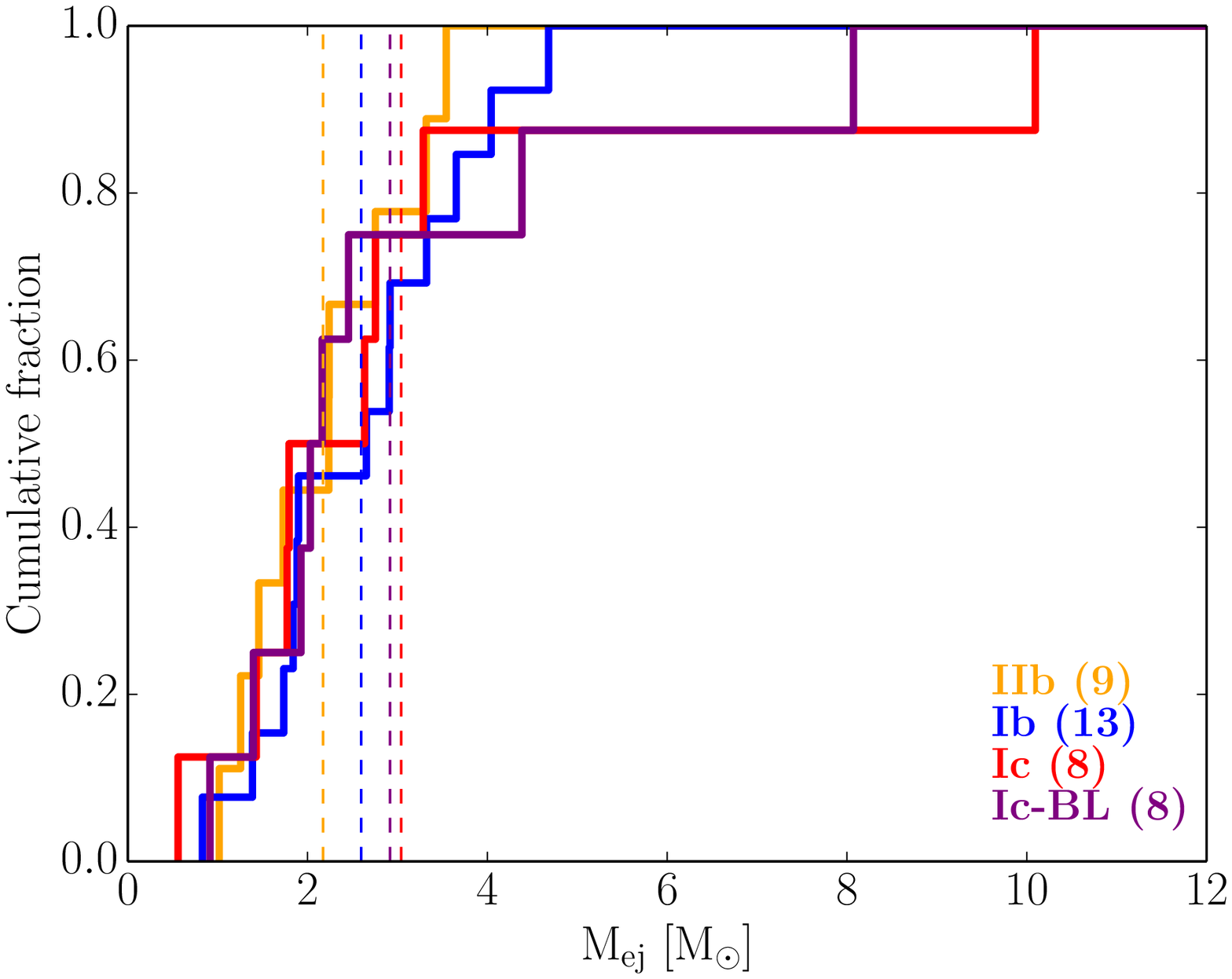}\\
   \includegraphics[width=\columnwidth]{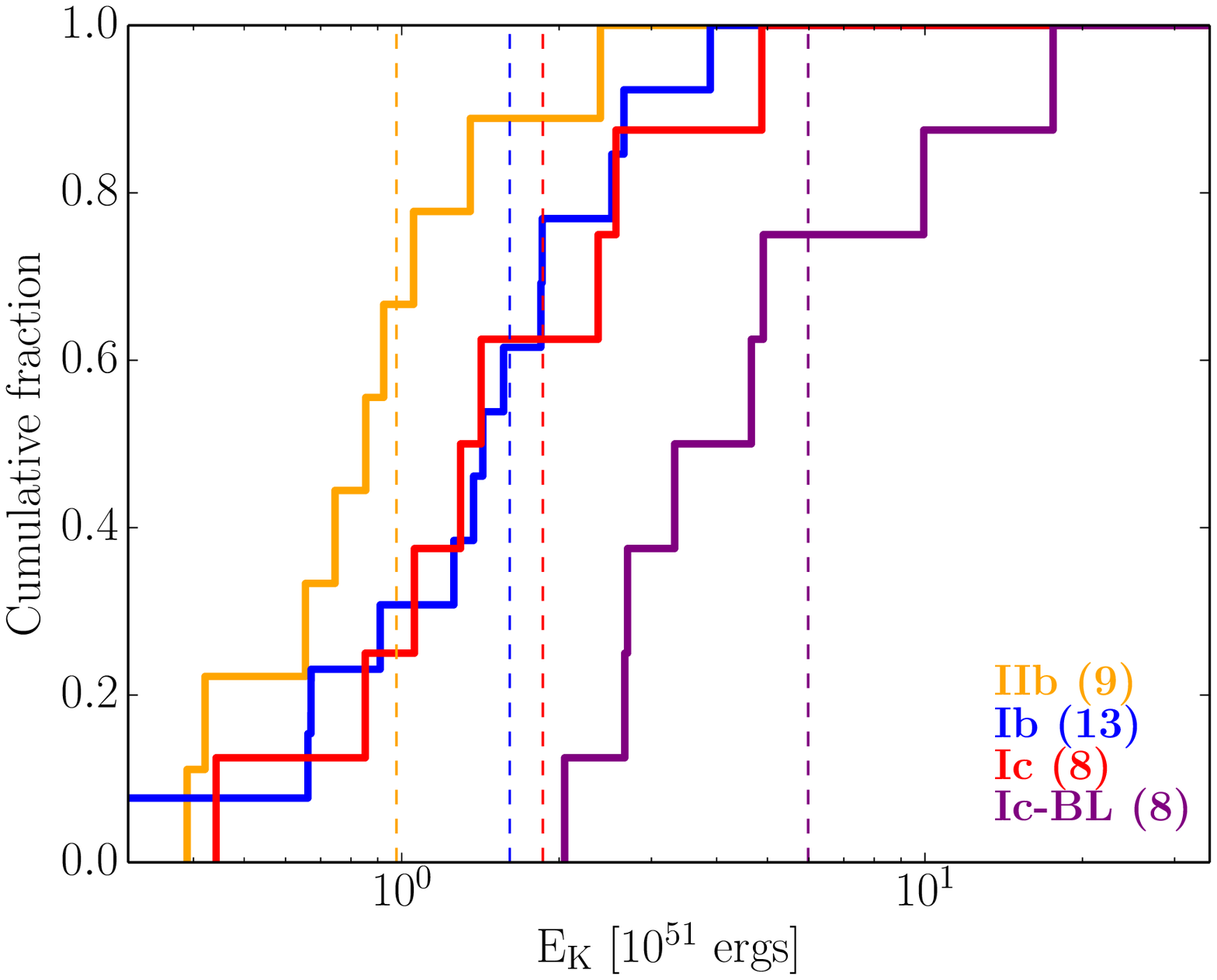}\\
\caption{Cumulative distributions for explosion parameters of SE~SNe (top: \mni{}, middle: \mej{}, bottom: \ek{}, note log-scale), divided by subtype. The average values for each SN type are indicated by the vertical dashed lines.}
\label{fig:sn_cdf_mod}
\end{figure}

With a sample of explosion parameters for many different SNe, the statistical distribution as a function of SN type can be investigated. The cumulative distributions of the parameters for each type are shown in \cref{fig:sn_cdf_mod}. This figure highlights the extreme nature of SNe~Ic-BL in \mni{} and \ek{}. SNe~Ic-BL are more energetic than any of the other subtypes (the least energetic SN~Ic-BL has an \ek{} value above the average value of any of the other subtypes), and also have much larger \mni{} values on average, although it should be noted that SNe~Ib and Ic can reach such high \mni{} values even though the bulk have much lower values. However, \mej{} distributions do not distinguish SNe~Ic-BL from other SE~SNe clearly. SNe Ib and Ic are indistinguishable in all three parameters. There appears to be a hint that SNe~IIb favour lower values of \mni{} and \ek{} cf. SNe~Ib and Ic, whereas their \mej{} values do not show this. We note good agreement in the relative average explosion parameters between SN subtypes compared to \citet{cano13}, although numerical factor differences in the models compromise, good absolute agreement. Additionally, the large fraction of events for which an average \vph{} (based on subtype) is assigned in \citet{cano13} limits the usefulness of direct comparisons. However, both studies find similarity in the ejecta masses of SNe Ib, Ic and Ic-BL, similarity between SNe Ib and Ic in each parameter, and that SNe Ic-BL exhibit generally larger \mni{} and \ek{} values than other SN subtypes.

The two-sample Kolmogorov-Smirnov test (\ks{}) was applied to each pair of SN types to ascertain the probability ($p$ value) that the two samples are drawn from the same parent population given the maximum difference, $D$, between their cumulative distributions, where a small $p$ value indicates that it is statistically unlikely the two samples are explained by a single population.\footnote{Tests were repeated with the two-sample Anderson-Darling test, with significances remaining at very similar levels.} The results of the \ks{} are given in \cref{tab:sn_kstest_mod}, which confirm the `by-eye' judgements on the distributions made above. The distribution of \mni{} values of (IIb, Ic-BL) are significantly different, with \mni{} values of (IIb, Ic), (Ib, Ic-BL) and (Ic, Ic-BL) distinguished at a lower significance of $\sim 2 \sigma$. 
The \ek{} distribution of SNe~Ic-BL is statistically distinguished from those of SNe~IIb, Ib and Ic. As expected, \mej{} distributions cannot be distinguished and all 4 subtypes are consistent with being drawn from any of the other distributions. On the whole, SNe~IIb, Ib and Ic are indistinguishable, although there is some marginal evidence of a difference in the \mni{} and \ek{} distributions of SNe~IIb to those of SNe~Ib and Ic.

An important caveat to consider when regarding comparison between the parameter distributions is that the sample was drawn from literature events. As such, many selection and observational biases are intrinsic to its creation. For example, it may be that we are skewing the \mni{} distribution for SNe~Ic-BL by preferentially including bright SNe~Ic-BL (i.e. high \mni{}) as SNe~Ic-BL appear intrinsically rarer in the very local universe than the other subtypes, whose distributions would therefore be less affected. Such caveats motivate a similar study on a more homogeneously created sample of SNe to further investigate the initial distribution comparisons of this study.

\begin{table*}
\begin{threeparttable}
     \caption{Results of explosion parameter modelling for SE~SNe}
     \begin{tabular}{ll@{\hskip 0.4cm}Hc@{\hskip -0.04cm}c@{\hskip -0.04cm}cccc@{\hskip 0.5cm}cccp{3.5cm}}
      \hline
    &   & \multicolumn{6}{c@{\hskip 0.5cm}}{This study} & \multicolumn{4}{c}{Literature values} \\   
SN name           & Type    & \taum{}     & $t_\textrm{peak}$-$t_0$ & Phase fitted & \mni{}    & \mej{}     & \ek{}          &  \mni{}     & \mej{}     & \ek{}            & Refs \\
                  &         & (days)      & (days)                  & (days)    & (\msun{}) &  (\msun{}) & $10^{51}$~ergs &  (\msun{})  &  (\msun{}) & ($10^{51}$~ergs) &           \\
\hline
1993J             & IIb     & $14.7-15.5$ & $18.3-19$    & $-10,10$  & 0.12$\substack{+0.01 \\ -0.01}$ &$  2.2\substack{+0.7\\-0.5}$ & $  0.9\substack{+ 0.4\\ - 0.3}$   & 0.06--0.14  & 1.9--3.5   & 1--1.6         & (1--3) \\
1994I             & Ic      & $ 5.6-7.3 $ & $8.6-10$     & $-5,10$   & 0.07$\substack{+0.01 \\ -0.01}$ &$  0.6\substack{+0.3\\-0.1}$ & $  0.4\substack{+ 0.2\\ - 0.2}$   & 0.07        & 0.9--1.3   & 1              & (4--6)  \\ 
1996cb            & IIb     & $12.0-15.3$ & $15.5-18.5 $ & $-10,10$  & 0.09$\substack{+0.03 \\ -0.02}$ &$  1.7\substack{+1.0\\-0.4}$ & $  0.7\substack{+ 0.6\\ - 0.3}$   & --          & --         & --             & --        \\ 
1998bw            & Ic-BL   & $13.2-13.2$ & $15.1 $      & $-8,10$   & 0.54$\substack{+0.08 \\ -0.07}$ &$  4.4\substack{+1.2\\-0.8}$ & $  9.9\substack{+ 3.8\\ - 2.2}$   & 0.4-0.7     & $\sim$10   & 20--50         & (7,8)      \\ 
1999dn            & Ib      & $13.9-17.4$ & $14-18.5 $   & $-4,15$   & 0.10$\substack{+0.01 \\ -0.02}$ &$  4.0\substack{+1.1\\-1.7}$ & $  2.7\substack{+ 1.1\\ - 1.3}$   & --          & --         & --             & --        \\ 
1999ex            & Ib      & $15.5-16.9$ & $18-19 $     & $-10,15$  & 0.15$\substack{+0.04 \\ -0.03}$ &$  2.9\substack{+0.9\\-0.7}$ & $  1.3\substack{+ 0.8\\ - 0.5}$   & $\sim$0.16  & --         & $\sim$2.7      & (9)       \\ 
2002ap            & Ic-BL   & $ 9.1-12.2$ & $11-13.9 $   & $-6,15$   & 0.07$\substack{+0.01 \\ -0.01}$ &$  2.0\substack{+0.8\\-0.7}$ & $  2.0\substack{+ 1.3\\ - 0.9}$   & 0.11        & 2.5        & 4              & (10)        \\ 
2003bg            & IIb     & $17.7-19.7$ & $22-23 $     & $-15,21$  & 0.15$\substack{+0.02 \\ -0.02}$ &$  3.5\substack{+1.1\\-0.8}$ & $  1.4\substack{+ 0.7\\ - 0.5}$   & 0.15--0.20  & 4--5       & 5              & (11)   \\ 
2003jd            & Ic-BL   & $11.9-13.2$ & $14.3-15.3 $ & $-6,10$   & 0.43$\substack{+0.09 \\ -0.07}$ &$  2.5\substack{+0.9\\-0.5}$ & $  2.7\substack{+ 1.1\\ - 0.7}$   & --          & --         & --             & --        \\ 
2004aw            & Ic      & $14.6-15.8$ & $15.8-17 $   & $-5,8$    & 0.20$\substack{+0.04 \\ -0.03}$ &$  3.3\substack{+0.9\\-0.8}$ & $  2.4\substack{+ 0.9\\ - 1.1}$   & --          & --         & --             & --        \\ 
2004dk            & Ib      & $15.9-19.0$ & $18.2-21 $   & $-7,12$   & 0.22$\substack{+0.04 \\ -0.04}$ &$  3.7\substack{+1.3\\-1.0}$ & $  1.8\substack{+ 1.1\\ - 0.7}$   & --          & --         & --             & --        \\ 
2004dn            & Ic      & $10.2-14.4$ & $14.3-17.3 $ & $-9,13$   & 0.16$\substack{+0.03 \\ -0.03}$ &$  2.8\substack{+1.0\\-1.2}$ & $  2.6\substack{+ 1.3\\ - 1.2}$   & --          & --         & --             & --        \\ 
2004fe            & Ic      & $ 9.3-12.5$ & $12.5-14.9 $ & $-7,10$   & 0.23$\substack{+0.04 \\ -0.04}$ &$  1.8\substack{+0.7\\-0.7}$ & $  1.3\substack{+ 0.6\\ - 0.6}$   & --          & --         & --             & --        \\ 
2004ff            & IIb     & $ 9.0-12.1$ & $11-15 $     & $-4,15$   & 0.18$\substack{+0.03 \\ -0.03}$ &$  1.5\substack{+0.7\\-0.5}$ & $  1.1\substack{+ 0.6\\ - 0.7}$   & --          & --         & --             & --        \\ 
2004gq            & Ib      & $ 9.5-12.7$ & $13-15.5 $   & $-5,1$    & 0.10$\substack{+0.05 \\ -0.04}$ &$  1.8\substack{+1.0\\-0.5}$ & $  1.9\substack{+ 1.1\\ - 0.7}$   & --          & --         & --             & --        \\ 
2005az            & Ic      & $13.0-16.9$ & $16-20 $     & $-7,30$   & 0.24$\substack{+0.05 \\ -0.04}$ &$  2.6\substack{+1.2\\-0.8}$ & $  1.4\substack{+ 0.9\\ - 0.6}$   & --          & --         & --             & --        \\ 
2005bf            & Ib      & $ 9.2-14.5$ & $13-17 $     & $-8,6$    & 0.07$\substack{+0.03 \\ -0.02}$ &$  0.8\substack{+1.2\\-0.2}$ & $  0.3\substack{+ 0.5\\ - 0.1}$   & 0.08        & --         & --             & (12)  \\ 
2005hg            & Ib      & $11.0-13.4$ & $15-18 $     & $-8,15$   & 0.66$\substack{+0.10 \\ -0.09}$ &$  1.9\substack{+0.6\\-0.6}$ & $  0.9\substack{+ 0.4\\ - 0.4}$   & --          & --         & --             & --        \\ 
2005kz            & Ic-BL   & $15.8-21.1$ & $17-25 $     & $-3,15$   & 0.45$\substack{+0.09 \\ -0.07}$ &$  8.1\substack{+3.7\\-2.6}$ & $ 17.6\substack{+11.1\\ - 7.9}$   & --          & --         & --             & --        \\ 
2005mf            & Ic      & $ 9.8-13.5$ & $12-20 $     & $-1,14$   & 0.17$\substack{+0.06 \\ -0.02}$ &$  1.4\substack{+1.0\\-0.4}$ & $  0.9\substack{+ 0.7\\ - 0.4}$   & --          & --         & --             & --        \\ 
2006T\tnote{a}    & IIb     & $10.9-12.6$ & $16-18 $     & $-5,15$   & 0.07$\substack{+0.03 \\ -0.01}$ &$  1.3\substack{+0.5\\-0.3}$ & $  0.4\substack{+ 0.2\\ - 0.2}$   & --          & --         & --             & --        \\ 
2006aj            & Ic-BL   & $7.7 $      & $9.6 $       & $-5,15$   & 0.28$\substack{+0.03 \\ -0.02}$ &$  1.4\substack{+0.4\\-0.2}$ & $  2.7\substack{+ 0.8\\ - 0.6}$   & 0.21        & 2          & 2              & (13)        \\ 
2006el            & IIb     & $14.2-16.6$ & $17.8-19.5 $ & $-10,12$  & 0.13$\substack{+0.03 \\ -0.02}$ &$  3.3\substack{+1.1\\-1.0}$ & $  2.4\substack{+ 1.0\\ - 1.5}$   & --          & --         & --             & --       \\ 
2006ep\tnote{a}   & Ib      & $10.8-17.3$ & $15-20 $     & $-8,14$   & 0.06$\substack{+0.03 \\ -0.01}$ &$  2.7\substack{+1.3\\-1.3}$ & $  1.4\substack{+ 0.8\\ - 0.8}$   & --          & --         & --             & --        \\ 
2007C             & Ib      & $ 8.7-12.9$ & $12.5-17 $   & $-7,10$   & 0.17$\substack{+0.04 \\ -0.04}$ &$  1.9\substack{+0.7\\-0.9}$ & $  1.4\substack{+ 0.6\\ - 0.8}$   & --          & --         & --             & --        \\ 
2007Y             & Ib      & $10.9-14.9$ & $13.6-17 $   & $-7,11$   & 0.04$\substack{+0.01 \\ -0.00}$ &$  1.4\substack{+1.3\\-0.4}$ & $  0.7\substack{+ 0.7\\ - 0.3}$   & 0.06        & --         & --             & (14)    \\ 
2007gr            & Ic      & $10.9-12.9$ & $13.5-15 $   & $-5,9$   & 0.08$\substack{+0.01 \\ -0.01}$ &$  1.8\substack{+0.6\\-0.4}$ & $  1.1\substack{+ 0.5\\ - 0.4}$   & --          & --         & --             & --        \\ 
2007ru            & Ic-BL   & $ 6.8-11.2$ & $9-14 $      & $-3,14$   & 0.41$\substack{+0.05 \\ -0.06}$ &$  2.2\substack{+1.1\\-1.1}$ & $  4.7\substack{+ 2.4\\ - 2.5}$   & --          & --         & --             & --        \\ 
2007uy            & Ib      & $12.5-14.7$ & $15-16.5 $   & $-6,6$    & 0.28$\substack{+0.04 \\ -0.04}$ &$  3.3\substack{+1.1\\-1.0}$ & $  3.9\substack{+ 1.5\\ - 2.0}$   & --          & --         & --             & --        \\ 
2008D             & Ib      & $15.4$      & $17.9 $      & $-10,12$  & 0.09$\substack{+0.01 \\ -0.01}$ &$  2.9\substack{+1.0\\-0.6}$ & $  1.6\substack{+ 1.3\\ - 0.5}$   & 0.07--0.1   & 4.3--7     & 3.5--8.5       & (15,16) \\ 
2008ax            & IIb     & $16.7-17.0$ & $19.3-19.6 $ & $-8,13$   & 0.15$\substack{+0.05 \\ -0.03}$ &$  2.8\substack{+1.0\\-0.6}$ & $  0.9\substack{+ 1.1\\ - 0.4}$   & 0.09--0.12  & 2.2--3.2   & 0.7--1.7       & (17,18) \\ 
2009bb            & Ic-BL   & $ 8.5-10.0$ & $11.2-12.4 $ & $-5,10$   & 0.25$\substack{+0.04 \\ -0.03}$ &$  1.9\substack{+0.6\\-0.5}$ & $  3.3\substack{+ 2.2\\ - 1.0}$   & --          & --         & --             & --        \\ 
2009jf            & Ib      & $18.4-21.8$ & $20.5-22.5 $ & $-10,15$  & 0.24$\substack{+0.03 \\ -0.02}$ &$  4.7\substack{+1.7\\-1.1}$ & $  2.5\substack{+ 2.2\\ - 0.9}$   & --          & --         & --             & --        \\ 
2010bh            & Ic-BL   & $ 4.9$      & $8.2 $       & $-5,10$   & 0.17$\substack{+0.03 \\ -0.02}$ &$  0.9\substack{+0.2\\-0.2}$ & $  4.9\substack{+ 1.3\\ - 1.0}$   & --          & --         & --             & --        \\ 
2011bm            & Ic      & $28.3-30.4$ & $30.3-32.3 $ & $-10,13$  & 0.62$\substack{+0.09 \\ -0.08}$ &$ 10.1\substack{+2.8\\-2.3}$ & $  4.9\substack{+ 2.2\\ - 1.7}$   & --          & --         & --             & --        \\ 
2011dh            & IIb     & $15.5-16.0$ & $19.2-19.8 $ & $-8,13$   & 0.08$\substack{+0.02 \\ -0.02}$ &$  2.2\substack{+0.6\\-0.5}$ & $  0.7\substack{+ 0.4\\ - 0.3}$   & 0.07--0.08  & 1.8--2.5   & 0.6--1         & (19,20) \\ 
2011hs            & IIb     & $ 8.8-12.1$ & $11.4-14.4 $ & $-5,10$   & 0.04$\substack{+0.01 \\ -0.01}$ &$  1.0\substack{+0.8\\-0.3}$ & $  0.4\substack{+ 0.3\\ - 0.2}$   & 0.04        & 1.8--2.5   & 0.8--0.9       & (21) \\ 
iPTF13bvn         & Ib      & $12.5-13.4$ & $15.6-16.2 $ & $-9,10$   & 0.06$\substack{+0.02 \\ -0.01}$ &$  1.7\substack{+0.5\\-0.4}$ & $  0.7\substack{+ 0.3\\ - 0.2}$   & 0.05-0.10   & 1.9--2.3   & 0.7-0.9        & (22,23) \\                
\hline
\end{tabular}
\label{tab:sn_results_mod}
\begin{tablenotes}
\item[a] Only Galactic extinction is accounted for, thus the \mni{} value has a large upper uncertainty arising from the possibility of significant, unaccounted for, reddening.
\item[s] Spectral modelling of SNe, other references relate to hydrodynamical modelling.
\end{tablenotes}
\vspace{0.3cm}
(1) \citet{utrobin94}; (2) \citet{woosley94}; (3) \citet{young95}; 
(4) \citet{iwamoto94}; (5) \citet{young95}; (6) \citet{sauer06}\tnote{s}; 
(7) \citet{iwamoto98}; (8) \citet{nakamura01}; 
(9) \citet{stritzinger02}; 
(10) \citet{mazzali07}\tnote{s}; 
(11) \citet{mazzali09}\tnote{s}; 
(12) \citet{maeda07}\tnote{s}; 
(13) \citet{mazzali06}\tnote{s}; 
(14) \citet{stritzinger09}\tnote{s}; 
(15) \citet{mazzali08}\tnote{s}; (16) \citet{tanaka09}; 
(17) \citet{hachinger12}\tnote{s}; (18) \citet{maurer10}\tnote{s}; 
(19) \citet{bersten12}; (20) \citet{shivvers13}\tnote{s}; 
(21) \citet{bufano14}; 
(22) \citet{fremling14}; (23) \citet{bersten14}; 

\end{threeparttable}
\end{table*}

\begin{table*}
 \centering
 \caption{Average \vph{} and explosion parameters for SE~SN types}

  \begin{tabular}{c@{\hskip 1cm}cc@{\hskip 1cm}cc@{\hskip 1cm}cc@{\hskip 1cm}cc}
      SN type & \multicolumn{2}{c@{\hskip 1cm}}{\vph{} (\kms)} & \multicolumn{2}{c@{\hskip 1cm}}{\mni{} (\msun)} & \multicolumn{2}{c@{\hskip 1.0cm}}{\mej{} (\msun)} & \multicolumn{2}{c@{\hskip 1.0cm}}{\ek{} ($10^{51}$~ergs)} \\
              & mean & std.\ dev.\                             & mean & std.\ dev.\                              & mean & std.\ dev.\                                & mean & std.\ dev.\ \\
      \hline
       IIb   & 8300  & 750  & 0.11 & 0.04 & 2.2 & 0.8 & 1.0 & 0.6 \\
       Ib    & 9900  & 1400 & 0.17 & 0.16 & 2.6 & 1.1 & 1.6 & 0.9 \\
       Ic    & 10400 & 1200 & 0.22 & 0.16 & 3.0 & 2.8 & 1.9 & 1.3 \\
       Ic-BL & 19100 & 5000 & 0.32 & 0.15 & 2.9 & 2.2 & 6.0 & 5.0 \\
      \hline
\label{tab:sn_av_mod}
\end{tabular}
\end{table*}

\begin{table*}
 \centering
 \caption{Results of two-sample Kolmogorov-Smirnov test on explosion parameters between SE~SN types}

  \begin{tabular}{ll@{\hskip 1cm}cc@{\hskip 1cm}cc@{\hskip 1cm}cc}
      \hline  
Sample 1  & Sample 2  & \multicolumn{2}{c@{\hskip 1cm}}{\mni{}} & \multicolumn{2}{c@{\hskip 1cm}}{\mej{}}  & \multicolumn{2}{c@{\hskip 1cm}}{\ek{}} \\
          &           &     $D$        &    $p$    &      $D$        &    $p$    &     $D$       &    $p$    \\
\hline
IIb       & Ib        & 0.308          & 0.608     & 0.291           & 0.680     & 0.504         & 0.089        \\
IIb       & Ic        & 0.639          & {\em0.034}& 0.194           & 0.992     & 0.528         & 0.125        \\
IIb       & Ic-BL     & 0.764          & {\bf0.006}& 0.250           & 0.915     & 0.889         & $\mathbf{8\times10^{-4}}$   \\
Ib        & Ic        & 0.365          & 0.428     & 0.269           & 0.800     & 0.164         & 0.998        \\
Ib        & Ic-BL     & 0.596          & {\em0.034}& 0.289           & 0.727     & 0.798         & {\bf0.001}   \\
Ic        & Ic-BL     & 0.625          & {\em0.050}& 0.250           & 0.929     & 0.750         & {\bf0.010}   \\
\hline
\label{tab:sn_kstest_mod}
\end{tabular}
\end{table*}

\section{Discussion}
\label{sect:discuss}

The results of this study provide for the first time a large sample of bolometric light curves of SE~SNe, with which the nature of the explosions of various SNe types has been investigated.

\subsection{Bolometric light curves}
\label{sect:discuss_lightcurves}

The bolometric light curves of SE~SNe are diverse in \mpk{} and \dmbol{} within each subtype and as a whole. Plotting the decline rate and peak of each light curve indicates that early bolometric light curves alone cannot be used to reliably distinguish between SE~SN types (\cref{fig:m15peak}). Properties such as colour evolution remain a more promising avenue for distinguishing SE~SNe in the absence of spectral information \citep[e.g.][]{poznanski02,galyam04,bianco14} -- a scenario that will be probable for the vast majority of SNe discovered by future SN surveys. The apparent \mpk{}--\dmbol{} inverse correlation for SNe~Ic-BL (excluding SNe~2002ap and 2005kz) in \cref{fig:m15peak} is akin to the suggestion of a possible Phillips' relation (i.e. brighter SNe have wider light curves, \citealt{phillips93}) for GRB-SNe that is discussed in \citet{schulze14}. Any possible relation will merit study as the sample of such events grows (as well as investigating the reasons behind outliers, should a relation present itself). Three of the events that appear to sit along the supposed relation did not show evidence of an associated high energy component (2003jd, \citealt{valenti08}; 2007ru, \citealt{sahu09}; 2009bb, \citealt{pignata11}, although strong radio emission suggests the presence of relativistic material in this SN, \citealt{soderberg10}). The standardising of GRB-SN light curves has been further studied by \citet{cano14,li14}, who indeed find relations based on light curve properties to allow their cosmological use, confirming the indications from the relatively cruder analyses in \citet{schulze14} and here.

\subsection{Photospheric velocities}
\label{sect:discuss_vphot}

Although we use empirical measurements of \vph{} for all but one of our sample (as opposed to relying on averages or fiducial values as has been done in similar previous studies) in order to reduce systematic biases, the data set used and methods employed are inherently heterogeneous. 

In particular one aspect to consider is the nature of the feature in SE~SNe around 6200\AA{}. We attribute this to Si\II{}~$\lambda$6355 in order to determine \vph{} for $\sim 20$~per~cent of our sample, but this identification has been widely debated \citep[e.g][]{branch02, folatelli06, parrent07, tanaka09, hachinger12, parrent15}. These studies (and others) argue this feature may be explained as being due to unburnt hydrogen via the \Ha{} line, or some combination of \Ha{} and Si\II{}~$\lambda$6355. Other species have also been proposed as being responsible for this feature such as detached He{\sc i} \citep{clocchiatti96} or C\II{} \citep{elmhamdi06}.

The nature of the absorption feature around 6200\AA{} is uncertain. We present some comparisons of velocities in the literature when it is attributed to Si\II{}~$\lambda$6355, compared to \vph{} for the same SNe (determined from Fe\II{} lines or via spectral modelling), in order to assess the impact of this potential misidentification. 
\citet{galyam02} determine a Si\II{} velocity (15000~\kms{}) at peak for SN~2002ap that is in good agreement with the Fe\II{} velocity determined here (13000~\kms{}) considering the slight difference in epoch measured.\footnote{The \citet{galyam02} Si\II{} velocity is at $B$-band peak whereas ours is just after \Lbol{} peak.} \citet{folatelli06} consider the absorption feature to be mainly or wholly due to \Ha{} for SN~2005bf since they find find \vph{} $\sim$~7500~\kms{} around the first peak, whereas the velocity of Si\II{} would be $\sim$~4800~\kms{}. \citet{tanaka09} determine a \vph{} for SN~2008D via spectral modelling of 9000~\kms{} and determine the velocity of Si\II{} as 9300~\kms{} \citep[][also find a similar Si\II{} velocity]{soderberg08}. \citet{pignata11} find their Fe\II{} ($\lambda$4924, 5018, 5169) velocity to be `a good match' to the one determined for Si\II{}~$\lambda$6355 for SN~2009bb near peak.

It should be noted that 5 of 8 of our \vph{} measurements that rely on using a Si\II{}~$\lambda$6355 velocity are SNe~Ic-BL (since the high velocities cause strong blending of the Fe\II{} features). For the SNe~Ic-BL 2002ap and 2009bb, where Fe\II{} velocities could be measured, these are found to agree very well with those of Si\II{} if the 6200\AA{} feature is so attributed. There appears to be the presence of a small amount of hydrogen in a mean SN~Ib peak-light spectrum \citep[see][]{liu15}, which may mean the influence of \Ha{} on the 6200\AA{} feature is stronger for this subtype, however, we only use a Si\II{}-determined velocity for one SN~Ib. Although there clearly may be significant departures from the true \vph{} when this feature is attributed to Si\II{}~$\lambda$6355, we only rely on this assignment for $\sim~20$~per~cent of events and a number of those are likely to produce consistent velocities to their \vph{}. We thus do not consider this potential misidentification to significantly affect overall conclusions and results, although of course it is an additional source of uncertainty for these individual events.

In agreement with the new results of \citet{liu15}, based on their analysis of Fe\II{} $\lambda$5169 in a large number of SE~SNe spectra, and with consideration to the above caveats, we find an increasing average \vph{} for the subtypes following SNe~IIb~$\rightarrow$~Ib~$\rightarrow$~Ic although the significant spread in values within any one subtype means there exists significant overlap between SNe~IIb, Ib and Ic, with SNe~Ic-BL (unsurprisingly, given the nature of their classification) at higher velocities.

\subsection{Explosion parameters}
\label{sect:discuss_exp_params}

The results of the simple modelling presented here agree for the majority of events where more detailed analysis has been performed to extract explosion parameters. Where results differ, it is generally the case that values are lower than other modelling. Differences for some SNe are likely to be due to the simplifications in the analytical model, which does not account for, e.g., velocity or opacity evolution, asymmetries in the explosion, or the presence of extended envelopes.
Nevertheless, these results require just two-filter optical observations and a single spectrum of each SN (compared to multi-band UV/optical/NIR photometry and at least several epochs of spectral coverage, required for more detailed modelling). The bolometric light curve creation method used \citepalias{lyman14} also appears robust to all well-observed SE~SNe thus far and hence no large uncertainties on the results are being introduced through its use (for example, the typical error in a host distance modulus is larger than the $\simlt 0.1$~mag rms on the BC fits). Even so, it is imperative that detailed modelling of SNe with much larger data sets continues apace. This is required not only for the intrinsic in-depth knowledge of SN explosions such studies afford, but also to act as a basis for assessing the consistency of coarser methods such as this for the future of data-starved SN studies, especially when including increasing samples of unusual events.

An interesting result is the similarity between SNe~Ib and Ic in each of the parameters explored. The SNe are very similar in their bulk properties, i.e. the exploding cores of these SNe have similar masses and produce explosions with similar amounts of \ek{} and \mni{}. Exploding pre-SN stars producing SNe~Ib must have a non-negligible mass of helium.
However, the presence of helium in the progenitors of SNe~Ic has been debated. The predictions of conflicting theoretical results, which argue for \citep{dessart12, piro14} and against \citep{frey13} the presence of helium in SE~SNe have been investigated by \citet{liu15}. These authors conclude that empirical differences in the spectra of SNe~Ib and Ic are inconsistent with the predictions made by studies that suggest helium is hidden in SNe~Ic, instead favouring very little to no helium being present. This is in agreement with results of radiative transfer models including non-local thermodynamical equilibrium effects of SE~SNe \citep[see][and references therein]{hachinger12}, which suggest that only $\sim$~0.1~\msun{} is required in the progenitor star to produce an observable signature in the spectra.
Assuming then SNe~Ic lack any significant helium envelope mass, this would mean the carbon-oxygen (CO) core mass will be lower for a SN~Ib with the same \mej{} as a SN~Ic. Since CO core mass increases with zero-age main sequence (ZAMS) mass of a progenitor, it follows that the ZAMS mass of a SN~Ib would be expected to be lower than that of a Ic, for equal \mej{}. The apparent similarity in the \mej{} distributions shown here would then hint towards lower CO core masses (and thus lower ZAMS mass) for SNe~Ib compared to Ic. However, as mentioned, the helium mass required to produce a SN~Ib spectrum may be only $\sim$~0.1~\msun{} \citep{hachinger12}.
Considering this, alongside the comparatively large uncertainties on our \mej{} values, the distributions cannot be used to rule conclusively on any potential mass sequence (or lack of) in the ZAMS masses of SNe~Ib and Ic.

Despite the modest sample sizes (from a statistical viewpoint), the SNe~Ic-BL manifest themselves as very different in two of the three explosion parameters determined here. Their \mni{} and \ek{} values are much larger on average than the distributions of any of the other SN types. However, unlike the \ek{} distributions, where even the least energetic SN~Ic-BL is more energetic than the majority of SNe IIb, Ib and Ic, the largest \mni{} masses of SNe~Ic-BL are matched by those of SNe~Ib and Ic. This indicates the presence of broad lines is not a certainty when a large amount of \mni{} is synthesised (since we see SNe~Ib and Ic with comparable \mni{}), and thus the peak brightness is not a uniquely determining factor. Additionally, although the high velocity nature of SNe~Ic-BL naturally implies a large \ek{}/\mej{} ratio (\cref{eq:vsc}), these large \ek{}/\mej{} ratios are occurring at very similar \mej{} values of the other SN subtypes (SNe~Ic-BL are indistinguishable from the individual or combined IIb/Ib/Ic distribution), favouring an energy source that is decoupled from a dependence on the mass of the exploding core.

\subsection{Explosion parameter correlations}
\label{sect:correlations}

\begin{figure}
\centering
 \includegraphics[width=\columnwidth]{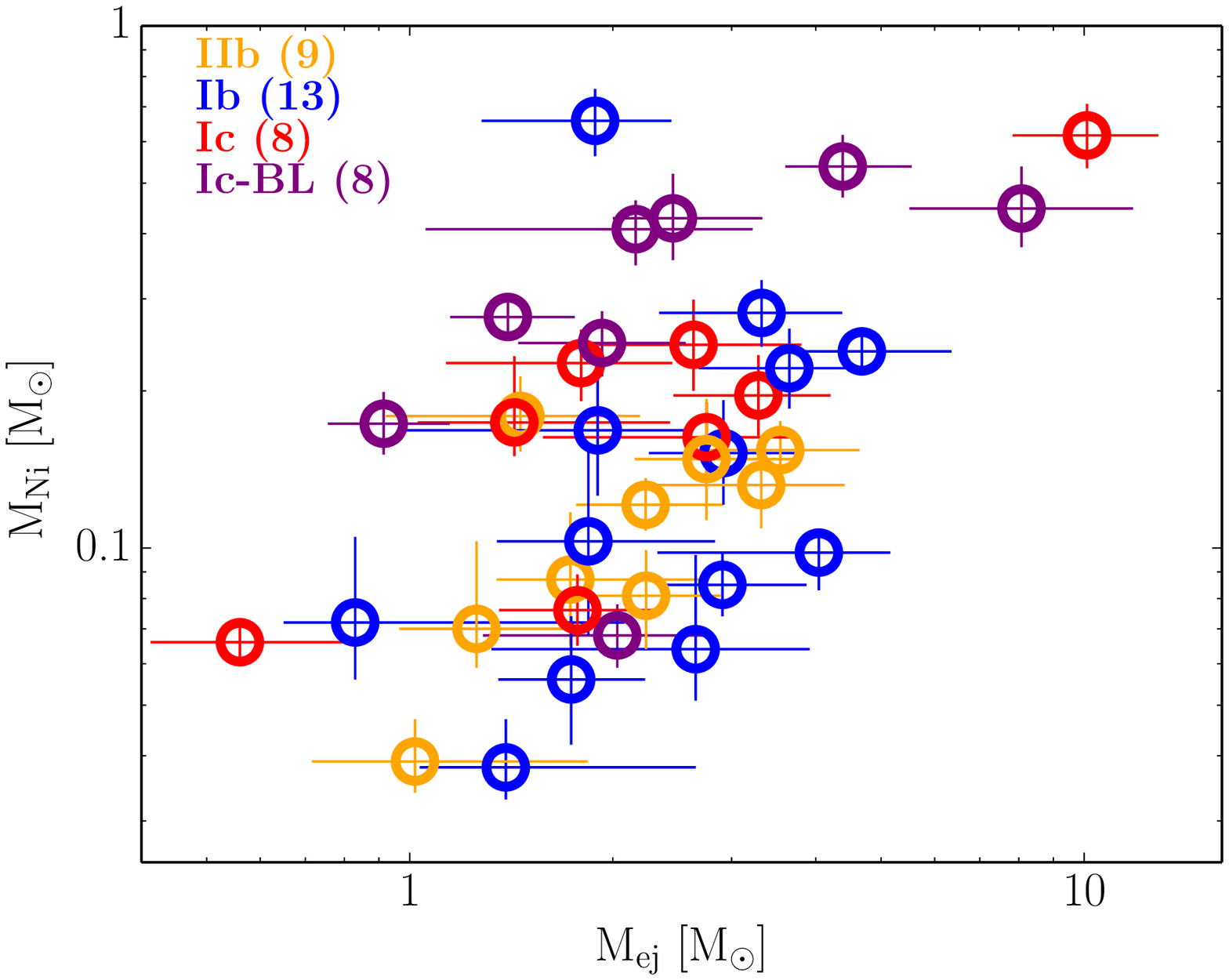}\\
  \includegraphics[width=\columnwidth]{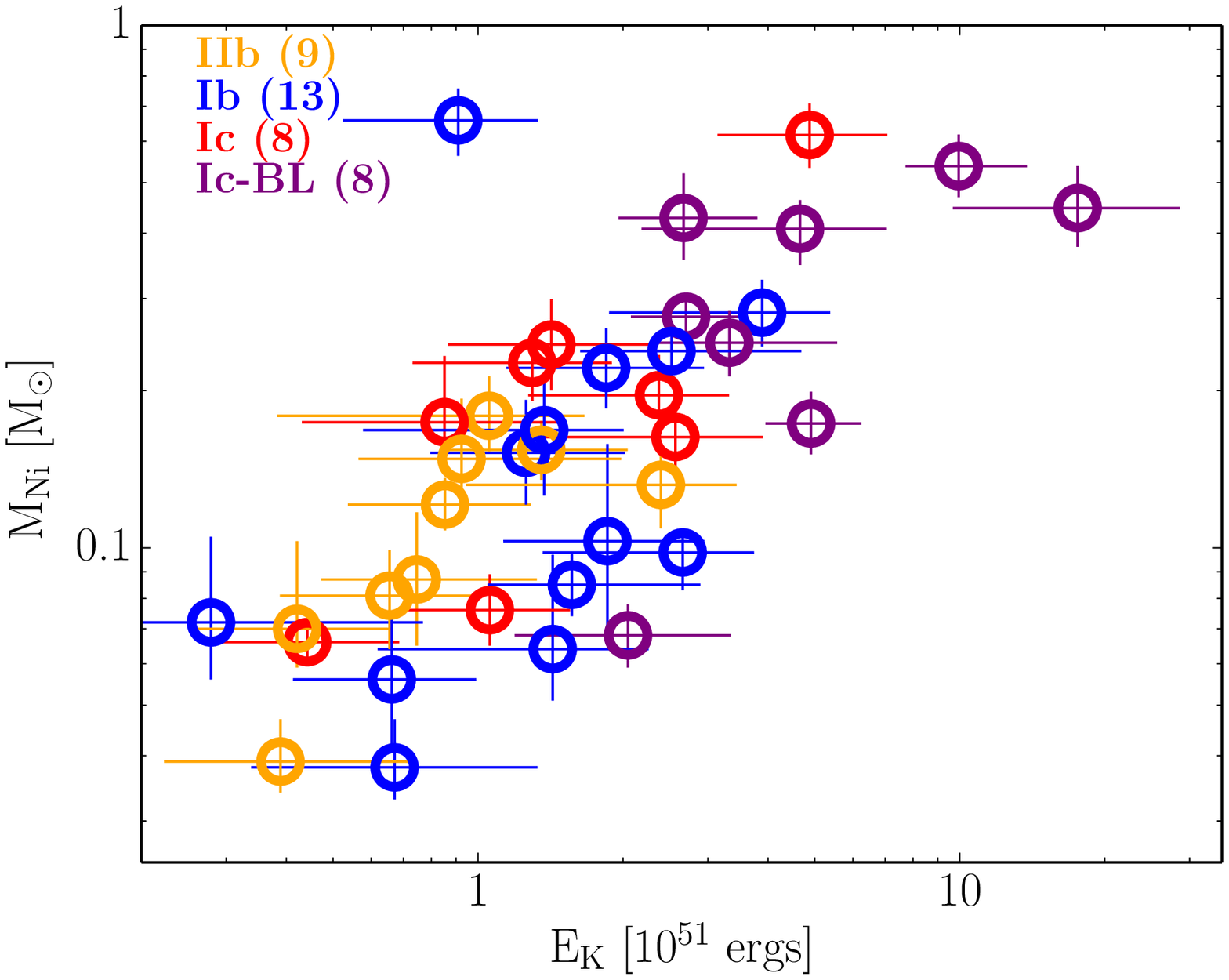}\\
   \includegraphics[width=\columnwidth]{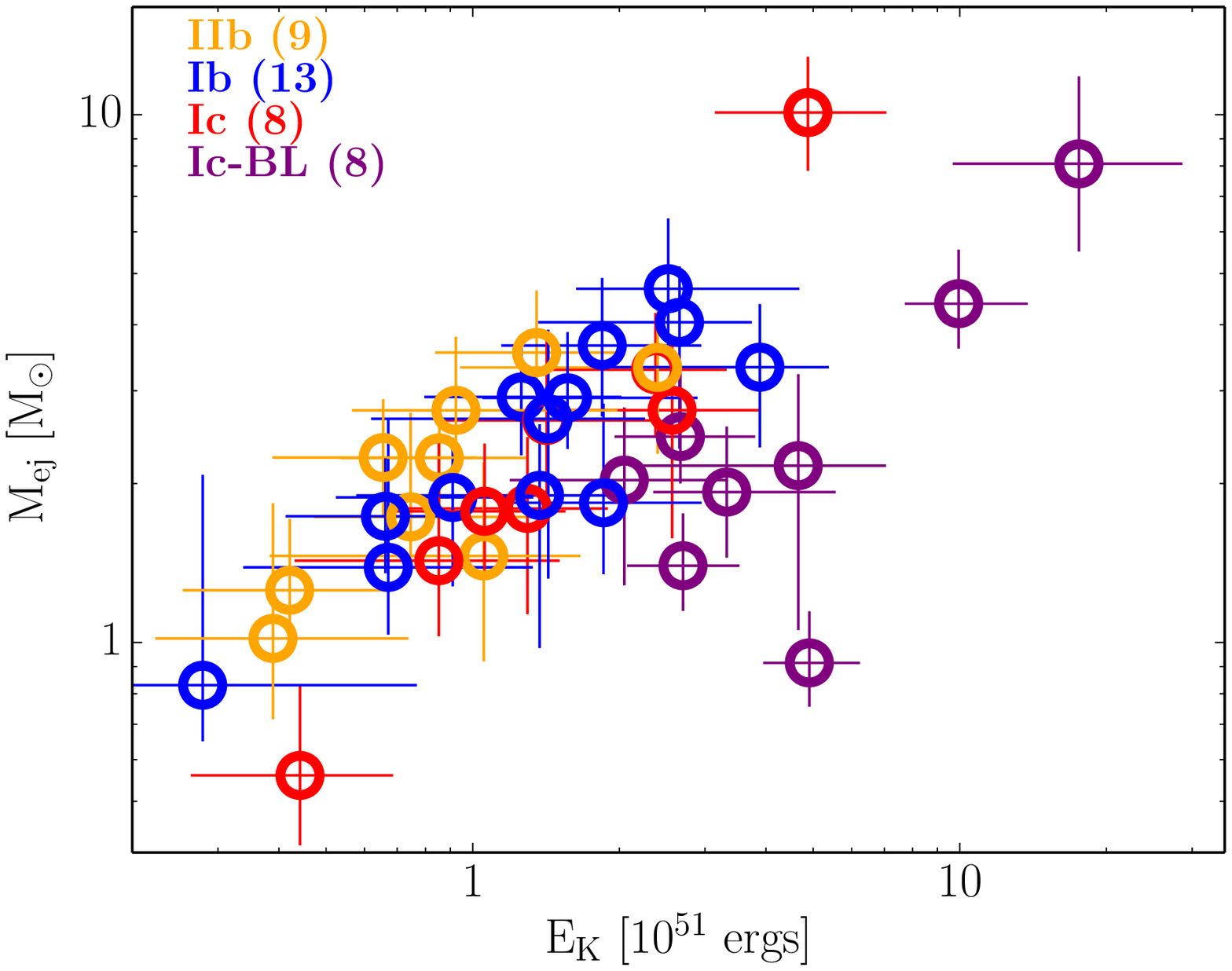}\\
\caption{Derived explosion parameters (\mni{}, \mej{} and \ek{}) of literature SE~SNe are plotted against each other. Data are given in \cref{tab:sn_results_mod}. SNe are colour-coded according to their type.}
\label{fig:sn_corr_mod}
\end{figure}

The parameters derived from the modelling are plotted against each other in \cref{fig:sn_corr_mod}. The bulk of SNe IIb, Ib and Ic appear to form a fairly tight correlation in the \mej{}--\ek{} plot, this is a result of the similar \vph{} values they exhibit (which, in turn, gives the \mej{}/\ek{} ratio). Conversely, SNe~Ic-BL, which can have very high velocities (\cref{tab:sn_vph_mod}), are found at larger \ek{}/\mej{} ratios, as dictated by \cref{eq:vsc}. Some splitting of SNe~Ic-BL occurs with the `hypernova branch' (i.e.\ very high \mej{} and \ek{} values, e.g. \citealt{mazzali13}) being populated by SNe~1998bw and 2005kz, whereas other SNe~Ic-BL sit at similar \mej{} values to other SN types, but with higher \ek{} values. SN~2011bm appears as an intermediate member of the hypernova branch in these plots, despite displaying very modest velocity, with \vph{}~$= 9000$~\kms{}. In this case the huge explosion parameter values found were due to the extremely slow evolution of the SN (\citealt{valenti12}, \cref{fig:m15peak}), and could point to an alternative signature of the explosion of a very massive star, perhaps without the angular momentum to produce an accretion-disk powered jet. Although SN~Ic-BL \mej{} values have a similar distribution to those of other SE~SN types, their \mni{} values (barring SN~2002ap) are much higher than the bulk of SE~SNe, indicating the production of \mni{} is much more efficient in these explosions.

SNe~IIb appear to be the most homogeneous subtype of SE~SNe as evident from the clustering of their bolometric light curve properties (\cref{fig:sn_bololc}) and explosion parameters (\cref{fig:sn_corr_mod}). This may be a result of a much more restrictive progenitor range for SNe~IIb \citep[e.g.][]{yoon10}.

\begin{figure}
\centering
 \includegraphics[width=\columnwidth]{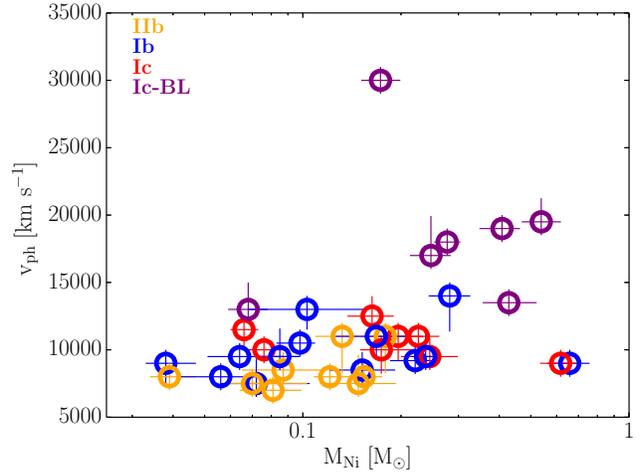}\\
\caption{The \mni{} and \vph{} values for SE~SNe found from the modelling presented here. Only SNe with a directly measured \vph{} are included. No clear indication of a single relation exists (cf. SNe~IIP, see \citealt{hamuy03} and \citealt{spiro14}). SNe are colour-coded according to their type.}
\label{fig:mni_vph}
\end{figure} 

\citet{hamuy03} show a correlation of increasing \ek{} with \mni{} for SNe~IIP and to a weaker extent this is also found for SE~SNe. Although there is a large amount of scatter and SN~2005hg is a prominent outlier, a correlation between \mni{} and \ek{} can be seen in \cref{fig:sn_corr_mod} -- no highly energetic and \mni{} deficient events are seen.
Similarly, a correlation between \mni{} and $v_{50}$ (\vph{} at 50~days after explosion) was shown for SNe~IIP by \citet{hamuy03}, which was found to extend to under-luminous SNe~IIP by \citet{spiro14}. \cref{fig:mni_vph} shows the analogous data for SE~SNe, with the velocity here being defined as that at peak light (note that only SNe with a directly measured \vph{} are included). There appears to be no strong dependence of \mni{} on \vph{}, and indeed the average behaviour of the SNe~IIb, Ib and Ic looks flat in \vph{} over a wide range of \mni{} values.

\subsection{The role of binaries as SE~SNe progenitors}
\label{sect:binaries}

\begin{figure}
\centering
 \includegraphics[width=\columnwidth]{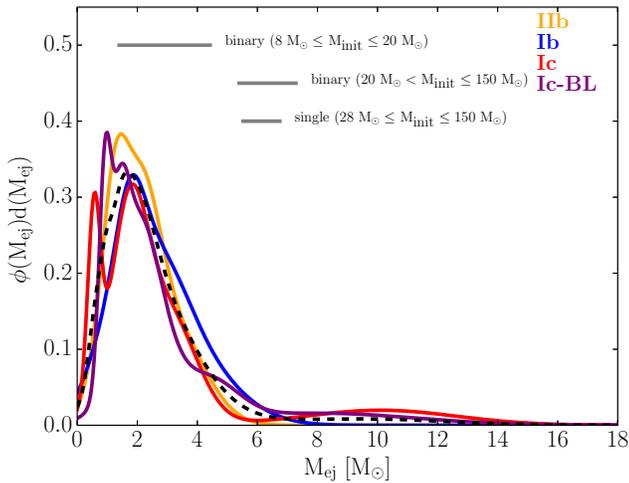}\\
\caption{Probability density functions of SE~SNe types for \mej{}, found by summing the individual SNe in each type assuming gaussian errors and normalising the integrated area to one. The entire sample is shown by the black dashed line. Note the lower \mej{} peak in the SN~Ic distribution occurs solely due to SN~1994I. Also plotted are the 1 standard deviation ranges for the \mej{} values for binary and single stars from the BPASS models at $Z=0.008$, values at solar metallicity are similar for M$_\textrm{init} \leq 20$~M$_\odot$ but larger on average for more massive stars. Note the values for \mej{} of stars M$_\textrm{init} > 20$~\msun{} are conservative, and are likely to extend to higher values (see text, \cref{fig:bpass_mej}).}
\label{fig:mej_pdf}
\end{figure}

The question of whether massive single stars or binary systems are the progenitors of SE~SNe is an area of active debate. The value of \mej{} for a SN can be used to infer the nature of the progenitor system, by comparing to results from stellar evolution models. The \mej{} values for each SN type were summed over the probability density functions of the \mej{} values of individual SNe (assuming gaussian errors) and normalised to give overall probability density functions for the SN types and entire sample. These functions are presented in \cref{fig:mej_pdf}, with \mej{} values of 1--3~\msun{} dominating. These low \mej{} values are incompatible with the distributions of \mej{} expected from massive, single WR stars ($Z$ = 0.008 and 0.02). In \cref{fig:mej_pdf} the observed distribution is compared to the stellar evolution models used in the binary population and spectral synthesis (BPASS) code \citep{eldridge08,eldridge09}.\footnote{\url{http://www.bpass.org.uk/}} The single star models in this set give no \mej{} values lower than 5~\msun{}. 
The spread of \mej{} values for stars of $20$~M$_\odot <$ M$_\textrm{init} \leq 150$~M$_\odot$ are shown in \cref{fig:mej_pdf} -- note although these seem low compared to some of the high \mej{} SNe, these are conservative estimates, as discussed later, meaning the distribution is likely to extend to larger \mej{} masses. Similarly large \mej{} values for single stars were found by \citet{groh13c}, who show for a initial progenitor mass of $\sim 30$~\msun{}, a SN~Ib/c has $\simeq 7-8$~\msun{} of material beyond the remnant mass upon explosion. One must invoke more moderately massive progenitors in binary systems in order to reproduce the observed \mej{} values. 
Binary models from the BPASS code with initial primary masses of 8 to 20~\msun{} produce \mej{} values that are in better agreement with the range of the observed distribution (\cref{fig:mej_pdf}).
More massive progenitors evolving in binaries converge on similar \mej{} values as single stars. Thus, although large \mej{} events such as SN~2011bm cannot be distinguished as residing in a binary or not from this analysis, the probability density functions show that moderately massive ($8$~M$_\odot \leq$ M$_\textrm{init} \leq 20$~M$_\odot$) binary progenitors are not only a necessary progenitor channel for each SE~SNe type, but also that they dominate; only 5/38 of the sample are consistent with \mej{}~$>$~5~\msun{}. Such a result is in agreement with the findings of \citet{eldridge13}. Furthermore, the selection effects associated with discovering and characterising SNe should make these high mass events favourable to observation (i.e. broad light curves) compared to the narrower, faster-declining SNe with lower-mass progenitors. This gives confidence that a population of high mass progenitor SNe are not being missed from the current SN survey strategies.
Although we again stress that some \mej{} values may be underestimated with this simple modelling scheme, even when considering only values in \cref{tab:sn_results_mod} from more detailed modelling, the same arguments hold: the majority of SNe having \mej{} $\sim$~1--3~\msun{} and only 2/11 with \mej{}~$\simgt$~5~\msun{}.

\begin{figure}
\centering
 \includegraphics[width=\columnwidth]{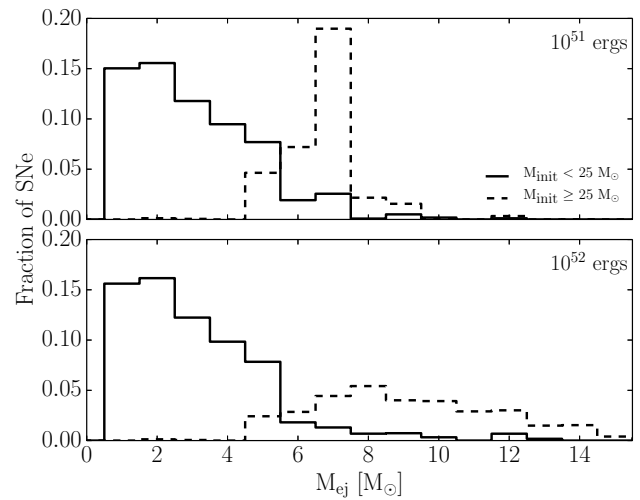}\\
\caption{BPASS model predictions for the \mej{} distribution of SE~SNe weighted by the stellar IMF. The energy of the SN used was $10^{51}$~ergs (top panel) and $10^{52}$~ergs (bottom panel). Increasing the explosion energy has minimal effect on lower mass stars (solid lines), since even the lower energy case essentially unbinds all the outer core beyond a 1.4~\msun{} compact remnant. For very massive stars (dashed lines), the additional energy allows more of the core to be ejected during the explosion, increasing the average \mej{} and spread. This distribution is a good match to the high \mej{} tail of the observed \mej{} distribution (\cref{fig:mej_pdf}), however it also contributes a much larger fraction of all SNe than is observed (see text).}
\label{fig:bpass_mej}
\end{figure} 

The model \mej{} values for the more massive stars shown in \cref{fig:mej_pdf} are conservative estimates. These were found using a canonical SN explosion energy of $10^{51}$~ergs with a simple treatment of integrating the binding energy of the envelope inwards until it reached this explosion energy or to a point where the interior mass was 1.4~\msun{}, which we take as the minimum mass of the compact remnant. However, this energy is somewhat modest compared to most SN, in particular for larger \mej{} SNe which exhibit $\simgt$ 3--10 $\times 10^{51}$~ergs. The effect of increasing the unbinding energy by a factor ten is shown in \cref{fig:bpass_mej} for a range of binary models over all masses. This has little influence on the lower mass progenitors where the ejecta mass is roughly that of the final mass minus 1.4~\msun{} for each energy used, and the spread of \mej{} values from the models remains in good agreement with the observed peak around 2~\msun{}. For more massive stars, the extra unbinding energy is able to liberate more of the outer core, reducing the size of the compact remnant (the effect is qualitatively the same for single massive stars).
An unbinding energy of $10^{52}$~ergs produces a spread of \mej{} values that covers the observed very high \mej{} events, and dilutes the strong peak at \mej{} $\sim$ 7~\msun{}, which is not observed. 
In order to determine the fraction of SNe producing each bin of \mej{}, the models were weighted by the IMF \citep{kroupa01}. As can be seen, this treatment predicts that high \mej{} events constitute a larger fraction of events than the observed distribution; for \mej{} $> 5$~\msun{}, the integrated observed probability density function is $\sim$10~per~cent, whereas the $10^{52}$~ergs models predict 35~per~cent. Thus, although the spread of model \mej{} values are good representations of the observed distribution, the quantitative divide between low and high \mej{} events is inconsistent with expectations from the IMF, and something must act to suppress the observability of SNe from progenitors that would otherwise produce large \mej{} explosions. Fall-back SNe, in which little or none of the mass is ejected or direct collapse to a black hole for very massive pre-SN progenitors could be possible solutions \citep[e.g.][]{woosley93,fryer99,heger03,fryer09,kochanek14}. This discrepancy will warrant further investigation with larger, more homogeneously-selected observed samples and improved modelling.

One may expect very low \mej{} systems to be more abundant, given some proportionality between the initial mass of the star and the exploding core mass, and considering the shape of the stellar IMF. For example, many more stars with final core masses of $\sim 2$~\msun{} (producing \mej{} $\sim 0.6$~\msun{}) are produced per galaxy than stars with final core masses $\sim 4-5$~\msun{}(producing \mej{} $\sim 2.6-3.6$~\msun{}). This is in contrast to our observations of a strong peak for SE~SNe at \mej{}~$\sim 2-3$~\msun{} and a dearth of low \mej{} values, \simlt{}~$1$~\msun{} (\cref{fig:mej_pdf}), although it should be noted that we are observationally biased against such quickly-evolving SNe, especially in regards to requiring observations at, or prior to, peak.
Low mass He-stars, however, are formed from stars with ZAMS masses at the low end of the range for a CCSN. It is very difficult to remove the hydrogen envelope in such low mass stars under normal circumstances \citep[e.g.][]{yoon10,eldridge13}, and they will explode as hydrogen-rich SNe, and thus not form part of this sample by definition.
Despite this, rapidly-fading SNe such as, e.g., SN~2005ek \citep[][see also \citealt{drout14}]{drout13} may represent very low \mej{} systems, which observationally are SE~SNe, indicating hydrogen-deficient explosions can occur within the lower mass range; modelling of `ultra-stripped' cores producing low \mej{} SNe~Ic as an explanation for these events has been performed by \citet{tauris13}. However, considering binary evolution of the progenitors, a second binary mass transfer episode can occur in the later stages of stellar evolution. This mass transfer can occur during helium core or shell burning for low mass He-stars, which would otherwise ostensibly produce a CCSN \citep{habets86} -- this can remove enough mass to prevent an explosion and instead result in a white dwarf \citep[e.g.][]{delgado81,law83,dewi02,podsiadlowski04}. He-stars with M \simgt{} 4~\msun{} do not grow to a `red giant' phase and do not undergo this mass transfer \citep{paczy71,delgado81,law83}. The more restrictive mass range in which a low mass He-star can retain a core mass above the Chandrasekhar mass after undergoing mass transfer may provide additional theoretical support to explain the lack of low \mej{} values, and the peak at \mej{} $\sim 2-3$~\msun{} is then attributed to more massive He-stars, which do not undergo this mass transfer. 

An outstanding issue is the use of a constant optical opacity, and its value, when performing analytical modelling for SE~SNe. The values of \mej{} are inversely proportional to the choice of \kopt{}, and so we can ask what the value of \kopt{} would need to be in order to shift the \mej{} distribution to the point where it becomes consistent with the \mej{} values of very massive stars (\simgt{}~20--25~\msun{}). From \cref{fig:mej_pdf,fig:bpass_mej}, a factor of 3 increase in the peak of the \mej{} distribution would place it at the lower bound of very massive star ejecta masses. This corresponds to using \kopt{}~$= 0.02$~cm$^2$~g$^{-1}$. Although this is comparable to the suggested value for \kopt{} from the study of \citet{wheeler15}, this is outside the bounding range, $\sim$0.04--0.1~cm$^2$~g$^{-1}$, that is typically found via detailed modelling \citep[e.g.][Mazzali et al., in prep]{mazzali00, chugai00, mazzali13}. We again further reiterate the inconsistency of the distribution of \mej{} with very massive star ejecta predictions (\cref{fig:bpass_mej}, \citealt{groh13c}) solely from results of other modelling (\cref{tab:sn_results_mod}), where more careful prescriptions of the opacity are made.

\subsection{The progenitors of SNe~Ic-BL and GRB-SNe}

SN~Ic-BL and GRB-SN have been suggested to be more massive (younger) than their lower velocity counterparts \citep[e.g][]{larsson07,raskin08,svensson10,sanders12,cano13}. Unfortunately, the sample of confirmed GRB-SN to which this method is applied is limited to three low-redshift events, SNe~1998bw, 2006aj and 2010bh, with the high energy component of SN~2006aj being an X-ray flash. Notwithstanding having only three objects in the sample, when extracted from the SN~Ic-BL sample, they are inconsistent with SNe~IIb, Ib and Ic distributions in \ek{} (\ks{} reveals $p \sim 1-2$ per cent), but cannot be distinguished in \mej{} or \mni{}. They also cannot be distinguished from the remaining SN~Ic-BL sample (5 events).  
\mej{} values of SNe~Ic-BL/GRB-SNe and SNe IIb, Ib and Ic, are indistinguishable, indicating SNe~Ic-BL/GRB-SNe have similar exploding core masses as other SN types, unless a large fraction of the core mass is not being ejected due to fall-back onto a compact remnant in SNe~Ic-BL/GRB-SNe. A complication of disentangling GRB-SNe is the prospect of off-axis jets, which would be missed; although radio detections can inform on the presence of strongly relativistic material (e.g. SN~2009bb, \citealt{soderberg10}) and potentially infer an off-axis jet, current detection limits, the prospect of other radio-emitting mechanisms, and the overlap between relativistic and non-relativistic SN radio light curves currently makes this very difficult \citep{bietenholz14}.

The extreme nature of GRB-SNe (and, to a lesser extent, SNe~Ic-BL) means the \ek{} and \mej{} estimates here may be underestimates as we make no account of the contribution from a denser, inner core of material that will reveal itself only in the late time light curves, and which may have a significant contribution in SNe~Ic-BL \citep{maeda03}. Indeed for SN~1998bw, for which we can compare results to more detailed modelling, we find a lower \mej{} and \ek{}, although for SN~2002ap our estimates are in reasonable agreement.

Other studies of GRB-SNe that have extracted explosion parameters have found similarly remarkable SNe accompanying the high-energy burst\footnote{Although there are two examples (GRBs 060505 and 060614) for which deep limits preclude all but extremely faint accompanying supernovae \citep[e.g.][]{fynbo06,dellavalle06}}. For example, SN~2003lw/GRB031203 was found to be best described by an explosion with \mni{}$\sim 0.55$~\msun{}, \mej{}$\sim 13$~\msun{} and \ek{}$\sim 60\times10^{51}$~ergs \citep{mazzali06b}, and SN~2012bz/GRB120422A had \mni{}$\sim 0.4-0.6$~\msun{}, \mej{}$\sim 6-7$~\msun{} and \ek{}$\sim 35\times10^{51}$~ergs \citep{melandri12,schulze14}. Both these events populate the hypernova subset of SNe~Ic-BL. Inclusion of such spectacular GRB-SNe would only serve to further distinguish them from `normal' SE~SNe and may begin to distinguish them from the more modest SN~Ic-BL \citep[see][]{cano13}. \citet{walker14} give explosion parameters for all SNe~Ic-BL from the literature, including several that did not meet the selection criteria for this sample, as well as those found for a new object, PTF~10qts. Their collection of explosion parameters generally agrees with those presented here for overlapping events. The \mej{} values display a similar distribution to that seen here for all SE~SNe (\cref{fig:mej_pdf}), i.e. predominantly events with a few \msun{} of ejecta, and a smaller fraction displaying much larger \mej{} that is indicative of a higher ZAMS mass progenitor (M$_\textrm{init}\simgt 25-30$~\msun{}).

\section{Summary}

A large sample of SE~SN bolometric light curves has been made though the use of BCs presented in \citetalias{lyman14}. Peak bolometric absolute magnitudes range from $-16.3$ to $-19.2$~mag, with both luminosity extremes occupied by a SN~Ib. \dmbol{} values range from 0.20 to 1.37~mag, with SNe~Ic making up the extremes of this distribution. The possibility of a Phillips-type relation for GRB-SNe, suggested by \citet{schulze14} and independently confirmed by \citet{cano14,li14}, is evident here for the bolometric light curves of the majority of the SNe~Ic-BL sample. 

The bolometric light curves were modelled using an analytical prescription utilising the velocity of the photosphere at peak light.
When directly comparing to other detailed modelling, there is general agreement in most parameters, but there are notable exceptions. For the cases where there is disagreement, limitations and assumptions in this modelling are likely to be compromising a good agreement by not accounting for the true nature of the explosion (e.g. extended supergiant progenitors, or strongly evolving photospheric velocities). 
Nevertheless, similar analysis on large numbers of SNe with relatively little follow-up can be used to further analyse populations of SE~SNe. We again stress the importance of detailed study of observationally favourable SNe to further quantify potential uncertainties arising from such a simple treatment of the explosions. 
Of great importance is to further test how valid the assumptions in such analytical models are for larger numbers of SNe, particularly in relation to using a single value to characterise each of the photospheric velocity and the opacity.

The extreme nature of SNe~Ic-BL was shown, with their \mni{} and \ek{} distributions being distinct from other SE~SNe types. Conversely, the \mej{} values for SNe~Ic-BL are very similar to those of SNe~IIb, Ib and Ic. When specifically comparing to SNe~Ic in \mej{} (i.e. where \mej{} will be that of the CO core minus the mass locked in a compact remnant, although see discussion in \cref{sect:discuss_exp_params}), it appears the mass of the core does not play a major role in determining the presence of broad-line features (i.e. large \vph{}), and this must be dictated by another property of the core (e.g. composition or angular momentum).

\mej{} values from all SN subtypes peak around 2~\msun{}; this is inconsistent with massive single star models, which predict \mej{} values $>$~5~\msun{}. Conversely, the introduction of a \emph{dominant} binary population of moderate mass progenitors ($8$~M$_\odot \leq$ M$_\textrm{init} \leq 20$~M$_\odot$) for SE~SNe explains the \mej{} distributions extremely well. This is additional support to direct imaging studies that appear to favour lower mass binary progenitors (e.g. SNe 1993J, 2011dh and iPTF~13bvn). The lack of very low \mej{} values also agrees with He-star binary evolution modelling, in which these low mass systems instead become a white dwarf due to mass transfer, or retain a hydrogen envelope and as such would not explode as SE~SNe. The lack of large \mej{} events is somewhat at odds from predictions of stellar models with a simple weighting from the IMF. Fall-back SNe or direct collapse to a black hole for very massive stars may alleviate this discrepancy by reducing the observability of the SNe of such stars. We additionally note that these arguments are also valid when considering only those \mej{} values derived from more detailed modelling (see \cref{tab:sn_results_mod}), in that $\sim$2 out of 11 SNe have large \mej{} determinations ($\simgt 5$~\msun{}) with the rest around 1--3~\msun{}.

The current small sample of GRB-SNe analysed here cannot be distinguished from SNe~Ic-BL. SNe IIb, Ib and Ic are all similar in each of the explosion parameters analysed, with some indication that SNe~IIb, the most homogeneous subtype in bolometric properties, are \mni{} deficient and have lower \ek{} values compared to SNe Ib and Ic. The average \mni{} and \ek{} values follow the same sequence of increasing value of IIb~$\rightarrow$~Ib~$\rightarrow$~Ic~$\rightarrow$~Ic-BL, however the average \mej{} values are very similar amongst the subtypes.

\section*{Acknowledgements}

We thank the referee for suggestions and comments that led to an improved manuscript. A peak-light spectrum of SN~2007uy and photometric data of iPTF13bvn were kindly provided by Rupak Roy and Christoffer Fremling, respectively. Philipp Podsiadlowski, Andrew Levan and Felipe Olivares are thanked for helpful discussions and comments. This research has made use of the NASA/IPAC Extragalactic Database (NED) which is operated by the Jet Propulsion Laboratory, California Institute of Technology, under contract with the National Aeronautics and Space Administration. The Weizmann interactive supernova data repository (www.weizmann.ac.il/astrophysics/wiserep) was used to obtain SN spectra. This research has made use of the CfA Supernova Archive, which is funded in part by the National Science Foundation through grant AST 0907903. JDL acknowledges support from the UK Science and Technology Facilities Council (grant ID ST/I001719/1) and the Leverhulme Trust, as part of a Philip Leverhulme Prize award. This work was partly supported by the European Union FP7 programme through ERC grant number 320360.

\newcommand{\araa}{ARA\&A}   \newcommand{\aap}{A\&A}
\newcommand{\aj}{AJ}         \newcommand{\apj}{ApJ}
\newcommand{\apjl}{ApJ}      \newcommand{\apjs}{ApJS}
\newcommand{\mnras}{MNRAS}   \newcommand{\nat}{Nature}
\newcommand{\pasj}{PASJ}     \newcommand{\pasp}{PASP}
\newcommand{\procspie}{Proc.\ SPIE} \newcommand{\physrep}{Phys. Rep.}
\newcommand{\apss}{APSS}
\newcommand{\solphys}{Sol. Phys.}
\newcommand{\actaa}{Acta Astronom}
\newcommand{\aaps}{A\&A Supp}
\newcommand{\iaucirc}{IAU Circular}
\newcommand{\pasa}{PASA}
\bibliographystyle{mn2e}
\bibliography{/home/jdl/references}

\appendix

\section{Template bolometric light curves}
\label{sect:templatelc}

\cref{tab:sn_temp} shows the data for the template bolometric light curves of the various SN types shown in \cref{fig:lc_templates}. The phases are with respect to the peak of \Lbol{} and the median luminosity and standard deviation are given (as determined from the spread of luminosities of SNe that have data at that phase).

\begin{table*}
\centering
\tiny
  \caption{Template bolometric light curve data for SE~SNe}
 \begin{tabular}{c@{\hskip 0.5cm}cc@{\hskip 0.5cm}cc@{\hskip 0.5cm}cc@{\hskip 0.5cm}cc}
\hline\\
      &\multicolumn{2}{c@{\hskip 0.5cm}}{{\normalsize IIb}}&\multicolumn{2}{c@{\hskip 0.5cm}}{{\normalsize Ib}}&\multicolumn{2}{c@{\hskip 0.5cm}}{{\normalsize Ic}}&\multicolumn{2}{c@{\hskip 0.5cm}}{{\normalsize Ic-BL}}\\ 
Phase & $\log_{10}$\Lbol{} & std.\ dev.\ & $\log_{10}$\Lbol{} & std.\ dev.\ & $\log_{10}$\Lbol{} & std.\ dev.\ & $\log_{10}$\Lbol{} & std.\ dev.\ \\
(days)&   (erg~s$^{-1}$)     &          &   (erg~s$^{-1}$)     &          &   (erg~s$^{-1}$)     &          &   (erg~s$^{-1}$)     &           \\
\hline
 -18 &  42.155 &   0.560 &  41.582 &   0.317 &  --     &   --    &  --     &   --    \\ 
 -17 &  42.113 &   0.452 &  41.810 &   0.264 &  --     &   --    &  --     &   --    \\ 
 -16 &  42.073 &   0.339 &  41.865 &   0.214 &  --     &   --    &  --     &   --    \\ 
 -15 &  41.899 &   0.281 &  41.924 &   0.177 &  --     &   --    &  --     &   --    \\ 
 -14 &  41.888 &   0.260 &  41.996 &   0.159 &  --     &   --    &  --     &   --    \\ 
 -13 &  41.965 &   0.211 &  41.965 &   0.340 &  --     &   --    &  --     &   --    \\ 
 -12 &  41.956 &   0.166 &  42.042 &   0.332 &  --     &   --    &  --     &   --    \\ 
 -11 &  42.038 &   0.144 &  42.108 &   0.324 &  --     &   --    &  --     &   --    \\ 
 -10 &  42.119 &   0.130 &  42.163 &   0.313 &  --     &   --    &  --     &   --    \\ 
  -9 &  42.157 &   0.202 &  42.209 &   0.302 &  --     &   --    &  --     &   --    \\ 
  -8 &  42.178 &   0.198 &  42.397 &   0.387 &  42.506 &   0.185 &  --     &   --    \\ 
  -7 &  42.233 &   0.183 &  42.309 &   0.356 &  42.498 &   0.264 &  --     &   --    \\ 
  -6 &  42.276 &   0.172 &  42.394 &   0.352 &  42.634 &   0.231 &  --     &   --    \\ 
  -5 &  42.314 &   0.166 &  42.417 &   0.352 &  42.580 &   0.258 &  42.883 &   0.281 \\ 
  -4 &  42.342 &   0.162 &  42.397 &   0.339 &  42.601 &   0.236 &  42.923 &   0.254 \\ 
  -3 &  42.428 &   0.184 &  42.450 &   0.328 &  42.581 &   0.205 &  42.926 &   0.254 \\ 
  -2 &  42.440 &   0.182 &  42.460 &   0.329 &  42.595 &   0.197 &  42.923 &   0.254 \\ 
  -1 &  42.447 &   0.181 &  42.467 &   0.329 &  42.603 &   0.193 &  42.915 &   0.254 \\ 
   0 &  42.450 &   0.181 &  42.469 &   0.329 &  42.641 &   0.181 &  42.903 &   0.254 \\ 
   1 &  42.448 &   0.181 &  42.467 &   0.329 &  42.634 &   0.182 &  42.887 &   0.255 \\ 
   2 &  42.442 &   0.181 &  42.460 &   0.329 &  42.625 &   0.184 &  42.869 &   0.254 \\ 
   3 &  42.433 &   0.182 &  42.448 &   0.329 &  42.615 &   0.189 &  42.849 &   0.254 \\ 
   4 &  42.404 &   0.183 &  42.432 &   0.329 &  42.603 &   0.194 &  42.828 &   0.254 \\ 
   5 &  42.370 &   0.184 &  42.413 &   0.330 &  42.590 &   0.201 &  42.806 &   0.254 \\ 
   6 &  42.333 &   0.186 &  42.390 &   0.330 &  42.577 &   0.209 &  42.784 &   0.253 \\ 
   7 &  42.295 &   0.188 &  42.365 &   0.331 &  42.561 &   0.217 &  42.763 &   0.253 \\ 
   8 &  42.256 &   0.190 &  42.338 &   0.331 &  42.545 &   0.226 &  42.741 &   0.252 \\ 
   9 &  42.219 &   0.192 &  42.312 &   0.332 &  42.528 &   0.235 &  42.714 &   0.252 \\ 
  10 &  42.185 &   0.194 &  42.286 &   0.332 &  42.510 &   0.244 &  42.689 &   0.251 \\ 
  11 &  42.155 &   0.196 &  42.261 &   0.333 &  42.491 &   0.253 &  42.665 &   0.251 \\ 
  12 &  42.129 &   0.197 &  42.238 &   0.333 &  42.472 &   0.262 &  42.642 &   0.251 \\ 
  13 &  42.106 &   0.198 &  42.215 &   0.332 &  42.446 &   0.271 &  42.621 &   0.252 \\ 
  14 &  42.088 &   0.199 &  42.192 &   0.332 &  42.418 &   0.279 &  42.600 &   0.253 \\ 
  15 &  42.073 &   0.199 &  42.169 &   0.331 &  42.390 &   0.287 &  42.605 &   0.272 \\ 
  16 &  42.060 &   0.199 &  42.144 &   0.330 &  42.363 &   0.294 &  42.579 &   0.275 \\ 
  17 &  42.049 &   0.199 &  42.126 &   0.340 &  42.338 &   0.301 &  42.554 &   0.279 \\ 
  18 &  42.038 &   0.200 &  42.103 &   0.339 &  42.313 &   0.307 &  42.530 &   0.283 \\ 
  19 &  42.026 &   0.200 &  42.139 &   0.337 &  42.290 &   0.313 &  42.508 &   0.288 \\ 
  20 &  42.014 &   0.201 &  42.117 &   0.336 &  42.269 &   0.318 &  42.486 &   0.293 \\ 
  21 &  42.001 &   0.202 &  42.096 &   0.333 &  42.249 &   0.323 &  42.465 &   0.297 \\ 
  22 &  41.986 &   0.203 &  42.132 &   0.348 &  42.230 &   0.327 &  42.445 &   0.299 \\ 
  23 &  41.948 &   0.182 &  42.112 &   0.346 &  42.211 &   0.331 &  42.464 &   0.293 \\ 
  24 &  41.935 &   0.183 &  42.093 &   0.344 &  42.194 &   0.334 &  42.443 &   0.297 \\ 
  25 &  41.922 &   0.185 &  42.075 &   0.342 &  42.177 &   0.338 &  42.456 &   0.166 \\ 
  26 &  41.911 &   0.188 &  42.058 &   0.340 &  42.161 &   0.340 &  42.434 &   0.170 \\ 
  27 &  41.901 &   0.190 &  42.042 &   0.338 &  42.145 &   0.343 &  42.413 &   0.173 \\ 
  28 &  41.892 &   0.193 &  42.027 &   0.336 &  42.128 &   0.345 &  42.393 &   0.176 \\ 
  29 &  41.885 &   0.196 &  42.013 &   0.334 &  42.112 &   0.347 &  42.373 &   0.179 \\ 
  30 &  41.878 &   0.199 &  41.999 &   0.333 &  42.096 &   0.349 &  42.335 &   0.140 \\ 
\end{tabular}
\label{tab:sn_temp}
\end{table*}

\label{lastpage}

\end{document}